\documentclass[twocolumn]{aastex63}

\usepackage[utf8]{inputenc}
\usepackage{natbib}
\usepackage[scaled]{helvet} 
\usepackage{graphicx}
\usepackage{txfonts} 
\usepackage{epsfig}
\usepackage{longtable}
\bibpunct{(}{)}{;}{a}{}{,}
\usepackage[normalem]{ulem} 
\usepackage{xspace}   

\newcommand{\Bcep}{$\beta$ Cephei~}

\newcommand{\cd}{d$^{-1}$\xspace}
\newcommand{\kms}{km\,s$^{-1}$\,}

\submitjournal{AAS Journals}

\shorttitle{Be star variability with TESS}
\shortauthors{Labadie-Bartz et al.}

\begin{document}

\title{Classifying Be star variability with TESS I: the southern ecliptic}

\correspondingauthor{Jonathan Labadie-Bartz}
\email{jbartz@usp.br}

\author[0000-0002-2919-6786]{Jonathan Labadie-Bartz}
\affiliation{Instituto de Astronomia, Geof\'{i}sica e Ci\^{e}ncias Atmosf\'{e}ricas, Universidade de S\~{a}o Paulo, Rua do Mat\~{a}o 1226, Cidade Universit\'{a}ria, 05508-900 S\~{a}o Paulo, SP, Brazil}

\author{Alex C. Carciofi}
\affiliation{Instituto de Astronomia, Geof\'{i}sica e Ci\^{e}ncias Atmosf\'{e}ricas, Universidade de S\~{a}o Paulo, Rua do Mat\~{a}o 1226, Cidade Universit\'{a}ria, 05508-900 S\~{a}o Paulo, SP, Brazil}

\author{Tajan Henrique de Amorim}
\affiliation{Instituto de Astronomia, Geof\'{i}sica e Ci\^{e}ncias Atmosf\'{e}ricas, Universidade de S\~{a}o Paulo, Rua do Mat\~{a}o 1226, Cidade Universit\'{a}ria, 05508-900 S\~{a}o Paulo, SP, Brazil}

\author{Amanda Rubio}
\affiliation{Instituto de Astronomia, Geof\'{i}sica e Ci\^{e}ncias Atmosf\'{e}ricas, Universidade de S\~{a}o Paulo, Rua do Mat\~{a}o 1226, Cidade Universit\'{a}ria, 05508-900 S\~{a}o Paulo, SP, Brazil}

\author{André Luiz}
\affiliation{Instituto de Astronomia, Geof\'{i}sica e Ci\^{e}ncias Atmosf\'{e}ricas, Universidade de S\~{a}o Paulo, Rua do Mat\~{a}o 1226, Cidade Universit\'{a}ria, 05508-900 S\~{a}o Paulo, SP, Brazil}

\author{Pedro Ticiani dos Santos}
\affiliation{Instituto de Astronomia, Geof\'{i}sica e Ci\^{e}ncias Atmosf\'{e}ricas, Universidade de S\~{a}o Paulo, Rua do Mat\~{a}o 1226, Cidade Universit\'{a}ria, 05508-900 S\~{a}o Paulo, SP, Brazil}

\author{Keegan Thomson-Paressant}
\affiliation{LESIA, Paris Observatory, PSL University, CNRS, Sorbonne University, Université de Paris, 5 place Jules Janssen, 92195 Meudon, France}

\begin{abstract}

TESS photometry is analyzed for 432 classical Be stars observed in the first year of the mission. The often complex and diverse variability of each object in this sample is classified to obtain an understanding of the behavior of this class as a population. 98\% of the systems are variable above the noise level, with timescales spanning nearly the entire range of what is accessible with TESS, from tens of minutes to tens of days. The variability seen with TESS is summarized as follows. Nearly every system contains multiple periodic signals in the frequency regime between about 0.5 -- 4 d$^{-1}$. One or more groups of closely-spaced frequencies is the most common feature, present in 85\% of the sample. Among the Be stars with brightening events that are characteristic of mass ejection episodes (17\% of the full sample, or 30\% of early-type stars), all have at least one frequency group, and the majority of these (83\%) show a concurrent temporary amplitude enhancement in one or more frequency groups. About one third of the sample is dominated by low frequency ($f < 0.5$ d$^{-1}$, and often much lower) variability. Stochastic signals are prominent in about 26\% of the sample, with varying degrees of intensity. Higher frequency signals ($6 < f < 15$ d$^{-1}$) are sometimes seen (in 14\% of the sample) and in most cases likely reflect p mode pulsation. In rare cases ($\sim$3\%), even higher frequencies beyond the traditional p mode regime ($f > 15$ d$^{-1}$) are observed.

\end{abstract}


\section{Introduction} \label{sec:intro}

Classical Be stars have been studied for over 150 years, yet key aspects of their nature remain veiled. Since their discovery in 1866 \citep{Secchi1866}, it has been established that classical Be stars (here after simply Be stars) are non-supergiant B-type stars with rapid, near-critical rotation, and are in general non-radial pulsators, which non-radiatively eject mass to form a viscous, near-Keplerian circumstellar ``decretion'' disk from which characteristic spectral emission features arise \citep[][and references therein]{Rivinius2013}. While significant progress has been made in the past many decades, 
the following questions regarding Be stars remain outstanding in general: How do they acquire such fast rotation? Why do they pulsate the way they do? How does rapid rotation influence internal processes such as angular momentum transport and chemical mixing? How are matter and angular momentum ejected at sufficiently high amounts? How does this ejected matter organize itself around the star on a geometrically thin and mostly Keplerian disk?

Mass, rotation, binarity, metallicity, and magnetic fields are the primary factors that dictate the life of a star.
Be stars span the entire spectral type range from late O to early A, do not host large-scale magnetic fields down to a detection limit of about 50 -- 100 Gauss \citep[][in a sample of 85 Be stars]{Wade2016}, are more common in lower metallicity environments \citep{Maeder1999,Wisniewski2006,Peters2020}, and are the most rapidly rotating non-degenerate class of objects known, on average rotating at or above 80\% of their critical rotation rate \citep{Fremat2005,Rivinius2013}. The binary fraction and binary parameters of Be stars as a population are somewhat uncertain, but there is evidence that the binary parameters are similar to that of B type stars in general \citep{Oudmaijer2010}, as well as suggestions that a very high fraction of Be stars exist in binaries \citep{Klement2019}. Be stars then represent critical test beds, especially for theories that explain the role of rotation in stars, which to date remain insufficiently developed for rapid rotators.

Space photometry has led to significant advances in the field of Be stars in recent years. Analysis of space-based photometry has revealed that pulsation is ubiquitous among classical Be stars, and that they pulsate primarily in low order g modes where gravity is the restoring force \citep{Rivinius2003}, similar to the class of Slowly Pulsating B (SPB) stars. Higher frequency p modes (pressure being the restoring force) are also observed in some Be stars, but are less common than g modes. Rossby waves (r modes, where centrifugal forces dominate) may also be present in Be stars \citep{Saio2013,Saio2018}.  Analysis of photometry from the MOST \citep{Walker2003}, BRITE \citep{Weiss2014}, Kepler \citep{Koch2010}, and CoRoT \citep{Baglin2006} satellites has shown that the frequency spectra of Be stars are often complex relative to other B-type main sequence pulsators (the $\beta$ Cephei and SPB stars), typically exhibiting multiperiodicity, groups of closely-spaced frequencies (as well as isolated frequencies), signatures of stochastic variability, and long-term (usually aperiodic) trends \citep{Baade2017, Baade2018, Rivinius2016, Semaan2018, Walker2005}. The fact that virtually all Be stars pulsate suggests that pulsation is an important aspect of these systems, and is also a potentially useful probe of their interiors via asteroseismology. High precision space photometry therefore represents a valuable tool to study the physics of Be stars and to learn about the role of rapid rotation in stellar structure and evolution.

The Transiting Exoplanet Survey Satellite \citep[TESS;][]{Ricker2015} mission, launched in 2018, has opened a new window into OB star variability. The Kepler spacecraft dramatically advanced the state of the art of stellar variability with its unprecedented photometric precision and long (4 years) observational baseline of a single field \citep{Borucki2010}. However, due to its observing strategy, only a small number of relatively faint OB stars were observed. While TESS has the same general goal as Kepler, the discovery of transiting exoplanets, its observing strategy is markedly different, at great benefit to the study of OB stars. In its prime mission, TESS covers $\sim$74\% of the sky in two years, with a large field of view that shifts approximately every 27 days. Unlike Kepler, the TESS sectors have significant overlap with the galactic plane, where the vast majority of OB stars are found. Whereas Kepler observed only 3 known Be stars \citep{Rivinius2016}, TESS is observing over 1000. Another benefit is that the TESS mission was designed for brighter stars (V$\lesssim$12), which means that the systems viewed by TESS have more comprehensive historical datasets and are more practical to observe from the ground.

Capitalizing on the strengths of TESS, the main goal of this work is to provide an overview of the variability seen in the population of over 400 Be stars observed in the first year of the TESS mission at a high precision and at short timescales (tens of minutes to weeks). Variability characteristics are ascribed to every star in the sample, providing insight to the behavior of the population as a whole and bringing to light patterns that exist according to spectral type and correlations between the different variability characteristics. In this context, the most typical signals are those of stellar pulsation which manifest as periodic signals in the flux of a given system. As has been reported in many studies, such periodic signals often form ``frequency groups'' in the observed power spectra \citep[\textit{e.g.}][]{Baade2018, Rivinius2016, Semaan2018, Walker2005}. These frequency groups are in general the most characteristic signature of Be stars observed with space photometry. In addition to photospheric signals, TESS is also sensitive to changes in the circumstellar environment close to the star, which can be associated with episodes of mass ejection (with the matter perhaps being inhomogeneously distributed in azimuth).

In section~\ref{sec:data}, the TESS satellite and its data products are introduced, and methods for data extraction are described. Section~\ref{sec:analysis} describes the analysis methods and shows example results of these methods for artificial light curves. Section~\ref{sec:features} introduces the characteristic features seen in the TESS data of Be stars, which are then used to describe each star in the sample. The results of the analysis of these signals are presented in Section~\ref{sec:results}, including discussion of each characteristic and the relevant astrophysical context. In Section~\ref{sec:discussion} a broad overview is given with an emphasis on correlations between the different variability classifications presented in Section~\ref{sec:results}, followed by concluding remarks in Section~\ref{sec:conclusions}.

\section{Data} \label{sec:data}
The NASA Transiting Exoplanet Survey Satellite \citep[TESS;][]{Ricker2015} is a photometric mission performing wide-field photometry over nearly the entire sky. The 4 identical cameras of TESS cover a combined field of view of 24$^{\circ}$ $\times$ 96$^{\circ}$. During the first year of TESS operations, nearly the entire southern ecliptic sky was observed in 13 sectors, with each sector being observed for 27.4 days. Some regions of the sky are observed in multiple sectors. TESS records red optical light with a wide bandpass spanning roughly 600 -- 1000 nm, centered on the traditional Cousins I-band. For optimal targets, the noise floor is approximately 60 ppm hr$^{-1}$.

Full Frame Images (FFIs) from TESS are available at a 30-minute cadence for the entire field of view, allowing light curves to be extracted for all objects that fall on the detector. Certain high priority targets were pre-selected by the TESS mission to be observed with 2-minute cadence, some of which were chosen from guest investigator programs. For the sample studied in this work, light curves were extracted from the FFIs for all systems, and 2-minute cadence light curves were also used whenever available. Generally the 30-minute and 2-minute light curves contain the same signals, which bolsters our confidence in the methods used for the FFI light curve extraction. However, there are some subtle differences which are further explained in the following subsections.

\subsection{Extracting light curves from TESS FFIs} \label{sec:FFI_extraction}
Light curves are extracted from the TESS FFIs using three different methods. This is done to increase our confidence in the results when these methods produce LCs that agree, and also decreases the incidence of false-positive detections of signals if they exist in only one version of the extracted LC, being perhaps caused by imperfect removal of systematic trends or blending from neighboring objects.

The first method used to extract LCs begins by using the \textsc{lightkurve} package \citep{Lightkurve2018} and \textsc{TESScut} \citep{Brasseur2019} to download a target pixel file (TPF) with a 50 $\times$ 50 pixel grid centered on the target star's coordinates for every available TESS sector. An aperture threshold of 10 sigma relative to the median flux level is used as a first step to automatically determine the aperture mask for the target star. The size of the target pixel mask is allowed to scale with stellar brightness. All pixels outside of a 15 $\times$ 15 pixel exclusion zone (centered on the target star) are then used as regressors in a principal component analysis (PCA) to remove common trends across this region of the CCD. The result is a PCA detrended light curve that is largely free from systematic trends. At the same time, an alternate version of the light curve is produced (second method), using only background removal, since in some instances a PCA detrending method will remove astrophysical variations in the target star with timescales of many days or longer.   

The third method uses the \textsc{Astroquery} \citep{Ginsburg2019} routine \textsc{Catalogs} to identify five stars of similar brightness on the same CCD as the target star within an annulus of an inner radius of 0.1 degree and an outer radius of 0.35 degrees (although this is allowed to vary if there are too few stars of sufficient brightness within the original annulus). \textsc{TESScut} is then used to extract a light curve for the target star and all of the identified neighboring stars using a 3 $\times$ 3 grid of pixels. The trend filtering algorithm (TFA) in the \textsc{VARTOOLS} light curve analysis package \citep{Hartman2012} is then used to identify and remove trends that are common to the set of the target star and its neighbors. This aperture is almost always smaller than that used in the PCA method, and is therefore less susceptible to blending (at the cost of achieving a lower SNR due to excluding some flux-containing pixels). 

In some cases one of the above methods will fail partially or completely (most often when a target is close to the edge of a detector, or in the most crowded fields).
Using different extraction methods allows us to determine when this is the case so that improperly reduced data is flagged as such. The first method using \textsc{lightkurve} tends to be more reliable and produces a light curve with higher signal-to-noise ratio (SNR), and all plots of 30-min cadence data shown in this work use this version of the light curve. 
The sampling rate of 30 minutes allows for the detection of frequencies up to 24 \cd. 

\subsection{2-minute cadence light curves} \label{sec:2-min_cadence}

About 65\% of the sample has 2-minute cadence light curves (LCs) available from the TESS office. When these LCs exist, the same analysis was performed, as was done for the LCs extracted from the FFIs. In nearly all cases, the results are virtually identical between different versions of the LC. We use the Pre-search Data Conditioning Simple Aperture Photometry (PDCSAP) flux from the TESS LCs, which is calibrated in a way that removes long term trends. This generally yields a cleaner version of the LC (compared to the Simple Aperture Photometry (SAP) flux) that is well suited for frequency analysis. 
The disadvantage is that the detrending process may remove astrophysical variability on relatively long timescales. These, however, can easily be identified in the 30-minute data. For that reason both sets of data were always analysed for all stars for which they are available.
The higher cadence 2-minute LCs allow for the detection of much higher frequencies relative to the data extracted from the FFIs, up to the Nyquist limit of 360 \cd.

\subsection{Sample selection} \label{sec:sample}
The Be Star Spectra (BeSS) database\footnote{\url{http://basebe.obspm.fr}} \citep{Neiner2011} was used to create a list of Be stars between V-mag of 4 -- 12.
The sample of Be stars from \citet{Chojnowski2015} was also included. 
Only targets in the southern ecliptic hemisphere were chosen, in order to match the fields observed by TESS in its first year of operations.
This initial list of 539 stars was reduced to 432 by rejecting systems known or strongly suspected to be something other than a classical Be star (\textit{e.g.} interacting binaries or B[e] stars), and stars for which there are no TESS data or the data is of insufficient quality (\textit{i.e.} when a star falls into a gap between sectors, or systematic effects or unusually high noise levels severely hamper our ability to analyze a given LC). Of these 432 classical Be stars, 218 (50\%) are of early type (B3 and earlier), 79 (18\%) are mid type (B4, B5, and B6), 112 (26\%) are late type (B7 and later), and 23 (5\%) are of unknown spectral type.

The vast majority of these targets also have years-long light curves from the ground-based Kilodegree Extremely Little Telescope \citep[KELT,][]{Pepper2007,Pepper2012}, where the data for some of the TESS sample are published and available in \citet{Bartz2017,LabadieBartz2018}. While the KELT data was not directly used in the analysis of these systems in this work, in some instances it was used to corroborate long term trends or periodic signals seen in TESS, or to confirm systems that are not Be stars (\textit{e.g.} binaries with ellipsoidal variation). Further works studying these systems should take advantage of the more comprehensive suite of time-series data provided by TESS, KELT, BeSS, APOGEE, and perhaps other sources.

\section{Analysis} \label{sec:analysis}

A variety of methods were used to analyze the sample. 
Standard Fourier methods lie at the core of the analysis, but the diversity of signals expressed by Be stars demands a careful study, including visual inspection of the light curves, separately considering different frequency regimes, measuring aperiodic variability, tracing variable amplitudes of periodic signals, and documenting correlations in time between different signals.
Consideration is also given to the degree of blending, and potential issues from saturation, systematic effects, and noise. Details about the data analysis are described in the following subsections.

\subsection{Removing outliers and bad data} \label{sec:outliers}
Because of the variety of signals in the data, especially in systems exhibiting longer term trends of relatively high amplitude, outliers cannot be removed through standard methods such as sigma clipping. Instead, outliers and sections of poor quality data were removed in the following way. After reconstructing the light curve with a sum of Fourier terms, the calculated fit was subtracted from the observed data, and points greater than five times the median absolute deviation from the median of the residuals were automatically identified as outliers. At this stage, the observed and calculated data and the residuals were manually examined, allowing for the possibility of removing additional sections of poor-quality data. These cleaned light curves, with outliers and sections of poor quality data removed, are used in all following analysis steps.

\subsection{Lomb-Scargle periodograms and pre-whitening} \label{sec:lstechnique}

Because of the generally complex nature of photometric variability of Be stars, there are signals with a wide range of timescales and behavior that we are interested in measuring which requires extra analysis steps. For example, in a system with significant aperiodic low-frequency variability, it is prudent to detrend against these signals when analyzing the data for higher frequency signals. 
Through iterative pre-whitening, each sector of data for each object was detrended against all signals lower than 0.5 \cd up to a false alarm probability of 10$^{-2}$ \citep[in a similar fashion as][]{Rivinius2016}. 
A frequency analysis was then performed for the entire set of TESS observations for each star, separately for the original and detrended versions of the light curves. The \textsc{timeseries.LombScargle} package \citep{VanderPlas2012,VanderPlas2015} of \textsc{Astropy} \citep{astropy2013, astropy2018} was used to compute these Lomb-Scargle Periodograms (LSPs).
The above procedure was applied to both the 2- and 30-minute cadence data, when available.

To determine the individual frequencies present in the data, the \textsc{Vartools} light curve analysis software with the \textsc{Lomb-Scargle} routine \citep{Zechmeister2009,Press1992} was used to detect and iteratively pre-whiten the data against each recovered signal (up to a false alarm probability of 10$^{-2}$), recording the frequency, phase, amplitude, SNR, and false alarm probability. The frequency and amplitude of these signals are used to plot the pre-whitened periodograms, and are also used to re-construct each light curve based on the recovered signals, which are visually compared to the original photometric data to ensure a good fit from the Fourier analysis. 
These methods are similar to what is typically done for space photometry of Be stars, \textit{e.g.} with MOST \citep{Walker2005}, Kepler \citep{Rivinius2016}, and CoRoT \citep{Semaan2018}.

\subsection{Wavelet plots} \label{wavelet}

It is common for Be stars to show photometric signals that vary in amplitude over time. Wavelet plots are a convenient way to visualize this, as they depict the frequency spectrum as a function of time (at the cost of a degraded frequency resolution). The Python package \textsc{scaleogram}\footnote{https://github.com/alsauve/scaleogram} was used to perform a wavelet analysis for each star, considering the original and low-frequency detrended (signals with frequencies $<$ 0.5 \cd removed) versions of the data separately.

\subsection{Testing Artificial Light Curves} \label{sec:artificial}

In order to better understand the signals that are recovered from TESS light curves of Be stars, tests were performed where artificial signals are injected into the data and then recovered. This process begins by taking a TESS light curve of a typical Be star from a single sector, and removing all signals through iterative pre-whitening, leaving a light curve that contains only noise. Signals are then injected, and the light curve is analyzed using the same methods that are applied to the real data. The injected signals were motivated by common features seen in the data for Be stars.

\begin{figure}[ht!]
 \centering
 \includegraphics[width=0.45\textwidth,clip]{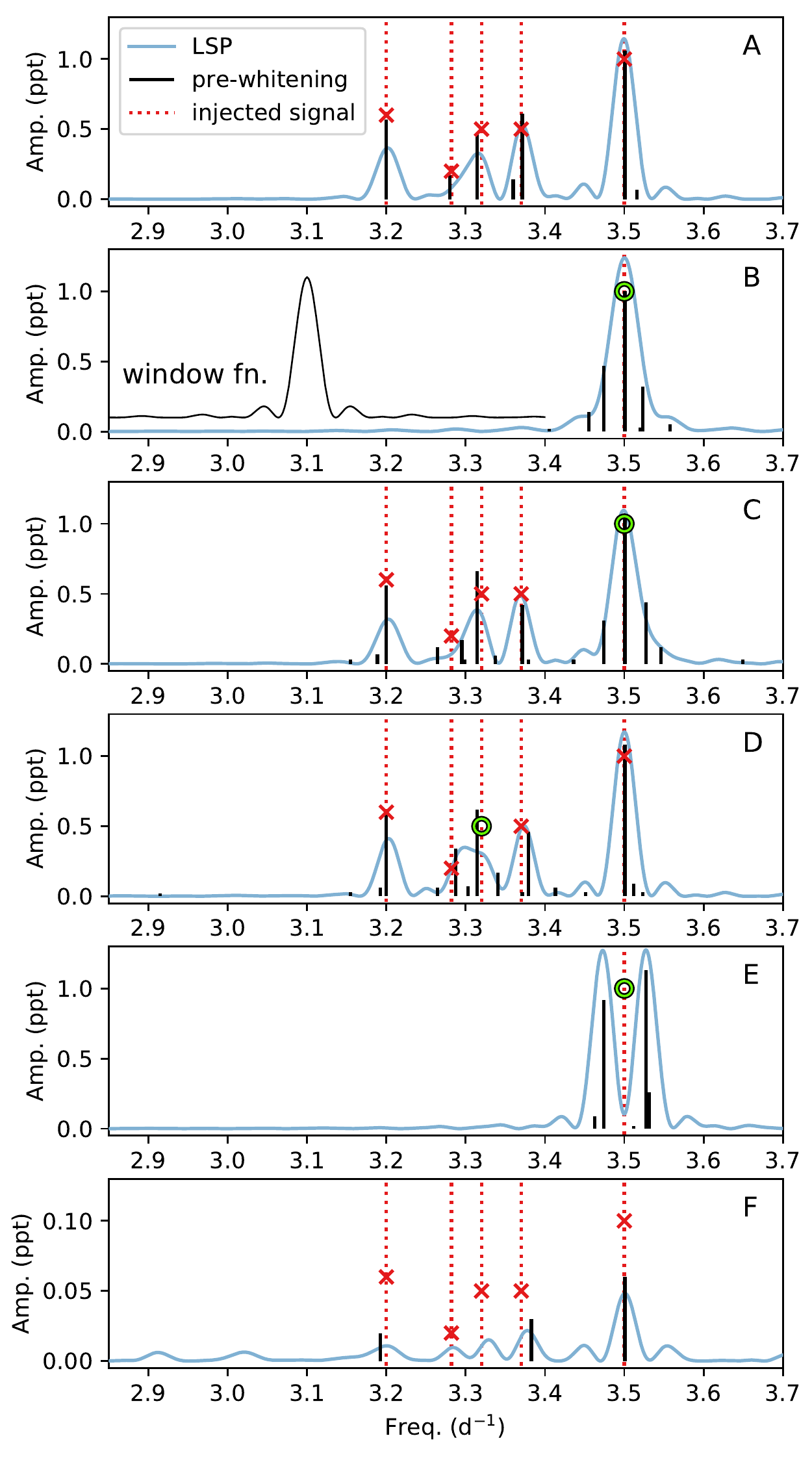}      
  \caption{
  Analysis of an artificial light curve with known signals injected. The location of injected signals are shown with red dotted vertical lines, with amplitudes of constant signals marked by a red x, and the maximum amplitude of signals that vary in strength with time is marked by a green circle. The Lomb-Scargle periodogram of the artificial light curve is shown in light blue, and the signals found through iterative pre-whitening of the artificial light curve are vertical black lines, where their height reflects the amplitude of the recovered signal. The window function (with arbitrary vertical scale), centered at 3.3 \cd, is shown in panel B. Note that each sector will have a slightly different window function, and for systems viewed in multiple sectors the window function becomes more narrow.
  \textit{A:} A frequency group where all injected signals have a constant amplitude. \textit{B:} A single frequency with a linearly decreasing amplitude is injected. \textit{C:} The same as panel A, but the strongest signal at 3.5 \cd has a linearly decreasing amplitude. \textit{D:} The same as the panel A, but the middle signal at 3.32 \cd has a linearly decreasing amplitude.  \textit{E:} The same as the panel B, but the amplitude is modulated by a sinusoid. \textit{F:} The same as panel A, but all amplitudes are smaller by a factor of ten. 
  }
  \label{fig:art_lcs}
\end{figure}

In the first trial, nine sinusoidal signals with different but constant amplitudes are injected into the pre-whitened light curve in two groups, with four signals centered around 2 \cd, and five signals centered around 3.3 \cd to imitate the frequency groups that are commonly seen in Be stars (the amplitudes of these signals are between 0.2 -- 1 ppt).

The artificial light curve is then analyzed through iterative pre-whitening. All of the injected signals are recovered, with the mean difference between the injected and recovered frequencies being 0.06\%, and the recovered amplitudes are precise to within 10\%. However, other signals are also found in the analysis.
These spurious signals are all close to the injected frequencies and lie within their respective groups. Panel `A' of Figure~\ref{fig:art_lcs} shows the Lomb-Scargle periodogram and the signals recovered through iterative pre-whitening of this first artificial light curve, with the location of the injected signals also marked. Only one frequency group is shown for clarity, as the behavior of the other group is qualitatively the same. In the following trials, only this group near 3.3 \cd is modified, and there is no discernible influence on regions of the frequency spectrum that are not shown in Figure~\ref{fig:art_lcs}. 

Next, a single frequency at 3.5 \cd is injected into the light curve with an amplitude that decreases linearly from 1 ppt to reach an amplitude of 0 at the end of the light curve. 
Panel `B' of Figure~\ref{fig:art_lcs} shows this trial. The injected frequency is recovered to within 0.02\% with an amplitude indistinguishable from that of the injected signal at its strongest.
Other signals are also found close to the injected frequency and with amplitudes decreasing with distance from this frequency. These signals are weaker than the injected single frequency, but are still significant relative to the noise level. The periodogram peak is also wider than it would be if the amplitude were not modulated. 

The original nine signals are again injected, with the strongest (at 3.5 \cd) being modulated in amplitude in the same way as the second trial. Again, there are spurious peaks near 3.5 \cd caused by the amplitude modulation, but now there are additional peaks in the group that did not exist in the first trial, and one of the injected signals (the second in the group, at 3.282 \cd) is not properly recovered. Panel `C' of Figure~\ref{fig:art_lcs} shows this.

This test is repeated, but with the amplitude of the 3.5 \cd signal being constant and the amplitude of the middle frequency of the group, at 3.32 \cd, linearly decreasing at the same rate as the second and third trials. Spurious peaks appear, and the precision with which the five injected signals of the group are recovered is degraded, as shown in panel `D' of Figure~\ref{fig:art_lcs}. 

A trial similar to the second one is done, but instead of a single frequency with a linearly decreasing amplitude, the amplitude is modulated by a sinusoid with a frequency of 0.0275 \cd. In this case, the original injected frequency at 3.5 \cd is not recovered, but rather two strong peaks appear in the periodogram (panel `E' of Figure~\ref{fig:art_lcs}). This is explained by the trigonometric identity $\sin(\alpha) \ times \sin(\beta) = [\sin(\alpha - \beta) - \cos(\alpha + \beta)]/2$. In other words, with a standard Lomb-Scargle analysis alone it is impossible to distinguish between a single frequency, $\beta$, whose amplitude is modulated by some lower frequency, $\alpha$, and two signals of constant amplitude located at $\beta - \alpha$ and $\beta + \alpha$. 
In principle this can be remedied by also considering phase information. However, such an analysis is not performed in this work.

Finally, the first trial was repeated but with all amplitudes being smaller by a factor of 10 (panel `F' of Figure~\ref{fig:art_lcs}). With these amplitudes approaching the noise floor of the data, the injected signals are poorly recovered. This demonstrates a practical, but approximate, lower limit on the amplitude of signals that can be reliably recovered in a typical TESS light curve is $\sim$0.05 ppt\footnote{A true lower limit depends on the brightness of the target, how successfully systematic trends are removed, the details of the frequency spectrum in the vicinity of a given signal, the number of sectors in which the target was observed, the choice of aperture used to extract the light curve, and perhaps other factors.}. 

\begin{figure}[ht!]
 \centering
 \includegraphics[width=0.50\textwidth,clip]{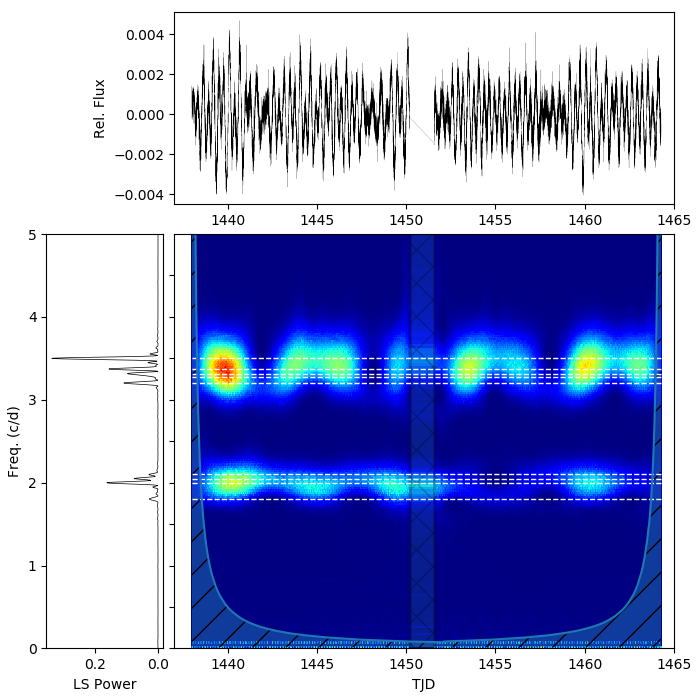}      
  \caption{Artificial light curve (top), LSP (left), and wavelet plot (middle) for the first trial discussed in Section~\ref{sec:artificial} with two frequency groups, with each signal having a constant amplitude. The wavelet plot shows both groups having apparently variable power, which is solely a consequence of the beating phenomenon. In the wavelet plot, horizontal dashed lines mark the injected frequencies, and regions occupied by hash marks denote gaps in data and regions where wavelet signals are otherwise unreliable due to edge effects.}
  \label{fig:art_wavelet}
\end{figure}

The first trial, where the injected signals that make up two groups all have a constant amplitude, is shown in more detail in Figure~\ref{fig:art_wavelet}. The wavelet analysis of this artificial light curve shows that the two groups are variable in power over time. However, this is solely a consequence of the beating phenomenon between signals in a given group, since each injected signal is known to be constant in amplitude. 
The beating can also be seen as a corresponding change of the overall amplitude in the light curve itself.

Numerous other similar tests were performed, but the above examples serve to demonstrate the overall results of attempting to recover signals in groups of closely spaced frequencies when amplitudes are allowed to vary (as is often the case with Be stars). The main conclusions of these tests are that the methods used to detect signals in the TESS data will almost always result in some spurious signals whenever there are groups of frequencies and that signals with varying amplitudes can, sometimes dramatically, compound this effect. While the frequency groups themselves, and usually the strongest frequencies comprising them, are reliably recovered, some degree of spurious detections is inevitable. Complex beating patterns can exist within a frequency group, causing apparent modulation of the strength of the group over time despite all individual signals having a constant amplitude. Therefore, caution must be exercised when analyzing TESS data for Be stars, and these limitations must be kept in mind when considering the often complex light variability seen in this sample. While these limitations are well known in general, the complex nature of Be star variability and the low frequency resolution of most TESS light curves exacerbates such issues.

\subsection{Interpreting observations of Be star systems} \label{sec:interpretation}

Certain observed variations can be confidently attributed to either the star or the disk. In terms of photometry, brightening or fading events that occur on timescales of months or years are understood to be due to disk growth or dissipation, while coherent, stable periodic signals on timescales of around one day and less are best attributed to stellar pulsation. There are, however, many cases where photometric signals are ambiguous in origin. The stellar rotation period, orbital period in the close circumstellar environment, and possible pulsational periods are all very similar. Since many factors can influence the total brightness of the system (and other observables) in often complex and time-variable ways, care must be taken in interpreting photometric data. 
In the following, examples are provided of variability that can be firmly connected to either the disk or star, but debatable cases are also discussed, in order to highlight both the great potential, as well as the complications ensuing from interpreting space photometry alone.

\section{Characteristic features of Be stars in TESS} \label{sec:features}
The main goal of this work is to assign characteristic variability features to each star in the sample, and to then use these characteristics to describe the sample as a population. This section introduces these features and shows examples. 

\subsection{Light curves and their frequency spectra}
\begin{figure*}[ht!]
 \centering
 \includegraphics[width=0.998\textwidth,clip]{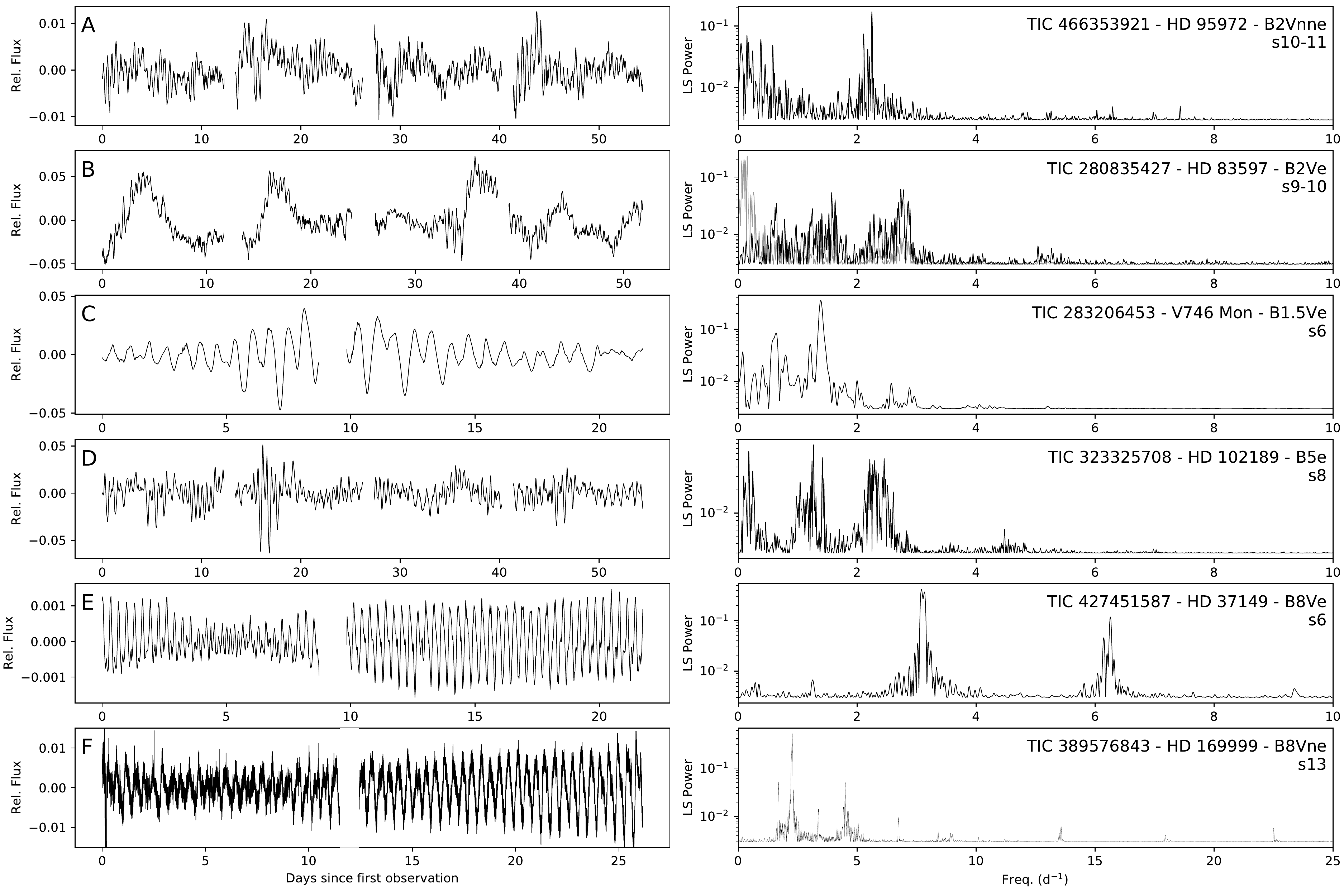}      
  \caption{
  TESS light curves (left) and Lomb-Scargle periodograms (right) for a representative selection of Be stars that show certain characteristic features, as described in Sec.~\ref{sec:features}. For panel B, the periodogram is re-calculated after removing the low frequency ($<$ 0.5 \cd) signals (which are shown in a lighter grey color). Panel F uses 2-minute cadence data, which better emphasizes the highest frequency signals. All other panels use 30-minute cadence data. The frequency axis of the periodogram in panel F is extended to include the high frequencies. Signals at frequencies higher than 10 \cd are absent in all other stars shown here. The TIC ID, common ID, spectral type, and TESS sectors are printed in the periodogram plots. } 
  \label{fig:examples}
\end{figure*}

Most of the characteristic features in the TESS data of Be stars can be inferred by inspecting the light curve and the Lomb-Scargle periodogram as a measure of the frequency spectrum. While each light curve and frequency spectrum is unique, there are certain features that are common among many members of the sample. Fig.~\ref{fig:examples} shows examples of Be stars that show these characteristic features, which are introduced and described here. 

In what follows, a convention is adopted where low frequencies are those less than 0.5 \cd, mid frequencies are between 0.5 -- 6 \cd, high frequencies are between 6 -- 15 \cd, and very high frequencies are greater than 15 \cd. While these distinctions are somewhat arbitrary, they are physically motivated. Low frequency signals are generally below those of typical g mode pulsation in Be stars, mid (high) frequencies span the typical range of g mode (p mode) pulsation in Be stars, and very high frequencies are above the typical p mode regime in Be stars \citep[\textit{e.g.}][and references therein]{Handler2013,Bowman2020Astero}. While the rapid rotation (and evolutionary stage) of Be stars can complicate this simplified scheme, it is useful to choose cutoffs to delineate these categories in order to classify the observed variability.

The characteristic features adopted are as follows:

\begin{itemize}

\item \textit{Flickers:} Features in the light curve whereby the brightness increases or decreases by a few percent over a few days, followed by a return towards baseline, are loosely defined as ``flickers''.
The largest amplitude features in panel B of Fig.~\ref{fig:examples} are examples of this. Flickers are not an oscillation around the mean brightness (like in panel C), but are rather a marked departure from the baseline brightness. 

\item  \textit{Low-frequency signals dominate:} Features with frequencies lower than 0.5 \cd are the most prominent type of variation in the data. Panel B in Fig.~\ref{fig:examples} is an example of this characteristic. This is usually apparent from the light curve, but can be more quantitatively determined if the strongest periodogram signals are in the low frequency regime.

\item \textit{Frequency groups:} Many closely-spaced frequencies often form groups in the frequency spectra of Be stars. The system in panel D of Fig.~\ref{fig:examples} shows three well defined groups near 0.05, 1.2, and 2.4 \cd, and panel E shows two groups near 3 and 6 \cd. Panel B and F also show two frequency groups each. There is one group in panel A (near 2 \cd) plus stochastic variation at lower frequencies. Panel C is more ambiguous as plotted, but shows two prominent groups near 0.6 and 1.4 \cd which are more easily identified through iterative pre-whitening and acknowledging that the periodogram peaks are wider than the window function, suggesting they contain multiple unresolved signals. 
It is common for a harmonic series of groups to extend from the typical $g1$ and $g2$, usually with decreasing amplitude.

\item \textit{Stochastic variation:} Non-periodic variability is a significant feature of the data. Stochastic signals can appear as extra ``noise'' in the frequency spectrum that is strongest at the lowest frequencies, and decreases towards higher frequencies. However, this ``red noise'' is astrophysical, and arises from genuine variability.
Panel A in Fig.~\ref{fig:examples} shows an example that includes stochastic variability. The forest of signals between 0 -- 2 \cd in the periodogram is stochastic in nature, while the frequency group just above 2 \cd is a separate (periodic) feature. Panel B likewise includes stochastic variability (manifesting as an underlying ``continuum'' of signals at low frequencies in the periodogram) in addition to frequency groups that clearly stand above the local noise.

\item \textit{Single, isolated frequencies:} In contrast to groups, some frequencies are singular and well-defined. There are many isolated frequencies in the periodogram of panel F in Fig.~\ref{fig:examples}, and also some in panel A (with low amplitudes). 

\item 
\textit{High and very high frequency signals:} 
Systems that exhibit periodic signals in the high (6 $< f <$ 15 \cd) and very high ($f > 15$ \cd) frequency regime are recorded.
Panel F in Fig.~\ref{fig:examples} shows a star with many of these high and very high frequency signals, while panel A also meets the criteria of having high frequency signals. Harmonics alone are not considered here, nor are near-harmonic sequences of groups that begin at mid frequencies. For example, if there is a signal at 4.0 \cd with an exact harmonic at 8.0 \cd, it is not classified as a high frequency signal since the presence of this harmonic may simply indicate that the fundamental signal is not perfectly sinusoidal. In panel E, the groups near 6 \cd and 9.5 \cd are the second and third groups in a series that begins with $g1$ near 3 \cd, and thus this star does not meet the criteria for having high frequency signals.

\item \textit{Harmonics of isolated signals:} Some frequency spectra show clear harmonics, where a signal is found at an integer number times the frequency of another signal. In some cases these harmonics are exact, while in others they are approximate. An exact harmonic is seen in panel F of Fig.~\ref{fig:examples}, where the lowest frequency, $f_{0}$ = 1.684 \cd has a first harmonic at 2$\times f_{0}$ = 3.368 \cd (and increasingly smaller amplitude second, third, and fourth harmonics). Frequency groups often have (multiple) harmonics, but those are considered separately.

\item \textit{No detected signals:} A small fraction of the stars in the sample show no variability above the TESS noise level. This is generally restricted to later spectral types, where it is well known that amplitudes are relatively low.

\end{itemize}

\subsection{Time variable signals}
\begin{figure}[ht!]
 \centering
 \includegraphics[width=0.48\textwidth,clip]{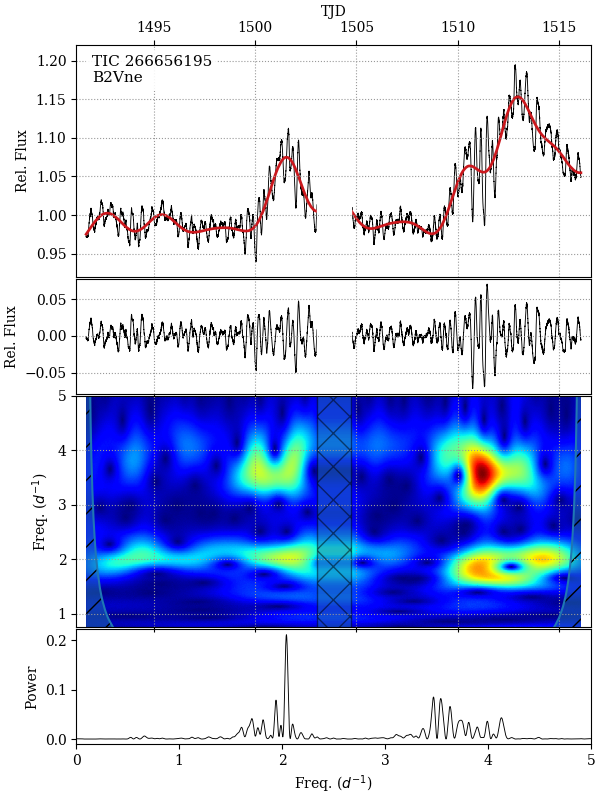}      
  \caption{A system where flickers correspond to the enhancement of the two main frequency groups. \textit{First:} 2-minute cadence LC (black), with the low frequency signals in red. \textit{Second:} The LC after removing low frequency signals. \textit{Third:} Wavelet analysis. \textit{Fourth:} Lomb-Scargle periodogram. The top three panels share the same x-axis (TJD).
  }
  \label{fig:wavelet_01}
\end{figure}

While traditional light curve and Lomb-Scargle analysis provide valuable information about the signals present in the data, it is clear from Fig.~\ref{fig:examples} that some signals are variable in time.
The most notable aspect of this is seen when flickers coincide with enhancement of the power of one or more frequency groups. This is clearly seen, for example, in Figure~\ref{fig:wavelet_01} where both main groups (near 2 \cd and 4 \cd) are strongest during the two flicker events. We limit our consideration of time-resolved signal analysis to only this situation, recording instances where groups are enhanced coincident with flickers. Further analysis is possible (\textit{e.g.} quantifying correlations between signal amplitudes over time) and will be explored in future works.

\subsection{Interpreting results of our analysis}

After identifying the characteristic features of interest, plots for the sample were visually inspected in order to determine which of the above characteristics can be attributed to each star and other information about the signals that are present (\textit{e.g.} the location and relative strength of frequency groups). 

There is some degree of subjectivity in assigning variability classifications. Each object was analyzed in detail independently by three authors of this work. If a consensus was reached regarding a given classification, then that classification was assigned to the system. If there was disagreement or uncertainty, then the object was inspected in more detail and a final decision was made regarding the classification in question. In some cases, it remained not possible to confirm or reject a given classification. 

An important aspect of this manual analysis was determining if a given object is not a classical Be star. To this end, a literature search was conducted, any available BeSS or APOGEE spectra were inspected, the SED was inspected, and any unusual features in the light curve were noted (\textit{e.g.} indications of a short period binary). Systems deemed to be something other than a classical Be star are not considered in any of the statistics in this work, and are briefly discussed in Appendix~\ref{sec:not-Be} and~\ref{sec:maybe-not-Be-rejected}. Systems suspected to not be classical Be stars, but without sufficient evidence to reject them from the sample, are noted in Appendix~\ref{sec:maybe-not-Be-not-rejected}.

\subsection{Determining the center of frequency groups} \label{sec:groupcenter}

A Python routine was developed to more objectively quantify groups and their properties from the pre-whitened frequency spectrum for each star. This clustering algorithm served to identify groups, and to provide numerical descriptions of their net amplitude and weighted center, thus allowing the location and relative strength of groups identified in a given light curve to be compared. Further details are provided in Appendix~\ref{sec:appendix_groups}.

\section{Results and discussion for each variability type} \label{sec:results}

The variability features for all Be stars in the sample were tabulated and the occurrence rates are shown in Table~\ref{tbl:var_rates}. In the following subsections, the variability types and characteristic examples are discussed. These results are discussed in a broader scope in Section~\ref{sec:discussion}.

\begin{table*}
 \centering
 \caption{Percentages showing variability classifications} \label{tbl:var_rates}
 \label{tbl1}
 \begin{tabular}{rrcrcrcrcrc}
    \hline
\multicolumn{1}{r}{variability }  & \multicolumn{2}{c}{all} & \multicolumn{2}{c}{early} & \multicolumn{2}{c}{mid} & \multicolumn{2}{c}{late} & \multicolumn{2}{c}{unknown} \\ 
\multicolumn{1}{r}{characteristic }   & \multicolumn{2}{c}{(432)} & \multicolumn{2}{c}{(218)} & \multicolumn{2}{c}{(79)} & \multicolumn{2}{c}{(112)} & \multicolumn{2}{c}{(23)} \\
\hline
flickers & 17.4\% & (75) & 29.8\% & (65) &  7.6\% & (6) &  0.9\% & (1) &  13.0\% & (3) \\ 
flickers + enhanced freq. groups & 82.7\% & (62) & 81.5\% & (53) &  100.0\% & (6) &  100.0\% & (1) &  66.7\% & (2) \\ 
low freq. dominated & 31.9\% & (138) & 47.2\% & (103) &  20.3\% & (16) &  13.4\% & (15) &  17.4\% & (4) \\ 
high freq. & 14.1\% & (61) & 15.6\% & (34) &  16.5\% & (13) &  10.7\% & (12) &  8.7\% & (2) \\ 
very high freq. & 3.0\% & (13) & 2.3\% & (5) &  2.5\% & (2) &  5.4\% & (6) &  0.0\% & (0) \\ 
freq. groups & 84.7\% & (366) & 88.5\% & (193) &  92.4\% & (73) &  73.2\% & (82) &  78.3\% & (18) \\ 
typical group configuration & 80.6\% & (295) & 84.5\% & (163) &  82.2\% & (60) &  72.0\% & (59) &  72.2\% & (13) \\ 
stochastic & 25.9\% & (112) & 33.9\% & (74) &  12.7\% & (10) &  19.6\% & (22) &  26.1\% & (6) \\ 
isolated freqs. & 31.9\% & (138) & 24.3\% & (53) &  35.4\% & (28) &  43.8\% & (49) &  34.8\% & (8) \\ 
harmonics of isolated freqs. & 7.6\% & (33) & 2.3\% & (5) &  10.1\% & (8) &  13.4\% & (15) &  21.7\% & (5) \\ 
\hline
 \end{tabular}
 \begin{flushleft}
  \footnotesize 
  Fraction of stars showing each type of variability, according to their spectral type, followed by the total number of systems showing the given characteristic. 
  \end{flushleft}  
\end{table*}

\subsection{Frequency groups} \label{sec:groups}

\subsubsection{Overview} \label{sec:groups_overview}

The existence of frequency groups in the power spectra is a common feature of Be stars observed from space \citep{Walker2005, Saio2007, Saio2013, Kurtz2015, Baade2018, Semaan2018}.
85\% (366/432) of this sample shows one or more frequency groups. 
According to spectral type, this percentage is 89\% (193/218) for early types (B3 and earlier), 92\% (73/79) for mid types (B4--B6), and 73\% (82/112) for late types (B7 and later). 
The presence of frequency groups is the most common characteristic signal seen in the Be stars of this sample. Multiple frequency groups like those seen in the majority of Be stars of this sample are not common in non-Be OB stars \citep[\textit{e.g.}][]{Burssens2020,Bowman2020}. Frequency groups are often a consequence of very rapid rotation in pulsators \citep{Saio2018,Lee2020}, and may be an important, albeit not essential, component of the Be phenomenon.

\begin{figure*}[ht!]
 \centering
 \includegraphics[width=0.99\textwidth,clip]{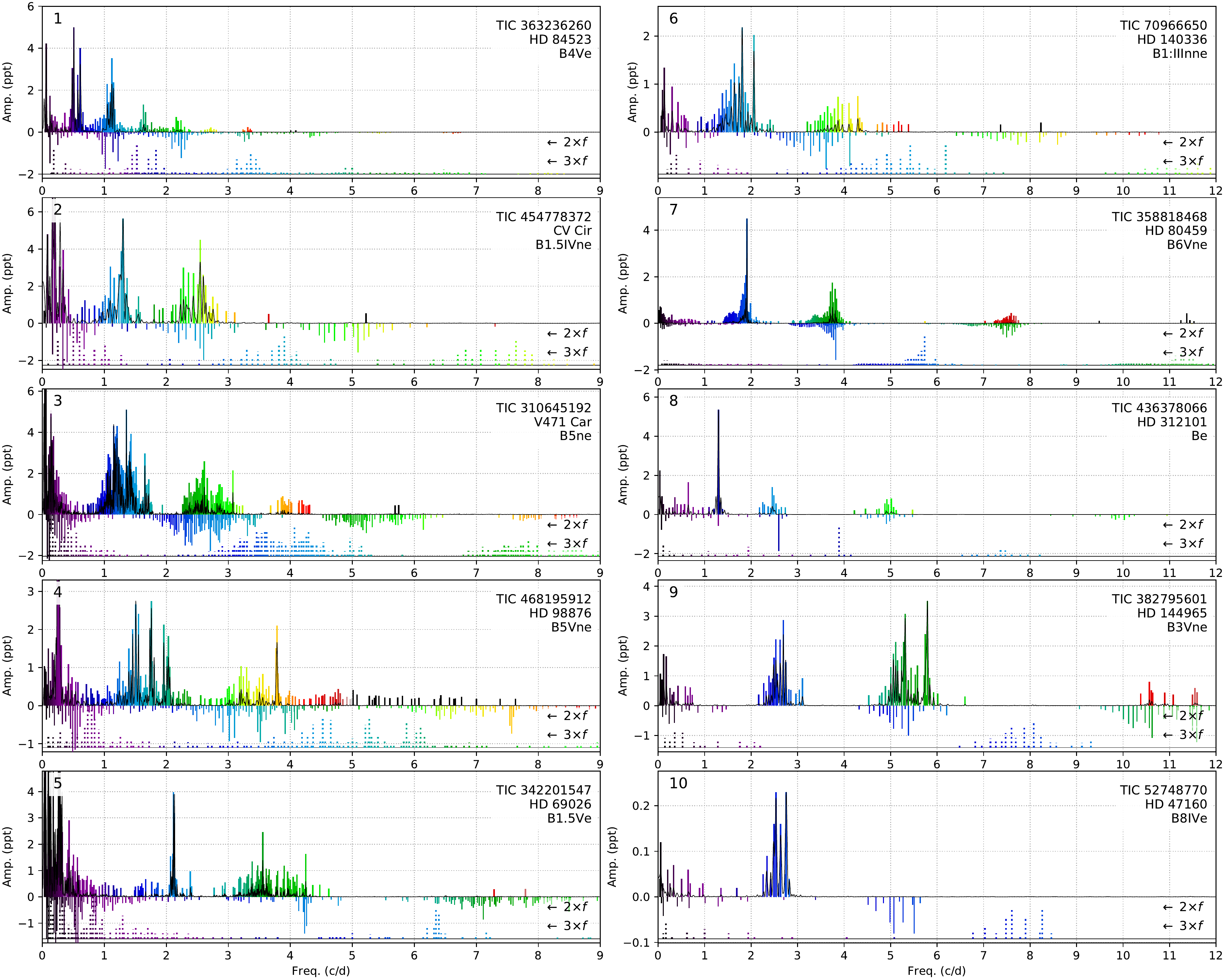}      
  \caption{A selection of frequency spectra showing various permutations of group configurations and morphology. Solid colored vertical lines are the signals recovered with iterative pre-whitening, and the solid black line is the Lomb-Scargle periodogram without any pre-whitening normalized to the highest amplitude at $f > 0.33$ \cd. The inverted solid lines show the same frequencies multiplied by two, and below that, the dashed lines show the frequencies multiplied by three to aid in visualizing the locations of the first and second harmonics of the recovered signals. \textit{1:} Closely spaced, narrow groups monotonically decreasing in strength. \textit{2:} $g1$ and $g2$ have similar strengths, without further harmonics. \textit{3:} Wider groups, with signals corresponding to the second harmonic of $g1$, but without significant signals corresponding to the first harmonic of $g2$ (or the third harmonic of $g1$). \textit{4:} $g1$ and $g2$ are wide, and the region beyond $g2$ is populated with signals that form a ``continuum'' corresponding to harmonics of signals in $g1$ and $g2$.   \textit{5:} $g1$ is narrow and dominated by a single signal, and $g2$ is relatively wide and centered at a frequency less than $2 \times g1$. \textit{6:} $g2$ is centered at greater than $2 \times g1$, and there are also signals located in the region occupied by the second harmonic of $g1$. \textit{7:} $g1$ and $g2$ are narrow and mirror each other in structure, but there are only signals that correspond to $2 \times g2$ and not $3 \times g1$. 
  \textit{8:} Two fairly typical groups, $g1$ and $g2$ being centered near 2.5 and 5 \cd, plus a strong apparently isolated signal at 1.30 \cd and its weaker sub-harmonic at 0.65 \cd.   
  \textit{9:} Widely separated groups, where $g2$ is stronger than $g1$, and with power located in the region corresponding to $2 \times g2$ and not $3 \times g1$. \textit{10:} Only one frequency group exists, along with weak low frequency signals. 
  }
  \label{fig:canon_freq_groups}
\end{figure*}

\subsubsection{Typical group configurations and relative strength} \label{sec:groups_configuration}

Although there is a wide range in the location, width, number and relative strength of individual frequencies making up the group, and relative strength of the groups themselves, there are some patterns that are common when considering the whole sample.
The most typical configuration includes three groups (although there may be further harmonics of these with decreasing amplitude). The lowest frequency group, $g0$, is centered at $<$ 0.5 \cd (and often much lower, around $\sim$0.05 \cd). The next group, $g1$, is centered at some intermediate frequency (typically between 0.5 -- 3 \cd), and $g2$ is located at approximately twice the frequency of $g1$. A variation of this configuration is seen when $g0$ is absent, but $g1$ and $g2$ still follow the same pattern. Figure~\ref{fig:canon_freq_groups} shows many examples of frequency groups, where all but panel 10 exhibit this typical configuration. In what follows the center of a typical frequency group is written as $f_{g1}$ or $f_{g2}$.

Interestingly, there are many cases where there are the typical groups $g1$ and $g2$, and a third group located at approximately $2 \times f_{g2}$ (but with no group near $3 \times f_{g1}$). Panels 7 and 9 of Fig.~\ref{fig:canon_freq_groups} show examples of this. This is in contrast to the Be stars whose groups roughly form a harmonic series past $g1$ and $g2$ (as most clearly seen in panels 1 and 3 of Fig.~\ref{fig:canon_freq_groups}). This may mean that in some cases, the signals that constitute $g2$ are non-sinusoidal. 

The frequency spectra of this sample have varying degrees of complexity, and there are some cases where multiple signals exist in the vicinity of each other. Panel 8 of Fig.~\ref{fig:canon_freq_groups} shows one such example, where the frequency spectrum does include two groups following the typical pattern ($f_{g1} \approx 2.45$ \cd, and $f_{g2} \approx 4.94$ \cd). There is also a strong isolated signal ($f \approx 1.30$ \cd) and its sub-harmonic, which are seemingly unrelated to the two frequency groups. Situations like this may reflect ``composite frequency spectra,'' where the lower frequency group and its sub-harmonic may indicate binarity or rotation (perhaps of a companion star or system), while the pair of higher frequency groups are formed by families of pulsation typical of Be stars. This example is purely illustrative -- such interpretation requires further data and analysis and is beyond the scope of this work.

\begin{figure*}[!ht]
\centering\epsfig{file=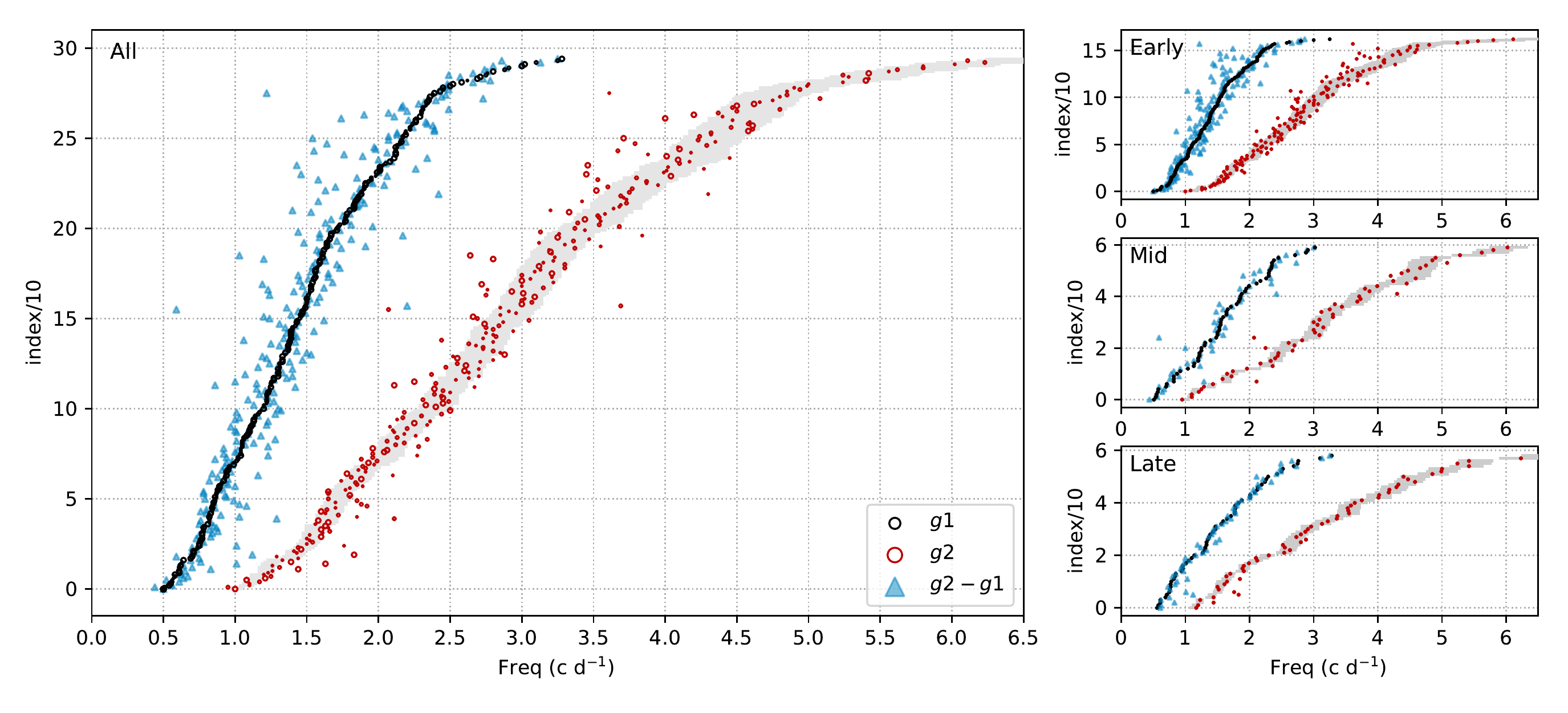,clip=,width=0.98\linewidth}
\caption{The center of the two main frequency groups are shown for each system with a typical frequency group configuration. The stars are ordered according to the location of $f_{g1}$, increasing upwards. Black (red) dots mark $f_{g1}$ ($f_{g2}$), and green triangles mark $f_{g2} - f_{g1}$. The grey lines are centered at $2\times (f_{g1} \pm 5\%)$. In the main panel, marker size is proportional to the relative group strength.
}
\label{fig:freq_groups_all}
\end{figure*}

\begin{figure}[ht!]
 \centering
 \includegraphics[width=0.49\textwidth,clip]{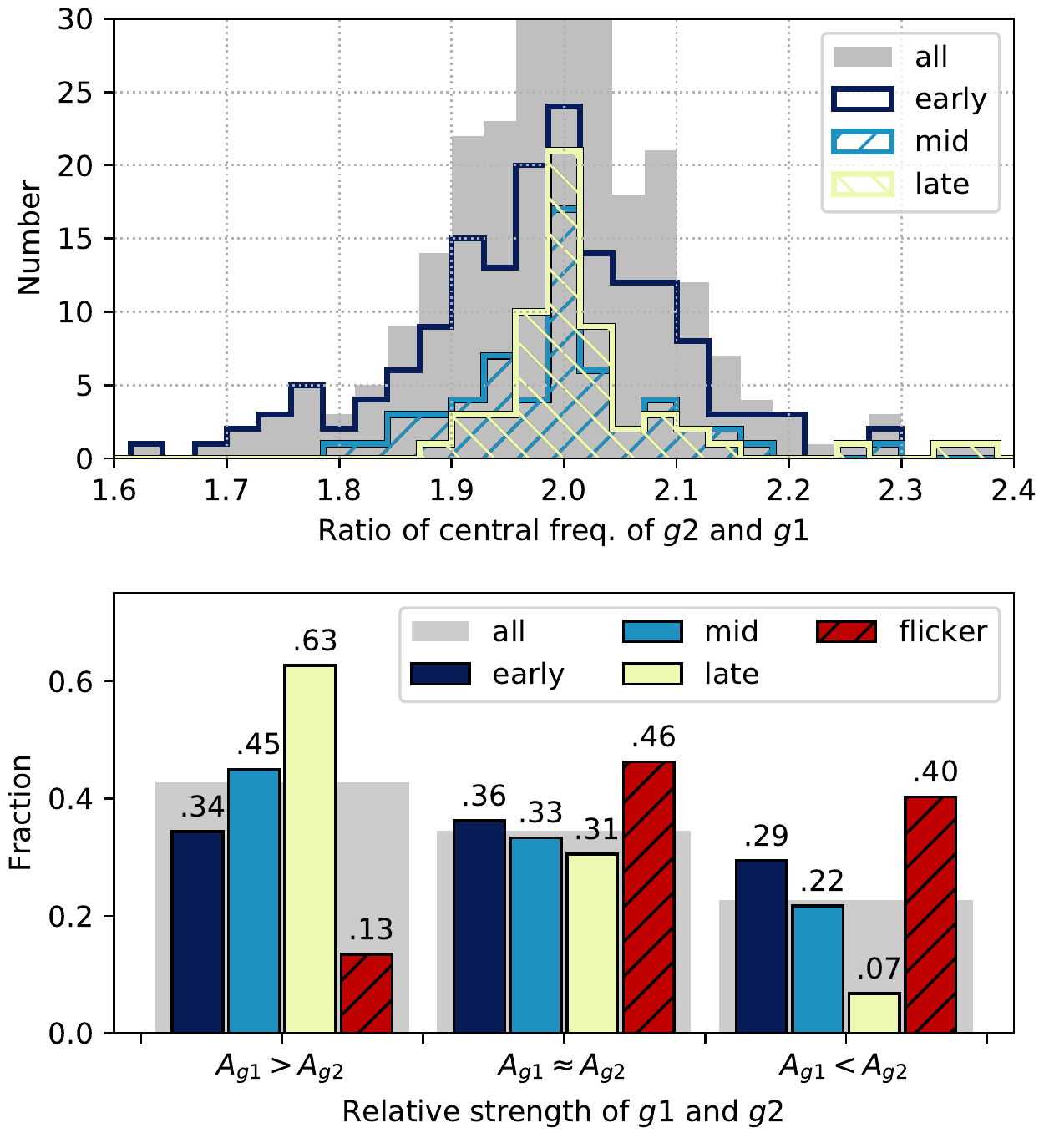}
  \caption{\textit{Top}: Histogram of the ratio of the weighted central frequency of $g2$ and $g1$ ($f_{g2}/f_{g1}$). \textit{Bottom}: Comparison of the relative strength of $g1$ and $g2$, with the fraction of each spectral sub-type bin falling into the three categories printed above its respective bar in the plot. A comparison of the relative strengths of $g1$ and $g2$ is also made for all systems with flickers (the majority being early-type).}
  \label{fig:comparison_g1g2}
\end{figure}

All cases where the typical group configuration is realized (and when it is not) and the relative strength and central location of $g1$ and $g2$ were recorded. No attempt was made to specify the center of $g0$ because the frequency resolution in this regime is poor owing to the short TESS observing baseline, except that, per the adopted definition, $g0$ is centered at $<$ 0.5 \cd.

Among the systems with frequency groups, 81\% (295/366) show the typical configuration with a well defined $g1$ and $g2$, with this fraction decreasing from early to mid to late spectral types (85\% (163/193), 82\% (60/73), and 72\% (59/82), respectively, but this decrease is likely impacted by instrumental sensitivity, as amplitudes are low for later spectral types). When a frequency spectrum does include groups, but does not follow the typical configuration, there is often just one group, or, less commonly, there are multiple groups that are clearly not harmonically related (\textit{e.g.} the second group is not located near twice the frequency of the first group). Panel 10 of Figure~\ref{fig:canon_freq_groups} is an example with atypical frequency groups, with a single group centered at 2.6 \cd without a second group at approximately twice this frequency. 
The lack of a typical group configuration is not evidence against a given system being (or including) a classical Be star \cite[\textit{e.g.}][]{Rivinius2020}. 

For all systems with groups in the typical configuration, the weighted central frequency of the two primary groups ($f_{g1}$ and $f_{g2}$), and their differences, are shown in Fig.~\ref{fig:freq_groups_all}, organized according to $f_{g1}$.
Fig.~\ref{fig:comparison_g1g2} also presents information about these groups, showing the ratio of the central frequency of $g2$ and $g1$ ($f_{g2}$/$f_{g1}$), 
and the fraction of systems having groups of similar strength ($A_{g1} \approx A_{g2}$), $g1$ being stronger than $g2$ ($A_{g1} > A_{g2}$), or $g2$ being stronger than $g1$ ($A_{g1} < A_{g2}$). 

From these plots, it is clear that the ratio $f_{g2}$/$f_{g1}$ is centered at 2.0 (or equivalently, $f_{g2} - f_{g1} \approx f_{g1}$), having a higher (lower) scatter for earlier (later) spectral subtypes. The distribution of the relative strength of $g2$ and $g1$ differs across the range of spectral sub-types, where later type stars are preferentially found to have a relatively strong $g1$ ($A_{g1} > A_{g2}$), with only 7\% having $A_{g1} < A_{g2}$. A similar, but more mild trend is seen in the mid-type stars, and there is a roughly flat distribution for early-type stars. Among the 67 systems with flickers and a typical group configuration, a different trend is seen where only 13\% have $A_{g1} > A_{g2}$. 
Comparing the strength of $g1$ and $g2$ is somewhat subjective, but this decision is aided by the algorithm used to identify groups and their central frequency and relative strengths described in Sec.~\ref{sec:groupcenter} and Appendix~\ref{sec:appendix_groups}.

\subsubsection{Frequency groups and their relevance in the Be phenomenon}

Frequency groups in classical Be stars are clearly complex in their nature. The majority of Be stars do exhibit two or more frequency groups, suggesting they are an important aspect of the Be phenomenon. However, a significant fraction of these systems show only one or zero frequency groups, suggesting that two or more frequency groups, at least in some systems, is not a necessary observable condition for the ejection of matter. 
This is contingent on the noise level (perhaps there are low amplitude frequency groups below the detection threshold), and it is also sometimes observed that the strength of groups and/or their most prominent frequencies can significantly vary over time \citep[][Labadie-Bartz et al. 2020, subm.]{Smith2006,Borre2020,LabadieBartz2020b}, so that the lack of observed groups at the time when a given star is observed by TESS does not necessarily mean that groups have not existed in the past (or will not become more prominent in the future). 

\subsection{Flickers} \label{sec:flickers}

\subsubsection{Background and overview} \label{sec:flickers_overview}
Outbursts (discrete episodes of mass ejection) are a well-known and common feature of Be stars which manifest in photometry as an increase in brightness as the growing quantity of circumstellar material emits and re-processes stellar light \citep{ Haubois2012, Sigut2013, LabadieBartz2018}. In the case of systems viewed at high inclination angles ($i \gtrsim 75$ degrees), the growth of a disk instead causes a net dimming, as the relatively cool circumstellar material obscures the stellar photosphere. The timescales of outbursts range from days to many years or even longer \citep{LabadieBartz2018, Rimulo2018, Bernhard2018, Ghoreyshi2018}. The term ``flicker'' can be used to describe outbursts with short timescales
\citep[days to weeks, ][]{keller2002}. With the relatively short baseline of TESS light curves, it is only possible to probe these flickers on short timescales. Due mostly to a lack of a sufficiently large sample with high-precision near-continuous photometry, flickers with timescales of days have so far been poorly studied.

As exemplified in the Be star HD 49330 observed with CoRoT \citep{Huat2009}, a typical flicker event can be broken up into four phases -- the relative quiescence phase (where the light curve is flat), the precursor phase (where the brightness gradually and mildly decreases), the outburst phase (where the brightness increases), and a relaxation phase (where the brightness returns towards the baseline level). 
These phases also have spectroscopic counterparts \citep[\textit{e.g.}][]{Rivinius1998}. 
Precursor phases are not always seen.

\subsubsection{Main results from TESS}
\begin{figure}[ht!]
 \centering
 \includegraphics[width=0.49\textwidth,clip]{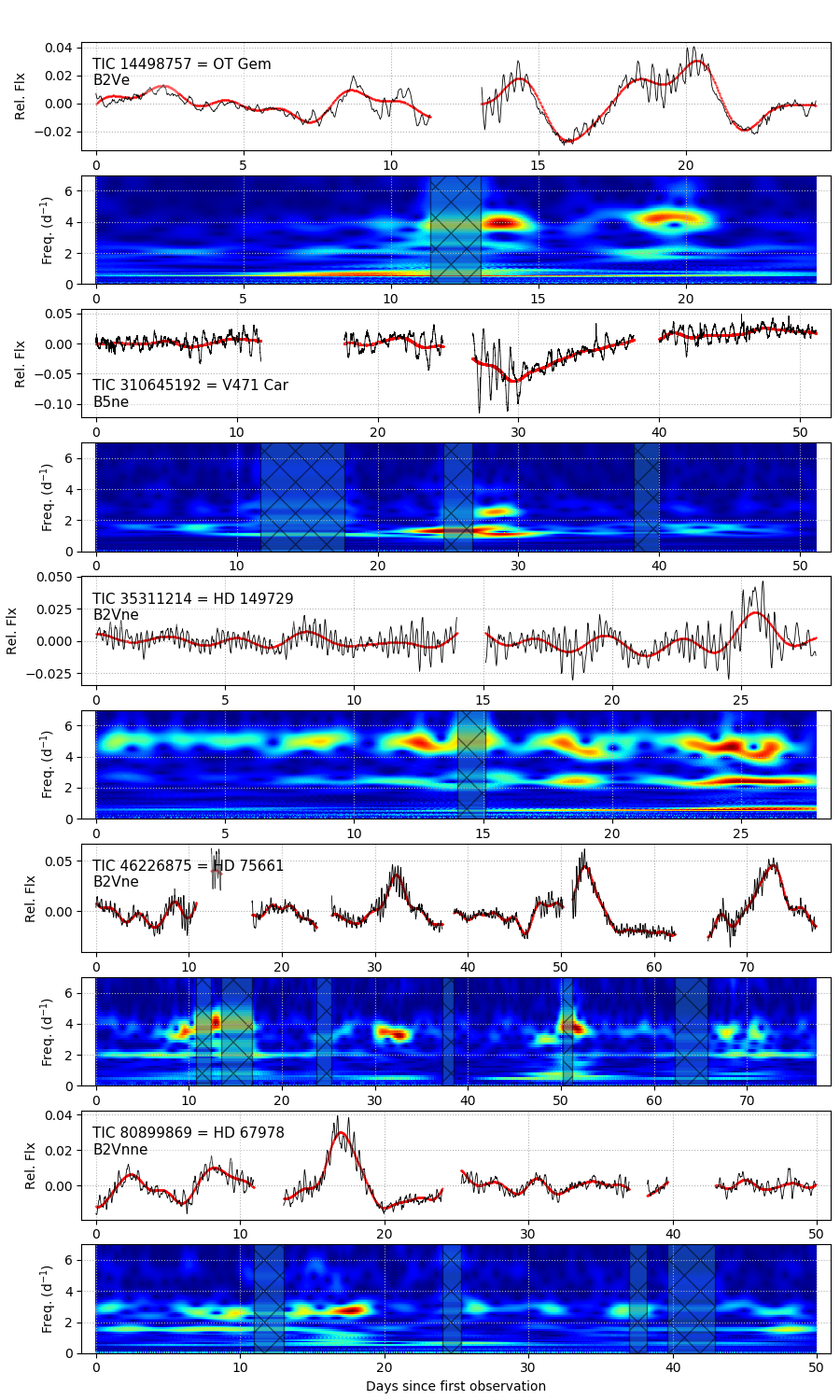}      
  \caption{ Examples of systems with flickers. Below each light curve is the wavelet plot showing the evolution of the frequency spectrum over time after subtracting the low frequency signals (the red curve). 
  }
  \label{fig:multi_flicker}
\end{figure}

Flickers were found in 17\% (735/432) of the sample. There is a strong dependence on spectral type, with 30\% (65/218) of early systems, and 8\% (6/79) and 1\% (1/112) of mid- and late-type systems showing flickers. This is consistent with other observational studies that find early Be stars to be significantly more variable and active, and with relatively large disk densities \citep[and therefore higher amplitude observables,][]{Vieira2017, LabadieBartz2018}. Figure~\ref{fig:multi_flicker} shows the light curve and wavelet plot for five examples of systems with flickers. 

In this work, the vast majority of flickers are identified by a net brightening.
Dimming flickers in shell stars are relatively difficult to confirm with photometry alone on the short timescales of TESS observations, largely because of the relatively small amplitude and naturally longer timescale of the change in optical continuum flux compared to equivalent events viewed at lower inclination angles \citep{Haubois2012}. Nevertheless, there are a few instances in this sample where dimming flickers are apparent. The second panel of Fig.~\ref{fig:multi_flicker} shows one example. 

Systems viewed at an inclination angle of $i \sim 70$ degrees may exhibit no detectable change in their optical flux associated with the growth or dissipation of a disk -- single-band photometry is largely blind to disk events at this intermediate inclination angle \citep{Haubois2012}. 
Some Be stars build up disks over timescales much longer than a TESS observing sector \citep[\textit{e.g.}][]{Rimulo2018,Bartz2017}, where the rate of change in brightness is gradual and difficult to detect in TESS. Mass loss happening on top of a strong pre-existing disk may impart little to no photometric excess, as Be disks can become saturated in certain observables, including visual continuum flux \citep{Haubois2012}. Weak mass loss events can occur, which may still contribute to the disk mass budget, but without a significant detectable photometric signature. 
For these (and possibly other) reasons, the incidence rate of flickers in our sample is underestimated, and a lack of flickers in a given system does not imply the absence of ongoing mass ejection.

\subsubsection{Photometric flickers as tracers of mass ejection} \label{sec:flickers_massejection}

\begin{figure*}[ht!]
 \centering
 \includegraphics[width=0.99\textwidth,clip]{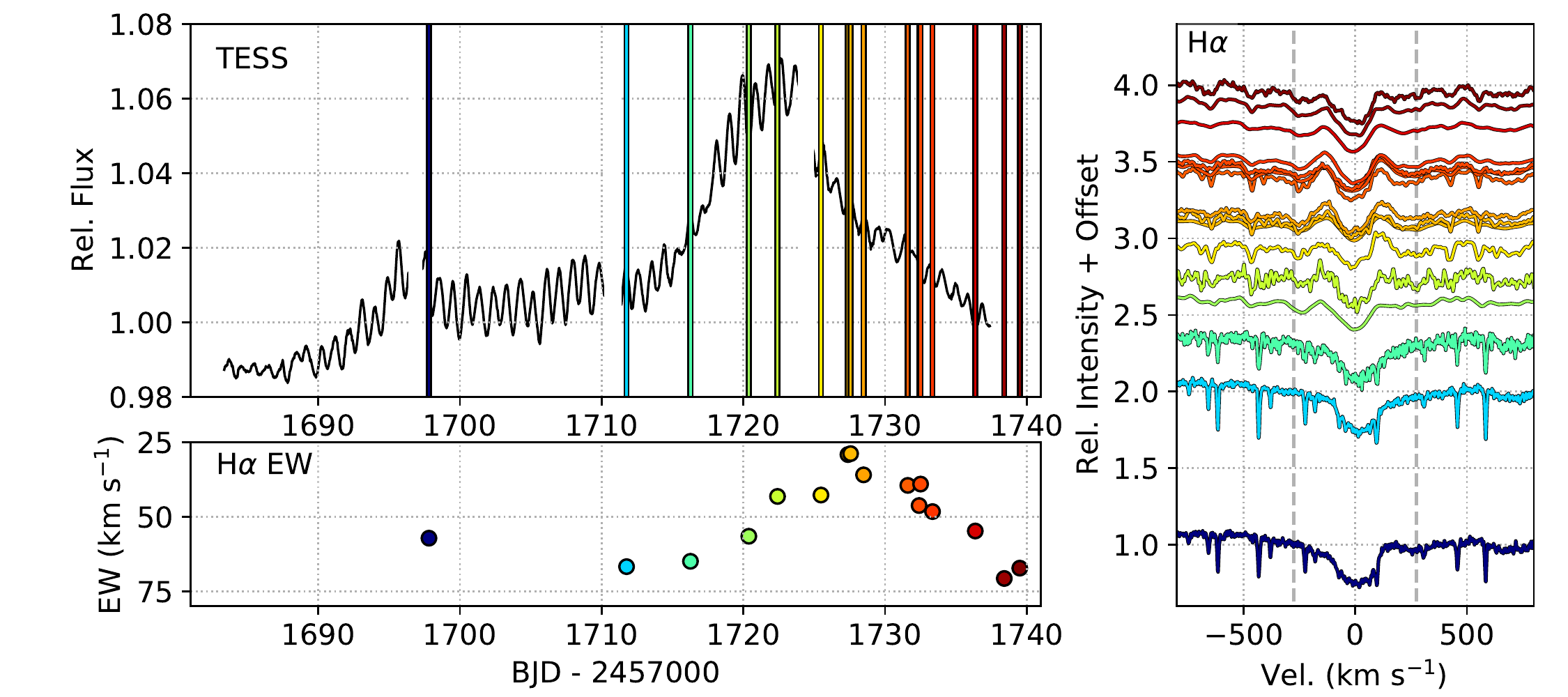}      
  \caption{ An example of a star with a typical flicker that was monitored spectroscopically during TESS observations. A portion of the TESS light curve is shown, along with the H$\alpha$ line from the spectroscopic observations. The vertical colored lines indicate the spectroscopic epochs, with the corresponding line profiles in the right panel (time increasing upward). H$\alpha$ emission begins to appear near TJD = 1720, indicating that material is being ejected into the circumstellar environment. Line asymmetries at the first two spectra may also indicate mass loss, albeit at too small rate to support disk formation. The H$\alpha$ equivalent width (EW) is plotted in the bottom panel. 
  }
  \label{fig:HD194779}
\end{figure*}

Fig.~\ref{fig:HD194779} demonstrates an example where the TESS brightness and H$\alpha$ line profile evolve together during a flicker for the Be star HD 194779, which was observed multiple times spectroscopically over the TESS observing baseline. In this system, the H$\alpha$ line indicates a transition from a disk-less to a disk-possessing state. This transition unambiguously indicates the ejection of material from the star into the circumstellar environment, which coincides with a flicker as seen in the TESS photometry. Although this system was observed in TESS Cycle 2, and is thus not included in the present sample, it serves to illustrate the connection between a photometric flicker and the ejection of stellar material. A spectroscopic campaign including HD 194779 and many other Be stars contemporaneous with TESS observations is ongoing and will be the subject of forthcoming works, with preliminary results reinforcing the interpretation of photometric flickers as mass ejection events.

\subsubsection{Flickers and frequency groups} \label{sec:flickers_groups}
Other works have presented evidence of the frequency spectrum changing before, during, and after outbursts \citep[\textit{e.g.}][]{Huat2009, Baade2018, Semaan2018}. This is also seen in the majority of flickering Be stars in this sample, as there are many cases where flickers coincide with the emergence or enhancement of frequency groups. About 83\% (62/75) of stars with flickers show this behavior. Under the interpretation that flickers correspond to mass ejection events, there is some ambiguity in interpreting the enhancement of frequency groups in these instances. Frequencies about 10\% lower than the dominant pulsation frequency were first discovered in the line profiles of Be stars actively ejecting material (and later seen in photometry of active Be stars), with the preferred interpretation being that they arise in the circumstellar environment due to an inhomogeneous distribution of recently ejected material orbiting the star \citep[{\v S}tefl frequencies;][]{Stefl1998,Baade2016}. The enhancement of the overall strength of a frequency group during times of outburst may then be related to increased pulsational amplitude and/or circumstellar variability. Circumstellar signals are expected to be stronger in systems viewed at higher inclination angles \citep{Baade2016}.

All systems exhibiting flickers also have one or more frequency groups, suggesting a strong link between these features.
In addition to the net change in brightness, which is the hallmark of a flicker, one or more frequency groups are often (but not always) enhanced near in time to the flicker event. Figure~\ref{fig:multi_flicker} demonstrates examples of this common trait. In systems with typical group configurations and flickers, it is more common for $g2$ to be more strongly variable than $g1$.

The temporary in-phase superposition of two or more modes in a frequency group has been in some instances found to correspond to mass ejection episodes \citep{Rivinius1998b,Baade2016,Baade2018,Baade2020}. In some of these cases, the resultant amplitude is greater than the sum of the base mode amplitudes, perhaps indicating nonlinear amplification (as opposed to beating). The enhancement in frequency group strength seen to commonly coincide with flickers in the TESS data may reflect this -- \textit{i.e.} non-linear amplification (or in the simplest case, linear beating) in the amplitude of frequency groups may be triggering some of the flickers.

\subsubsection{Comparison to other studies} \label{sec:flickers_comparison}
Flickers on the short timescales that are seen in TESS data have been captured in other Be stars with space photometry. In Kepler data for 3 Be stars, flickers are found in two systems with timescales of about 10 -- 50 days \citep{Rivinius2016}. Using data from CoRoT, \citet{Semaan2018} analyzed light curves for 15 Be stars, highlighting 6 with flicker events with total durations between about 10 -- 60 days. However, until this work, there has been a lack of analysis of a large set of space photometery for Be stars. 
The large dataset provided by TESS opens a window into studying these events in a larger number of individual systems, providing insight into the types of systems showing small-scale flickers and how frequently they occur. A detailed analysis of the flickering behavior of Be stars will be the subject of a forthcoming paper.

\subsection{High and Very High Frequencies} \label{sec:high-freq}

\subsubsection{Main results for high frequencies ($6 < f < 15$ \cd)}
A subset of the sample (14\%, 61/432) has high frequency signals that are not simply harmonics of the typical frequency groups (or isolated signals at lower frequencies), but are rather individual frequencies (or they can exist in groups, but with a different morphology or configuration than the typical lower frequency groups).
It should be kept in mind, however, that the short baseline of TESS implies a frequency resolution of $\approx0.04$\,\cd, so it is possible that what is here detected as single frequencies might be two or more unresolved frequencies.

\Bcep stars are early B-type (roughly B0 -- B2.5) that pulsate in p modes with typical frequencies between 3 -- 12 \cd \citep{Stankov2005}. While Be stars most typically show low frequency pulsation similar to the SPB stars \citep{DeCat2002}, some classical Be stars are also observed to have \Bcep pulsation \citep[\textit{e.g.}][]{Naze2020, Huat2009, Walker2005b,2020AJ....160...32L}. It is likely that some fraction of the Be stars that show these high frequency signals in TESS also belong to the class of Be stars that are SPB/\Bcep hybrid pulsators. 

However, signals in the traditional \Bcep regime are not necessarily p modes, because a pulsational signal in the SPB regime in the co-rotating frame can be pushed to higher observed frequencies in rapidly rotating stars via the equation $f_{\rm obs} = f_{\rm corot} - m ~\Omega$, where $\Omega$ is the rotational frequency of the star, and $m$ is the azimuthal order of the pulsation mode (being negative for prograde modes, and positive for retrograde modes). However, such a situation usually appears in the frequency spectrum as a harmonic series of groups (\textit{e.g.} panels 7 and 9 of Figure~\ref{fig:canon_freq_groups}), and these signals would therefore not be considered as (isolated) high frequencies. There is also the possibility of combination frequencies as described in \citet{Kurtz2015}, which may be an important ingredient in the observed frequency spectra.

Perhaps surprisingly, the presence of high frequency signals is not limited to early type stars. This is counter-intuitive if these signals reflect \Bcep pulsation, since the \Bcep phenomenon (\textit{i.e.} p modes excited by the $\kappa$ mechanism) is restricted to early B-type stars. We find high frequency signals in 16\% (34/218) of early-type, 17\% (13/79) of mid-type, and 11\% (12/112) of late-type stars.

\subsubsection{Main results for very high frequencies ($f > 15$ \cd)}

Very high frequencies are uncommon, with only 3\% (13/432) of this sample have signals beyond the typical \Bcep regime with frequencies greater than 15 \cd. The highest frequency signals detected are near 75 \cd in the system TIC 427400331 = HD 290662. Signals at the lower end of this very high frequency regime may be typical p mode pulsation shifted to higher observed frequencies due to rapid rotation, as is suggested for the signals at 17.27 and 19.31 \cd in the Oe star $\zeta$ Ophiuchi \citep{Walker2005b}. Or, in some cases, combination frequencies (and harmonics, which are explicitly not included in this aspect of this work) can produce signals at very high frequencies, and therefore some very high frequency signals may not be independent, but rather a sum of two lower frequency signals \citep[or perhaps a more complex combination,][]{Kurtz2015,Burssens2020}.

However, towards the higher end of these very high frequencies, and also for Be stars of later spectral type, it is unlikely that they arise in the Be star. Instead, these may indicate the presence of a stellar companion pulsating at these relatively high frequencies. This may be supported by the higher incidence of very high frequency signals in later spectral types (2\% (5/218), 3\% (2/79), and 5\% (6/112) in early, mid, and late types, respectively) which is an expected trend if these signals do indeed indicate a companion since the contrast ratio between the Be star and a hypothetical companion will be less severe for Be stars of later spectral type. However, the number of systems with very high frequencies is far too small to draw meaningful conclusions.

All 13 Be star systems (plus two systems rejected from the sample) with very high frequencies are briefly discussed in Appendix~\ref{sec:appendix_HF}, including plots of their frequency spectra. These should be studied further to confirm or reject their classical Be nature, and to attempt to resolve the origin of these very high frequency signals.
For stars with 2-minute cadence light curves, frequencies out to 360 \cd can be probed, as opposed to 24 \cd for 30-minute cadence data. This limits our ability to detect the highest frequency signals to the $\sim$65\% of systems with 2-minute cadence data, and so the percentage of systems in which very high frequencies are detected is likely slightly underestimated. 
Among the 13 Be stars with frequencies $>$ 15 \cd, only 3 have signals above 24 \cd.

\subsubsection{Binarity and composite frequency spectra}
Binarity is common in the massive star population, with the majority of B-type stars being members of a multiple star system \citep[\textit{e.g.}][]{Kouwenhoven2007}. \citet{Oudmaijer2010} find essentially the same binary fraction and properties when comparing Be vs. B stars. While the binary fraction of Be stars is poorly constrained (due largely to their broad absorption lines and high luminosity), many Be stars are known to have a binary companion. The frequency spectra of binary systems will then contain signals from two (or more) sources if they are variable. In fact, the prototype of the \Bcep class, $\beta$ Cephei, is a binary with a slowly rotating \Bcep primary and a rapidly rotating mid B-type classical Be star \citep{Wheelwright2009}. Often in Be binaries the companion to the Be star is an evolved object, such as an sdOB star \citep[][and references therein]{Gies1998, Chojnowski2018, Wang2018}, a neutron star \citep[\textit{i.e.} a Be X-ray binary system;][]{Ziolkowski2002}, or even a black hole \citep{Casares2014}. Even though a Be companion need not be significantly evolved, a recent study \citep{Bodensteiner2020} found no firm evidence for main sequence companions around Be stars
(note, however, that low-mass main sequence companions are difficult to detect because the total flux of the system tends to be dominated by the Be star, and that this study does not include the aforementioned $\beta$ Cephei system). Some unknown fraction of this TESS sample of Be stars are in binaries. It is therefore possible that some of the high and very high frequency signals detected in the TESS data arise not from the Be star, but from a companion. Compounded by the large pixel size of TESS, it is not possible to resolve these cases and determine which signals originate in the Be star from just TESS photometry.

\subsubsection{Potential pulsating companions}

Some examples of stars with pulsation at greater than 15 \cd include the main sequence A-type roAp stars with periods between 6 -- 20 minutes ($f \sim$ 72 -- 240 \cd) and amplitudes of a few mmag \citep{Kurtz1982}, and the (usually, but not always, main sequence A/F type) $\delta$ Scuti stars with typical periods between 0.02 - 0.25 d ($f \sim$ 4 -- 50 \cd) and typical amplitudes between 3 -- 10 mmag, but sometimes up to and in excess of 0.3 mag \citep[the high amplitude $\delta$ scuti stars;][]{Breger2000, Garg2010}.

Perhaps most relevant for Be star systems are those with hot subdwarf (sdOB) companions, where at some earlier time the initially more massive star expanded and donated mass and angular momentum to the present-day Be star, shedding its envelope and leaving behind its core (the sdOB star). Such Be+sdOB systems typically have (near) circular orbits with periods of roughly 100 days \citep[\textit{e.g.}][]{Bozic1995,Bjorkman2002,Chojnowski2018,Peters2016}. 
sdOB stars are also known to pulsate in this high to very high frequency regime. In sdOB stars, p-mode pulsations typically have periods between 2 -- 10 minutes ($f \sim $ 144 -- 720 \cd), and amplitudes of about 1\% \citep[$\sim$10 mmag,][]{Kilkenny1997}, and g modes oscillate with periods between about 45 minutes -- 2 hours ($f \sim $ 12 -- 32 \cd) and amplitudes typically between 0.1 -- 0.5 \% \citep[$\sim$1 -- 5 mmag,][]{Green2003, Kawaler2010}. However, in some cases the amplitudes can be much larger, and an sdOB star can pulsate in both p and g modes. For example, the sdO star PG 1605+072 shows over 55 modes, the strongest of which has a period of 8.03 minutes ($f = 180$ \cd) and an amplitude of over 50 mmag \citep[about 5\%, ][]{Pereira2004}. \citet{Sahoo2020} analyze TESS data for three hot subdwarfs, finding frequencies between about 9 -- 300 \cd, with most of the signals being between about 15 -- 30 \cd. 

In most cases, the Be star dominates the visible flux in Be binary systems. This can severely dilute any signals originating in a companion, to the point of being undetectable even with space photometry. However, there are many possible configurations where such a signal in a companion can rise above the detection threshold despite the contaminating flux from the Be star. For example, in the well known B0.5Ve+sdO binary $\phi$ Per \citep{Gies1998, Poeckert1981}, the sdO star contributes approximately 3\% of the total visible flux \citep{Mourard2015}. If the sdO star were to pulsate with an amplitude of 50 mmag (like PG 1605+072), this signal would be diluted down to an amplitude of $\sim$1.5 mmag, which is easily detectable in space photometry provided the observing cadence is such that the frequency of the signal is not beyond the Nyquist limit and the exposure time is sufficiently short. The flux ratio between the Be star and its companion can be even lower for later type Be primaries, such as the B5e primary in the 7 Vul binary system \citep[which is suspected to have an sdOB companion;][]{Vennes2011}, although it should be noted that most confirmed Be+sdOB binaries have primaries with spectral types between B0 -- B3 \citep{Wang2018}. 2-minute cadence TESS light curves are much better suited to this task (with a Nyquist limit of 360 \cd) compared to the 30-minute light curves (Nyquist limit of 24 \cd).

These systems with very high frequencies are therefore good candidates for further study to search for evidence of binarity. This can be done through radial velocity monitoring of the Be star and its disk \citep{Miroshnichenko2002,Bjorkman2002}, direct detection of spectral features (and their radial velocity) of a companion \citep{Wang2018, Chojnowski2018}, modeling of the radial structure of the Be star disk to infer truncation from a binary companion \citep{Klement2019}, and to observe the radial and azimuthal structure of the disk to detect density waves that propagate at the orbital period, which are caused by tidal forces from a binary companion \citep{Panoglou2016,Panoglou2018,Panoglou2019,Chojnowski2018,Cyr2020}. Direct interferometric detection may also be possible. 

This is an important topic because the binary fraction of Be stars is poorly constrained, and one of the proposed evolutionary channels by which a Be star achieves its near-critical rotation is through binary interaction \citep{Pols1991, deMink2013}, as opposed to, or in addition to, outward angular momentum transfer from a contracting core in a single stellar evolution scenario \citep{Ekstrom2008, Granada2013}.

\subsection{Stochastic Variability} \label{sec:stochastic}

\subsubsection{Overview and main results from TESS}
\begin{figure*}[ht!]
 \centering
 \includegraphics[width=0.99\textwidth,clip]{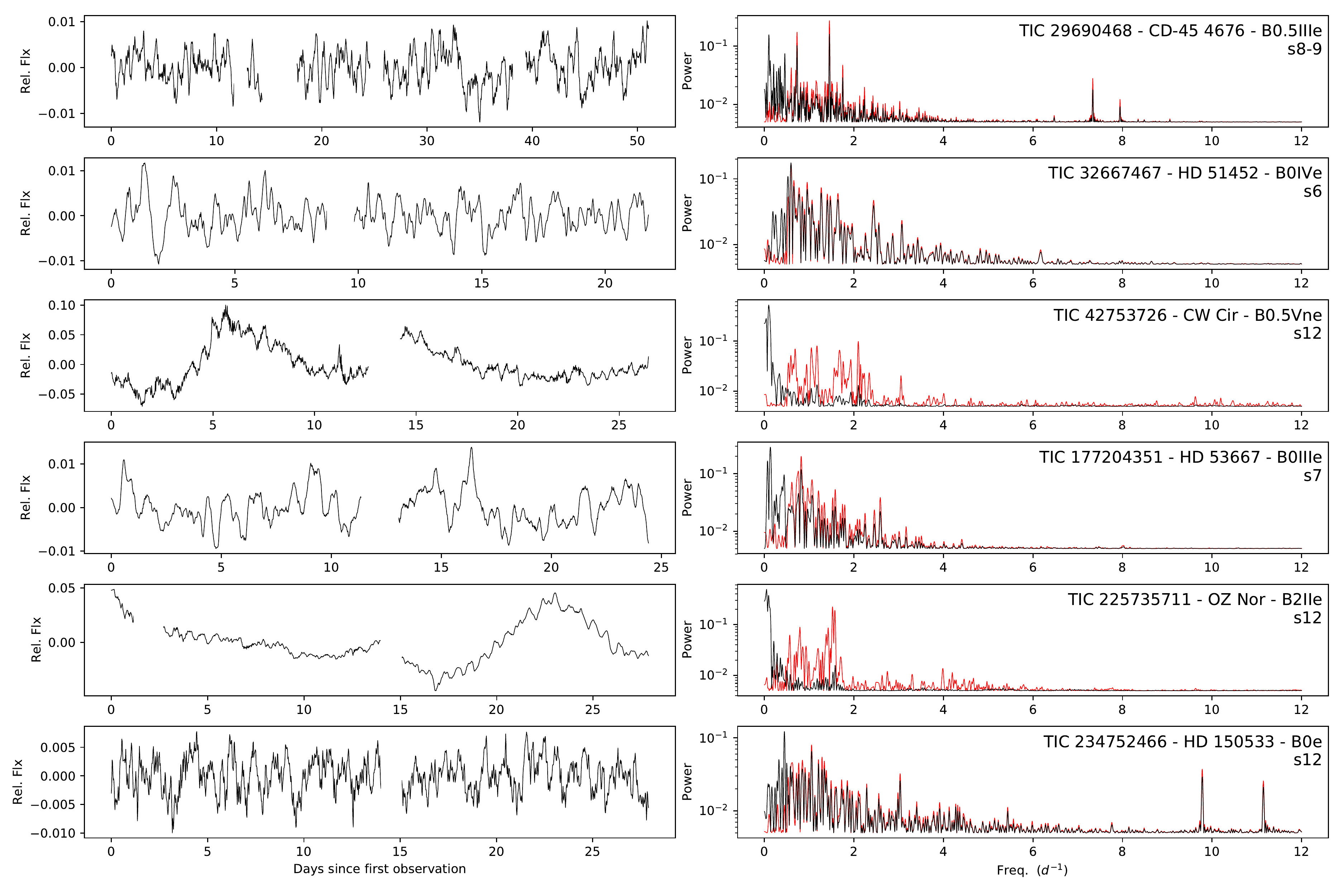}      
  \caption{Light curves (left) and frequency spectra (right) of a few example systems exhibiting stochastic variability. In the frequency spectra plots, the black curve is computed from the raw data (shown in the light curve plots), and the red curve is computed after removing signals with frequencies below 0.5 \cd. 
  }
  \label{fig:multi_stochastic}
\end{figure*}

Stochastic variability is sometimes a prominent feature in the frequency spectra of Be stars, as has been pointed out for some Be stars observed from space \citep[e.g. $\eta$ Cen,][where the star is apparently nearly continually losing mass]{Baade2016}.
26\% (112/432) of this sample has stochastic variability as a prominent feature. Stochastic variability is prominent among early (34\%, 74/218), mid (13\%, 10/79) and late (20\%; 22/112) spectral types, with amplitudes generally being highest in early type stars. Figure~\ref{fig:multi_stochastic} shows 6 systems where stochastic variability is a prominent feature.

In this work, the designation of stochastic variability is applied qualitatively. An important difference between this work and the work and terminology of, \textit{e.g.}, \cite{Bowman2019,Bowman2020}, is that a system is considered to show stochastic variability only if it is obviously a dominant aspect of its light curve and frequency spectrum (viewed in a linear scale). Otherwise, without more in-depth analysis, this label becomes meaningless as virtually every star in this sample is expected to have some degree of stochastic variability (\textit{i.e.} astrophysical correlated red noise) as it is defined in other works like \citet{Bowman2019,Bowman2020}. A proper analysis and parameterization of the stochastic low frequency excess of the TESS Be star sample can and should be undertaken and compared to the results for the more slowly rotating OB star sample of \citep{Bowman2020}. 

\subsubsection{Underlying mechanism(s) and relevance to the Be phenomenon}
Stochastic variability is a ubiquitous feature of OB stars \citep[\textit{e.g.}][]{Bowman2019,Bowman2020}, with a wide range in amplitude and characteristic timescale. There are many physical processes that can plausibly give rise stochastic variability in classical Be stars, including inhomogeneities in the stellar surface or wind combined with rotation  \citep{Moffat2008,Aerts2018,SimonDiaz2018}, sub-surface convection layers \citep{Cantiello2009,Cantiello2011,Cantiello2011b,Cantiello2019}, and internal gravity waves generated at the interface between the convective core and the radiative envelope, which propagate outwards \citep{Rogers2013,Horst2020,Edelmann2019,Bowman2019}. Although stochastic variability does not contain any coherent periodic signals, there is still valuable diagnostic potential in the profile and amplitude of the frequency spectrum of these stochastic signals. For example, \citet{Bowman2020} analyzed TESS data for 70 OB stars with TESS, concluding that there is strong evidence for internal gravity waves by comparing measurements of the stochastic frequency spectrum to models of wave propagation in stellar interiors which originate at interface of the convective core and radiative envelope. 

These internal gravity waves may be an important aspect of Be stars, since they are efficient at transporting angular momentum from the stellar interior outward \citep{Rogers2013,Edelmann2019}. While it is beyond the scope of this work to quantitatively analyze the stochastic variability seen in the TESS Be star sample, it is important to note that these features, although appearing random and incoherent, can still be used as a diagnostic to infer the physical origin(s) of the signals. 

Furthermore, internal gravity waves can drive g mode pulsation in Be star envelopes. This is especially important in early type Be stars, which are generally too hot for the $\kappa$ mechanism to excite g mode pulsations \citep{Dziembowski1993,Neiner2012, Neiner2020}. However, even in the hottest Be stars the $\kappa$ mechanism may still act in some limited capacity to drive g modes if certain conditions are met \citep{Pamyatnykh1999,Neiner2012}, such as enhanced metallicity or if the star is evolved to near the terminal age main sequence (TAMS), but this alone cannot explain the observed distribution of frequency groups among the earliest stars in this sample. 

Since virtually all of the Be stars with frequency groups have $g1$ in the traditional low frequency g mode regime (including the early spectral types, Fig.~\ref{fig:freq_groups_all}), the excitation of these modes by internal gravity waves may even be fundamental to the Be phenomenon in these systems. As outlined in \citet{Neiner2020}, internal gravity waves can serve the dual purpose of transporting angular momentum to the envelope (causing it to spin up and lowering the barrier to mass ejection) and can also excite groups of g mode pulsation, where the activity and processes in the surface layers may then meet the conditions required to trigger an outburst whereby mass and angular momentum is transported to the disk and ultimately out of the system.

\subsection{Low frequency dominated light curves} \label{sec:lowfreq}

The light curves of 32\% (138/432) of this sample are dominated by low frequency variability ($f < 0.5$ \cd), which may or may not be periodic or cyclic in nature. Like most other types of variability, the light curves of early type stars are more likely to display this behavior (47\%; 103/218) compared to mid (20\%; 16/79) and late (13\%; 15/112) types. Flickers stand out as the most remarkable type of behavior that causes a light curve to be dominated by low frequency signals, and in these cases the low frequencies are (at least in part) tied to occasional episodes of mass ejection. 

However, not all systems with low frequency signals exhibit flickers, and systems in which low frequencies dominate are not necessarily ejecting mass during TESS observations. These slow signals can reflect pulsation -- in particular the non-linear coupling of two or more pulsation modes, as a simple beating pattern of two frequencies will not produce power at their beating period in a frequency spectrum.

In the notation of this study and other works, often multiple signals form a low frequency group ($g0$) below 0.5 \cd. In space photometry with longer observational baselines (\textit{e.g.} BRITE, CoRoT, Kepler, and SMEI), it is sometimes found that one or more of the signals in $g0$ are related to prominent signals in $g1$ through difference frequencies. That is, a signal in $g0$ is found at the difference between two signals in $g1$. Since there are often many frequencies in $g1$, $g0$ can be populated by multiple difference frequencies. Unfortunately, the TESS observational baseline is usually too short to confidently measure the position of signals in $g0$, so in the present work no attempt is made to relate the frequencies in $g0$ to those in $g1$.

\subsection{Isolated signals and possible harmonics} \label{sec:isolated}

Many of the frequency spectra (32\%, 138/432) of this sample contain isolated, individual signals that do not obviously belong to a group. These signals favor cooler stars, as 24\% (53/218), 35\% (28/79), and 44\% (49/112) of early, mid, and late type stars, respectively, show these isolated signals. Again, this may be related to instrumental sensitivity, where a frequency group may exist, but only the strongest signal rises above the noise level (an effect expected to be more pronounced in later type stars with lower intrinsic amplitudes).
The prototypical Be star, $\gamma$\,Cas, despite having a spectral type of B0.5\,IVe and being intensively observed for well over 100 years, suffered from this observational bias where only isolated signals were identified \citep[\textit{e.g.}][]{Smith2006,Borre2020} until being observed by TESS, which finally revealed the presence of groups \citep[Labadie-Bartz et al., subm.,][]{Naze2020}. Nevertheless, it is clear that in many cases isolated signals rise well above the noise floor and are apparently unrelated to groups.

Harmonics of isolated signals appear in about 8\% (33/432) of cases (or, in 24\% of stars with isolated signals; 33/138), again being more common towards later spectral types (where 9\% (5/53), 29\% (8/28), and 31\% (15/49) of early, mid, and late type stars with isolated signals have clear harmonics). These signals may reflect individual pulsation modes, or, at lower frequencies, may be related to rotation (either of the Be star, or perhaps a companion), a close binary (possibly in a hierarchical triple configuration with a relatively distant Be star), or are simply the strongest signal in an otherwise undetected group. 

In the adopted characterization scheme, isolated signals must be roughly constant in amplitude over the observing baseline of TESS. Otherwise, they would appear as two or more signals in the frequency spectrum (not necessarily of equal amplitude) as demonstrated in Section~\ref{sec:artificial}. It is, however, certainly possible that seemingly isolated signals can vary in strength over longer time baselines. The case of $\gamma$ Cas exemplifies this, where an apparently isolated signal at 0.82 \cd was known for many years, but with a decreasing amplitude to the point where it is no longer present in space photometry of the system \citep{Borre2020, Henry2012}. The converse is also possible, where a seemingly isolated signal in TESS reflects a pair of frequencies with a beat period greatly exceeding the observational baseline.

\subsection{No variability detected} \label{sec:no-var}
There are 9 systems where no variability is detected. Eight (one) of these systems are of late (mid) spectral type. This does not necessarily mean that these systems do not have variability, as signals below the detection threshold remain possible. Very low photometric amplitudes attributed to pulsation are often found in late type Be stars. For example, a few signals with photometric amplitudes less than 0.05 mmag exist in Kepler data for the Be star ALS10705 \citep{Rivinius2016}, and the Be star CoRoT 102595654 has signals with amplitudes less than 0.2 mmag \citep{Semaan2018}. Further, in some of these cases there is enough spectroscopic data to confirm the presence of a disk that is variable over time \citep[\textit{e.g.} in TIC 247589847 = BD+13 976 = ABE-083, APOGEE spectra show a double-peaked, slightly variable Br11 emission feature over a 393 day baseline,][]{Chojnowski2015}.

\subsection{Non-classical Be Star Systems} \label{sec:non-Be}
A total of 44 systems are rejected from the sample, with 29 of these being confirmed as something other than a classical Be star (see Appendix~\ref{sec:not-Be}, and 15 strongly suspected of being some other type of system (see Appendix~\ref{sec:maybe-not-Be-rejected}). An additional 22 systems are suspected to not be classical Be stars, but lack sufficient evidence and are therefore not rejected from the sample (see Appendix~\ref{sec:maybe-not-Be-not-rejected}).
Some of the types of systems that masquerade as Be stars in this sample include OB stars with strong magnetic fields or chemically peculiar stars, B[e] stars, interacting and/or close binaries, Herbig Ae/Be stars, supergiants, and OB stars embedded in nebulae.
Sections~\ref{sec:not-Be}, ~\ref{sec:maybe-not-Be-rejected}, and ~\ref{sec:maybe-not-Be-not-rejected}  in the Appendix list and briefly discusses these systems. 

Of the original sample of 539 stars, 61 were either not observed by TESS or the data are problematic and were therefore not analyzed further. Then, out of the 478 stars with TESS data that were studied in this work, the 44 objects rejected from the sample suggest a ``contamination fraction'' of $\sim$9\%. Because there are many observational signatures shared by classical Be stars and other types of systems, there are numerous avenues by which a system may have been erroneously classified as a Be star. Some of the objects rejected from this sample have been included in works that aim to describe Be star populations \citep[\textit{e.g.}][]{Fremat2005,Bartz2017}, underscoring the need to carefully vet samples of supposed classical Be stars against contamination from other types of astrophysical systems. This contamination fraction may be slightly underestimated if additional systems, like those in Appendix~\ref{sec:maybe-not-Be-not-rejected}, are revealed to be imposter Be stars.

\section{Discussion of overall results} \label{sec:discussion}

Virtually all of the Be stars observed by TESS in sectors 1 -- 13 are variable. Their photometric signals carry information about the underlying physical processes causing this variability, and a careful study of these observations can elucidate the nature of Be stars as a population especially in regards to pulsation and mass ejection episodes. While there remains much to be explored with these data, the overview of Be star variability seen by TESS presented in this work provides a solid background of their photometric behavior as a population on short timescales and is a first step towards fully leveraging the unique dataset provided by TESS for these and similar objects. A discussion for each characteristic variability pattern is given in Section~\ref{sec:results}. The remainder of this section highlights correlations between different classes of variability.

\subsection{Correlations between variability classifications} \label{sec:correlations}

\begin{figure*}[ht!]
 \centering
 \includegraphics[width=0.99\textwidth,clip]{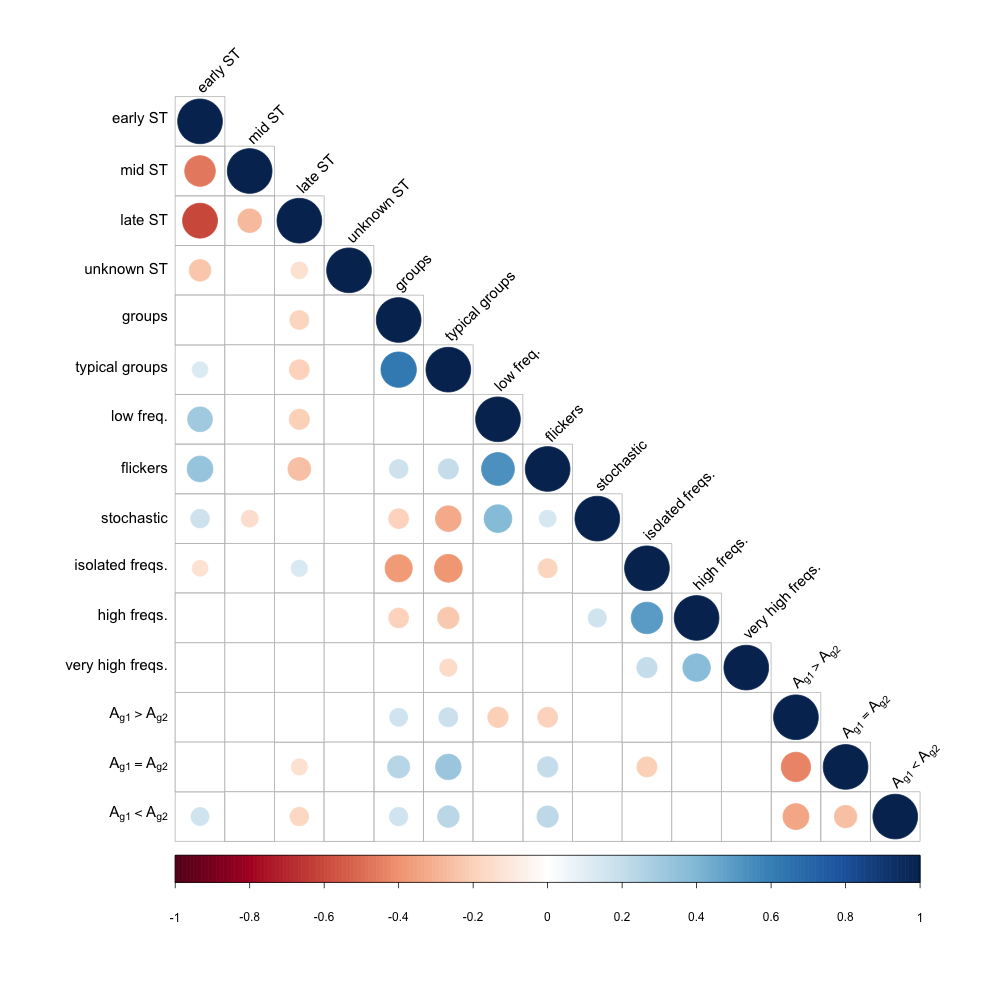}      
  \caption{Correlations between various characteristics of the sample (blue (red) = positive (negative) correlation). The size of each dot is proportional to the strength of the (anti-) correlation. From top to bottom, ``early ST,'' ``mid ST,'' ``late ST,'' and ``unknown ST'' refer to the spectral type bins. ``Groups'' include stars with one or more frequency groups, and ``typical groups'' refer to those with the typical group configuration. ``Low freq,'' ``flickers,'' ``stochastic,'' and ``isolated freqs.'' refer to the variability classifications introduced in Section~\ref{sec:features}. ``High freqs.'' and ``very high freqs.'' refer to systems with high (6 $< f <$ 15 \cd) and very high ($f >$ 15 \cd) frequencies. The remaining categories describe the relative strength of $g1$ and $g2$ in systems with the typical group configuration, where $g1$ is stronger (``$A_{g1} > A_{g2}$''), where both groups are of similar strength (``$A_{g1} = A_{g2}$''), and where $g1$ is weaker than $g2$ (``$A_{g1} < A_{g2}$'').}
  \label{fig:correlations}
\end{figure*}

Figure~\ref{fig:correlations} shows correlations between the observed characteristics of the sample (including spectral type). One of the most interesting correlations is between flickers and frequency groups. If there were no correlation, then 89\% of the early-type stars with flickers are expected to also have frequency groups (58 out of 65), yet \textit{every} star with flickers exhibits frequency groups (and, in 82\% of systems with flickers there is an increase in the amplitude of groups coinciding with the flicker event). This is suggestive of a physical link between frequency groups and the short-duration mass-loss events that flickers are interpreted to represent. 

This is not to say that frequency groups are necessary for a Be star to build up or sustain a disk. For example, HR 6819 (TIC 118842700) does not contain frequency groups, but rather is dominated by low frequency and stochastic variability and unambiguously supports a disk \citep{Rivinius2020}\footnote{HR 6819 is a multiple system that includes a narrow-lined B3 III star, which contributes about 50\% of the total visible flux.}. A caveat is that, given the transient nature in many cases of Be star variability patterns, it remains possible that groups sometimes exist in HR 6819 outside of the short observing baseline provided by TESS. Similar examples include HD 84567 \citep[TIC 11972111,][]{Shokry2018}, HD 53667 (TIC 177204351), HD 254647 (TIC 291385725), HD 44637 (TIC 438103655), and omi Pup \citep[TIC 127493611,][]{Koubsky2012}. 

In stars with flickers that have the typical group configuration, there is a difference in the relative strength of $g1$ and $g2$ when compared with the full sample (as shown in Figure~\ref{fig:comparison_g1g2}). In the 67 stars with flickers and typical groups, 9 have $g1$ stronger than $g2$ (13\%), 31 (46\%) have $g1$ and $g2$ being of similar strength, and 27 (40\%) have $g2$ being stronger than $g1$. In other words, in stars with flickers there is a tendency for $g2$ to be relatively stronger compared to the relative group strengths of the full sample. 
Although not quantified at present, there is a qualitative trend of $g2$ seeing a more dramatic enhancement than $g1$ during flicker events (\textit{e.g.} Figures~\ref{fig:wavelet_01},~\ref{fig:multi_flicker}), but this is not always the case.  

In systems where stochastic variability is a prominent feature, there is a negative correlation with frequency groups, and a positive correlation with high frequency signals, as well as positive correlations with flickers and being dominated by low frequencies. These (anti-) correlations are apparent in the examples shown in Figure~\ref{fig:multi_stochastic}.

There is an anti-correlation between late-type stars and most aspects of the variability of the sample. The exception is a positive correlation between late-type stars and isolated signals.

\section{Conclusions} \label{sec:conclusions}

Following in the footsteps of decades of observations from the ground, and in more recent years also from space, the TESS mission now, for the first time, provides high precision space photometry for the majority of known Galactic Be stars, allowing for a systematic study of their variability on timescales from minutes up to tens of days and down to amplitudes of approximately 100 parts-per-million. Analysis of TESS data for 432 classical Be stars observed in its first year of operation confirms that virtually all Be stars are variable (98\% of this sample is variable at the level of precision available with TESS). The stars in this sample show a variety of characteristic signals, the rates of incidence of which are summarized in Table~\ref{tbl:var_rates}.  Understanding the cause of these features and their incidence rates and patterns are important steps towards elucidating the physical processes in Be stars and other rapid rotators.

In general, Be stars show a higher level of pulsational variability than non-rapidly rotating stars of the same spectral type \citep{Diago2009}, and there is mounting evidence that rapid rotation enhances pulsational amplitudes and the number of excited frequencies \citep{Neiner2012b,Neiner2020,Rieutord2009}.
It is becoming increasingly clear that pulsation is a common, and likely ubiquitous, element of Be stars \citep{Rivinius2013,Semaan2018}, which is supported by this study. 
In particular, the commonality of frequency groups may have important consequences in regards to the Be phenomenon, as this is the most common feature of the sample studied in this work. While it is possible that rotation contributes in some degree to the observed variability of the sample (via an inhomogeneous stellar surface and/or co-rotating clouds), rotation alone is insufficient to explain the majority of the observed signals.
Rather, non-radial pulsation (NRP) has properties that can explain much of the observed variability in this sample. NRP in rapid rotators naturally form frequency groups \citep{Saio2013,Saio2018b,Kurtz2015}, NRP modes transport angular momentum to the surface layers which can assist in the triggering of outbursts \citep{Bowman2019,Bowman2020,Neiner2020}, and NRP modes can couple in the resonant cavity in which they oscillate to produce combination frequencies with amplitudes higher than the sum of the parent modes \citep[which can explain the lowest frequency group, $g0$ in many Be stars, and higher order groups][]{Kurtz2015,Baade2018}. It is clear that high frequency signals (taken to be $f > 6$ \cd, but excluding harmonics of lower-frequency signals) are related to pulsation, and not rotation (as such a rotational frequency would exceed the critical rotation rate). Further, stochastically-driven internal gravity waves (a type of non-periodic pulsation) have emerged as a plausible explanation for the low frequency excess and stochastic variability observed in many OB stars \citep{Bowman2019,Bowman2020}.

Conversely, it is also important to study the Be stars which do not exhibit groups, but instead often are dominated by stochastic variability and/or one or more isolated signals. Understanding the physical properties of the stars belonging to these two categories (with and without groups), especially their rotation rate and evolutionary status, may lead to an improved understanding of how different Be stars eject mass. For example, a hypothesis that can be tested with spectroscopic or polarimetric time-series data is that systems with a high degree of stochastic variation, but without groups, may feed a disk in a more continuous fashion, compared to the episodic mass-loss episodes commonly seen in systems with well defined groups. \citet{Semaan2018} find that both of the Be stars that are evolved slightly past the main sequence, out of their sample of 15, show only stochastic variability, in contrast to the majority of the sample which exhibits frequency groups (12 out of the 13 main sequence stars). Constraining the evolutionary status of the systems observed by TESS may therefore be of great importance in interpreting the observed variability patterns

A non-negligible fraction of the Be stars in this sample (17\% of the total, and 30\% for early type stars) exhibit flickers, which are interpreted as short-lived episodes of mass ejection (see Figure~\ref{fig:HD194779} for a proof of concept, and Figs.~\ref{fig:wavelet_01} and~\ref{fig:multi_flicker} for additional examples). TESS is uniquely suited for capturing such events, which can last for only a few days and change the net brightness of the system by a few percent. Similar events have been observed from space in many Be stars prior to TESS \citep{Semaan2018,Rivinius2013,Baade2016,Baade2017}, but the large number of Be stars observed with TESS finally allows for a substantial sample to be studied. Future work will aim to better quantify the occurrence rate, shape, timing, amplitude, duty cycle, dependence on inclination angle, and dependence on the strength of any pre-existing disk of these events. Hydrodynamic and radiative transfer codes will allow for the modelling of these events, providing estimates of the flux of mass and angular momentum from the star, which will inform estimates as to the degree which these relatively small-scale mass ejection events contribute to the total mass budget of the disk. 

It is well known that mass loss episodes in Be stars can last for decades, years, months, weeks, or days \citep{Rimulo2018,Bartz2017,LabadieBartz2018}. Whatever mechanisms are responsible for opening the mass-loss valve in Be stars must be able to account for this large range in timescales. The increased amplitude of frequency groups coinciding with flicker/outburst events that is commonly seen in this sample (frequency group enhancement coinciding with flickers is seen in 83\% of systems displaying these events) is suggestive of pulsation playing a key role, at least in the relatively short-lived events captured by TESS.
However, we caution that in some situations the enhancement of frequency groups can instead be a consequence of mass ejection if the variability is in some part circumstellar (or perhaps a temporary increase in the amplitude of r modes). TESS, with its short observational baseline, is not sensitive to variability with timescales of tens of days and longer. 

The contribution to massive star science with TESS has only just begun, yet is already producing important results for large samples \citep[\textit{e.g.}][]{Bowman2020,Balona2020,Burssens2020,David-Uraz2019}. Where other space photometry missions have paved the way towards studying Be stars at high photometric precision, TESS is continuing these opportunities for by far the largest sample yet of bright OB stars. The results presented in this paper are intended to be an overview of the types of variability that are seen in Be stars with TESS, and more in depth results regarding the nature of these signals and their significance will be explored further in forthcoming works by our group, and undoubtedly others. Especially exciting, given the brightness of TESS targets, is the relative ease of monitoring these stars with other observational techniques (spectroscopy, polarimetry, multi-band photometry, and interferometry) from the ground which, when combined with TESS data, can greatly improve our understanding of these objects.

\clearpage

\acknowledgments*

J.L.-B. acknowledges support from FAPESP (grant 2017/23731-1).
A.C.C. acknowledges support from CNPq (grant 311446/2019-1) and FAPESP (grant 2018/04055-8). T.A. acknowledges support from FAPESP (grant 2018/26380-8 ). A.R. acknowledges support from CAPES (grant 88887.464563/2019-00). A.L. acknowledges support from CAPES (grant 88882.332925/2019-01). P.S. acknowledges support from FAPESP (grant 2020/04445-0).
The authors are grateful to the amateur spectroscopy community, whose observations have directly supported this and forthcoming works, in particular Stephane Charbonnel, Christian Buil, Paolo Berardi, Tim Lester, Alain Maetz, Robin Leadbeater, Erik Bryssinck, Olivier Garde, Franck Houpert, Olivier Thizy whose data was used in this publication. This work makes use of observations from the LCOGT network. This paper includes data collected by the TESS mission, which are publicly available from the Mikulski Archive for Space Telescopes (MAST). Funding for the TESS mission is provided by NASA's Science Mission directorate. TESS Guest Investigator program G011204 provided the 2-minute cadence data. This project makes use of data from the KELT survey, including support from The Ohio State University, Vanderbilt University, and Lehigh University, along with the KELT follow-up collaboration. This work has made use of data from the European Space Agency (ESA) mission {\it Gaia} (\url{https://www.cosmos.esa.int/gaia}), processed by the {\it Gaia} Data Processing and Analysis Consortium (DPAC \url{https://www.cosmos.esa.int/web/gaia/dpac/consortium}). Funding for the DPAC has been provided by national institutions, in particular the institutions participating in the {\it Gaia} Multilateral Agreement. This research has made use of NASA's Astrophysics Data System. This research has made use of the SIMBAD database, operated at CDS, Strasbourg, France. This work has made use of the BeSS database, operated at LESIA, Observatoire de Meudon, France: http://basebe.obspm.fr. This research made use of Lightkurve, a Python package for Kepler and TESS data analysis \citep{Lightkurve2018}. This research made use of Astropy,\footnote{http://www.astropy.org} a community-developed core Python package for Astronomy \citep{astropy2013, astropy2018}.
This work used the BeSOS Catalogue, operated by the Instituto de Física y Astronomía, Universidad de Valparaíso, Chile : \url{http://besos.ifa.uv.cl} and funded by Fondecyt iniciación No 11130702, using the PUCHEROS instrument \citep{Vanzi2012}. The page is maintained thanks to FONDECYT No 11190945.

\vspace{5mm}
\facilities{TESS, Gaia, KELT}

\software{astropy \citep{astropy2013, astropy2018},  
          Period04 \citep{Lenz2005},
          VARTOOLS \citep{Hartman2012}
          }

\clearpage 

\appendix

Appendix~\ref{sec:appendix_groups} describes the clustering algorithm used to characterize frequency groups. The remaining sections provide remarks on individual systems, beginning with those having very high frequencies (Appendix~\ref{sec:appendix_HF}). Systems confirmed to not be classical Be stars are then briefly discussed in Appendix~\ref{sec:not-Be}, followed by systems strongly suspected to not be Be stars (and are thus rejected from the sample; Appendix~\ref{sec:maybe-not-Be-rejected}), and systems weakly suspected of not being classical Be stars, but kept in the sample due lacking solid grounds for rejection (Appendix~\ref{sec:maybe-not-Be-not-rejected}). Appendix~\ref{sec:appendix_sample} includes a table (Tbl.~\ref{tbl:table_sample}) for all the stars in the sample and the variability characteristics assigned to them.

\section{Frequency group clustering algorithm} \label{sec:appendix_groups}

In order to determine groups of frequencies and their ``center of mass'', a python routine was developed to automatically identify these properties from the iterative pre-whitened frequency spectrum for each star. This clustering algorithm served to identifying groups and provide a number to describe their weighted center and net amplitude. 

Broadly speaking, this routine starts with identifying the signals with the highest SNR (the ``seed'' signal for that group), and then moves outwards from these frequencies to identify signals that belong to the same group, while retaining information about the strength of each signal and ultimately computing the group center (with weights given in proportion to the SNR of each signal belonging to a given group). As the algorithm moves outwards from the highest SNR signals, each frequency is compared to the seed signal of the group by the equation

\begin{equation} \label{eq:clustering_algorithm}
    SNR(f_{\rm a}) + SNR(f_{\rm b}) \geq \frac{|f_{\rm a} - f_{\rm b}|^2}{d} ,
\end{equation}
where $SNR (f)$ is the SNR of the frequency $f$, $f_{\rm a}$ and $f_{\rm b}$ are the frequencies being compared and $d$ is a weight parameter, which is calculated by 
\begin{equation} \label{eq:weight_parameter}
    d = \frac{\alpha \times N_f \times I}{SNR_{\rm total}} ,
\end{equation}
where $\alpha$ is a free parameter, $N_f$ is the total number of frequencies in the spectrum, $I$ is the interval of frequencies considered and $SNR_{\rm total}$ is the sum of the SNR of all $N_f$ frequencies. If equation \ref{eq:clustering_algorithm} is false for the given frequencies, the frequency is not included in the group. Otherwise, the compared frequencies are merged inside the group by replacing both with a new signal having a frequency of 
$$f_{\rm new} = \frac{f_{\rm a} \times SNR(f_{\rm a}) + f_{\rm b} \times SNR(f_{\rm b})}{SNR(f_{\rm a}) + SNR(f_{\rm b})}$$ 
and a SNR of
$$SNR(f_{\rm new}) = SNR(f_{\rm a}) + SNR(f_{\rm b}),$$
which is subsequently compared to the remaining signals in the spectrum. After all frequencies are analyzed in this way, the routine returns the resulting clusters of frequencies centered at the weighted average for the group, and with a characteristic strength computed from the sum of the SNRs of the original signals determined to belong to the group. 

Because of the diversity of frequency spectra in the full sample, there is no single value for the $\alpha$ parameter that returns reliable results for all stars. Instead, this algorithm was applied to each star with three empirically determined values of $\alpha$ (1$\times 10^{-3}$, 5$\times 10^{-4}$, and 1$\times 10^{-4}$), where the authors then choose from these three outputs the results that best reflect the behavior of a given frequency spectrum. In practice, often all three values of $\alpha$ return virtually the same group information, but there are cases where a certain value of $\alpha$ is clearly preferable (\textit{e.g.} in densely populated frequency spectra when groups are close together, a high value of $\alpha$ may erroneously merge two adjacent groups). The results from applying this algorithm are what is used to determine the location of each frequency group, and to determine the relative strengths of $g1$ and $g2$ for a given star.

\section{Notes on individual stars}

\subsection{Stars with very high frequencies ($f > 15$ \cd)} \label{sec:appendix_HF}

\begin{figure*}[ht!]
 \centering
 \includegraphics[width=0.95\textwidth,clip]{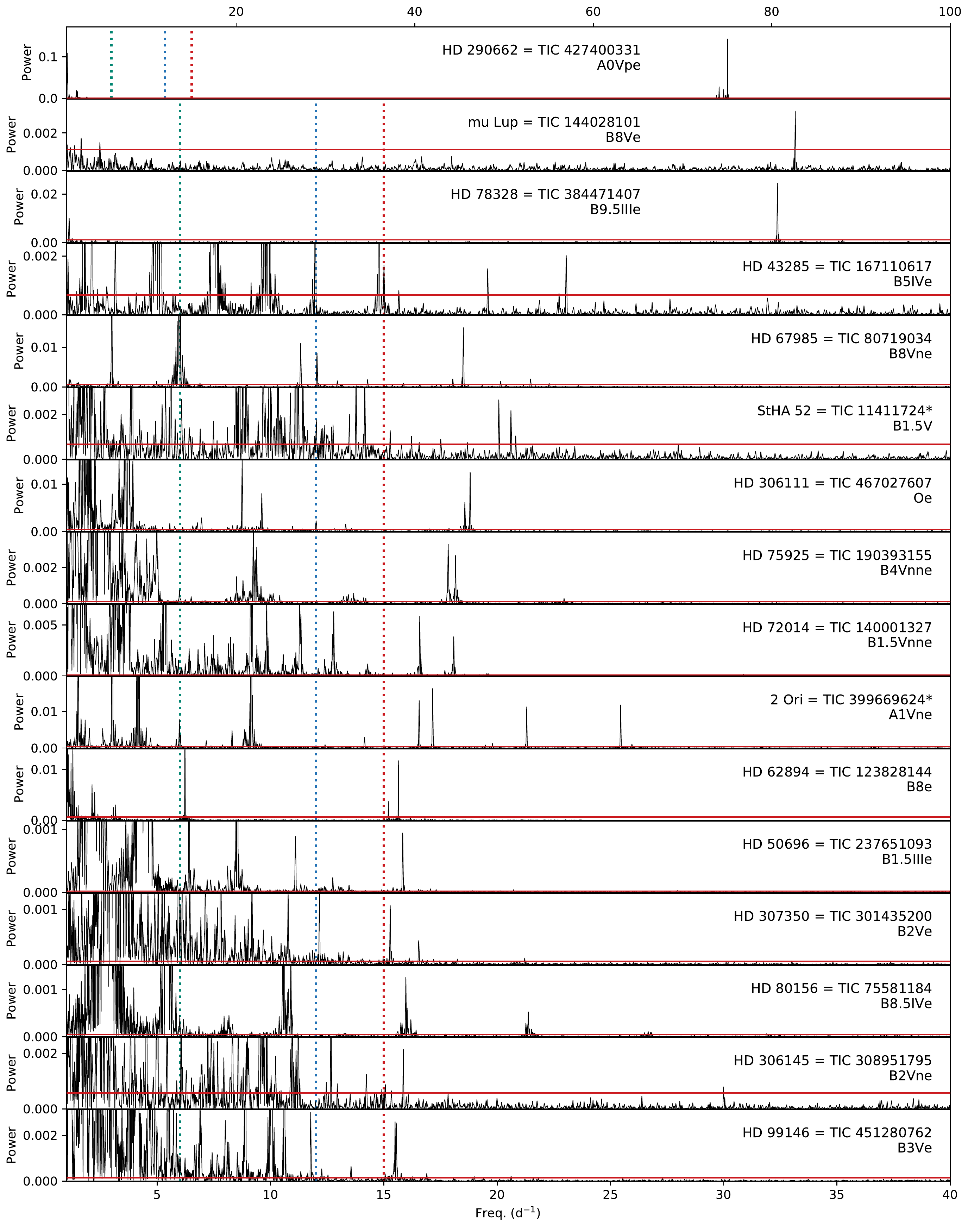}      
  \caption{Frequency spectra of systems with very high frequencies. Vertical dotted lines are at 6, 12, and 15 \cd for reference. In the top panel the x-axis extends to 100 \cd, while the remainder share the same axis, out to 40 \cd. The horizontal red line is 10$\times$ the mean periodogram power for $f > 15$ \cd, shown only for reference. TIC IDs marked with an asterisk are rejected as classical Be stars (see Sec.~\ref{sec:not-Be} and ~\ref{sec:maybe-not-Be-rejected}).
  }
  \label{fig:very_high_freq}
\end{figure*}

All of the Be star systems (plus a few rejected from the sample) with very high frequencies are briefly discussed below. Their frequency spectra are shown in Figure~\ref{fig:very_high_freq}.

{\it TIC 11411724 = StHA 52:} B1.5V. Likely not a classical Be star (see Sec.~\ref{sec:maybe-not-Be-rejected}). Embedded in strong reflection nebula. There is a pair of frequencies around 20 \cd, and 4 signals of similar strength centered around 8.75 \cd.  \\

{\it TIC 75581184 = HD 80156:} B8.5IVe. The most prominent feature is a group centered at 2.70 \cd. Harmonics of this group seemingly extend out to even 50 \cd. It is most likely that these very high frequency signals represent harmonics of lower frequency signals, but this system is unique in the extent of the apparent harmonic series. A possible explanation is that the dominant low-frequency signals are extremely non-sinusoidal. \\

{\it TIC 80719034 = HD 67985:} B8Vne. Two typical groups near 3 and 6 \cd, and a strong low-frequency signal near 0.36 \cd. There are many signals between 11 and 23 \cd with amplitudes between 0.05 -- 2 ppt.  \\

{\it TIC 123828144 = HD 62894:} B8e. There is a triplet centered around 15.6 \cd, with amplitudes between about 0.05 -- 0.2 ppt, and an isolated signal near 6.22 \cd. The dominant signals in the LC are a relatively strong (2 ppt) signal at 0.95 \cd, and its subharmonic (at 0.48 \cd, with an amplitude of 0.4 ppt), plus some lower-level stochastic variability. Blending is a concern, as there is a neighbor of equal brightness (CD-42 3473) at a distance of 24'', which is completely blended in the TESS pixels. It should first be determined if these very high frequency signals arise from the neighbor.  \\

{\it TIC 140001327 = HD 72014:} B1.5Vnne. Many signals between 9 -- 20 \cd. There are also two main frequency groups, short flickers, stochastic variation, and the LC is dominated by low-frequency variation. \\

{\it TIC 144028101 = mu Lup:} B8Ve. Very low amplitude signals in TESS at frequencies of 1.53 and 2.44 \cd (amp = 0.07, 0.06 ppt, respectively), and also two small groups near 14.8 and 17.8 \cd. Nice symmetric double-peaked H$\alpha$ profile in BeSOS in two epochs of 2015 (Feb. and July). The high and very high frequency signals are convincing. mu Lup itself is a close visual double (separated by about 1'') where both components are approximately the same brightness and spectral type \citep{Fabricius2002}. Further, an early A-type neighboring star, HD 135748 (approximately 2 magnitudes fainter), lies 23'' away. These three sources sharing a common proper motion \citep{Gaia2018}. With TESS data alone it is impossible to distinguish the origin of the various photometric signals. \\

{\it TIC 167110617 = HD 43285:} B6Ve. Very unusual frequency spectrum. There are 3 prominent pairs or triplets evenly spaced at 5, 7.5, 10 \cd, and weaker signals between 12 -- 24 \cd, along with clear low-freq signals consistent with SPB pulsation seemingly unrelated to the higher frequency signals. A BeSS spectrum from 2020 March 31 shows H$\alpha$ in pure absorption, with no shell features. Three professional BeSS spectra with FEROS and ELODIE show weak H$\alpha$ emission, so the Be classification seems reasonable. This is a strange system given the LC and relatively late spectral type.  \\

{\it TIC 190393155 = HD 75925:} B4Vnne. Pair of frequencies near 18 \cd. However, these may be combinations of signals in a group that is near 9 \cd. The strongest feature in the LC is a group centered at 2.5 \cd. Blended with the star SAO 220579, which is about 2 magnitudes fainter and 13'' distant. Included in the \citet{Renson2009} catalog of Ap and Am stars, with a note saying that the star is of ``doubtful nature''.  \\

{\it TIC 237651093 = HD 50696:} B1.5IIIe. Has a single frequency near 16 \cd, and other high frequency signals near 6.5, 8.6, and 11.2 \cd. There is typical, symmetric, double-peaked H$\alpha$ emission in one BeSS spectrum from 2001, but more recent spectra do not show evidence of H$\alpha$ emission, consistent with a Be star with a variable disk. \\

{\it TIC 301435200 = HD 307350:} B2Ve. Has a pair of signals near 15.3 and 16.5 \cd, and many high frequency signals between 6 and 15 \cd, plus two pairs of frequencies centered near 3.0 and 3.6 \cd. \\

{\it TIC 308951795 = HD 306145:} B2Vne. High and very high frequency signals between about 9 -- 30 \cd. Possibly not a classical Be star -- see Appendix~\ref{sec:maybe-not-Be-not-rejected}. \\

{\it TIC 384471407 = HD 78328:} B9.5IIIe. Has a single very high frequency signal near 32  \cd, and a group near 0.6 \cd. One low-resolution BeSS spectrum shows H$\alpha$ in weak emission. \\

{\it TIC 399669624 = 2 Ori:} A1Vne. Many high and very high frequencies out to 23 \cd. All signals are rather low amplitude, with the strongest being at 0.22 ppt. Many spectra from BeSS show a weak, symmetric, double-peaked H$\alpha$ emission profile that is roughly stable over 13 years. However, this is not a classical Be star, but is rather a $\lambda$ Boo star. Discussed further in Sec.~\ref{sec:not-Be}.

{\it TIC 427400331 = HD 290662:} A0Vpe. Has a very high-freq group at 75 \cd with 0.6 ppt amplitude, plus regular low-freq groups with amplitudes of around 0.25 - 1.5 ppt. Possibly not a classical Be star -- see Sec.~\ref{sec:maybe-not-Be-not-rejected}. \\

{\it TIC 451280762 = HD 99146:} Has a very closely-spaced pair of signals centered at about 15.5 \cd, plus many other high frequency signals. The frequency spectrum is remarkably rich. Discussed further in Appendix~\ref{sec:maybe-not-Be-not-rejected}. \\

{\it TIC 467027607 = HD 306111:} Oe. Has a pair of signals near 19 \cd, and also a pair centered near 9.25 \cd. There are two typical and wide low frequency groups, centered at 1.8 and 3.7 \cd, and the LC is dominated by low-frequency variation. \\

\subsection{Firmly non-classical Be stars rejected from the sample} \label{sec:not-Be}

{\it TIC 14709809 = RY Gem:} A known ``moderately interacting Algol-type eclipsing binary'' \citep{Plavec1987}. TESS data likewise indicate this is an EB with a period of about 9 days, where the primary eclipse reaches a $\sim$60\% depth. Many BeSS spectra show H$\alpha$ with a clear double-peaked profile and shell absorption, with variable V/R ratios, likely formed in an accretion disk around the A0V (present day) primary.  \\

{\it TIC 30562668 = HD 76838:}. Short-period EB ($\sim$4 d) with $\beta$ Cephei pulsations and reflection effect. Embedded in a nebula, which may have been formed in a past binary interaction that resulted in the current binary configuration. \\

{\it TIC 53063082 = HZ CMa:} Not a classical Be star. The TESS light curve looks much more like rotation or binarity P = 15 or 30 days. Strong single-peaked H$\alpha$ emission in a single high-resolution BeSS spectrum from 2009, with an emission level relative to the continuum (E/C) $\sim$2.3. This system is included in many studies of Be stars \citep{Touhami2011,Bartz2017,Wang2018,Klement2019,Zhang2005,Sterken1996}. This is a binary with an accretion disk and ellipsoidal variability with P$_{orb}$ = 28.6 d, and apparently has a Roche-lobe overfilling K giant companion \citep{Sterken1994}, where the authors interestingly note that the high $v$sin$i$ value \citep[300 km$^s$][]{Meisel1968} of the B star may be a consequence of accretion -- \textit{i.e.} this could be a Be star in the making (but currently has an accretion disk as it is being spun up). Listed as B6IVe+A in \citet{Slettebak1982}. Besides the orbital modulation, the TESS data contains only stochastic variability. \\

{\it TIC 90296023 = FY Vel:} KELT data show this to be a 33.75 d period EB ellipsoidal variable, as reported in \citep{Bartz2017}, and is a known Beta Lyrae type EB \citep{Thackeray1970}. Not a classical Be star. Presumably there is mass transfer via Roche-lobe overflow, leading to the strong H$\alpha$ emission seen in the 4 available BeSS spectra, which show H$\alpha$ E/C $\sim$8 -- 10. \\

{\it TIC 123545883 = V743 Mon:} An unclassified B[e] star, or potentially a Herbig Ae/Be star \citep{Varga2019}, where severe stochastic variability is its main feature as in TESS. \\

{\it TIC 141973945 = bet Hya:} The TESS light curve resembles rotation or binarity with a dominant signal at 0.43 \cd (12.5 ppt), and lower-amplitude harmonics. 4 BeSS spectra all show a shell signature with weak emission. Simbad lists this as an alpha2 CVn type variable. A spectral type of kB8hB8HeA0VSi is given in \citet{Garrison1994}, and is therefore not a classical Be star. \\

{\it TIC 151300497 = V1075 Sco:} Stochastic variability is prominent in TESS, as are low frequency signals. There are narrow ill-defined groups that stand out above the noise. Not a classical Be star, despite H$\alpha$ emission in BeSS spectra. The GOSS survey gives a spectral type of O7.5V((f))z(e) \citep{Sota2014}. \\

{\it TIC 187458882 = HD 57682:} Many spectra in BeSS show the H$\alpha$ emission as being single peaked, very variable in strength, narrow, and seemingly variable in position relative to the main H$\alpha$ absorption. The MiMeS survey find this to be a strongly magnetic O9.5 IV star \citep{Grunhut2017}, and is therefore incompatible with a classical Be (or Oe) designation. Its only feature in TESS is strong stochastic variability. There is also a diffuse nebulae surrounding the star in the WISE red band, and a strong SED excess at the longest radio wavelengths. \\

{\it TIC 207176480 = HD 19818:} High resolution spectroscopy acquired by our group shows this is an SB2, with an A0 main sequence star and a cooler giant. Rotational modulation is apparent in TESS and KELT with the same period of 3.3 days. Hydrogen and other emission features are variable on short timescales and does not resemble a Keplerian disk. We suggest that, due to the rapid rotation of the cool giant star, a strong magnetic dynamo exists causing the observed rotational modulation (due to spots), and the highly energetic flares seen in the TESS data \citep{LabadieBartz2020}. Chromospheric activity is likely the cause of the transient emission features (and the strong X-ray flux). \\ 

{\it TIC 220322383 = 15 Mon:} B1Ve. Very low amplitude signals in TESS, the strongest of which is 0.5 ppt at 12.5 \cd (with a nearby signal at 11.9 \cd). This is a known X-ray source and spectroscopic binary \citep[O7V+B1.5/2V,][]{Skiff2013}, is embedded in a nebula, and is a cluster member. Also classified as an O7V((f))z variable \citep{Sota2014}. \\

{\it TIC 224244458 = bet Scl:} The TESS data more closely resembles rotational modulation with a period of about 2 d. Spectral type of B9.5IIIpHgMnSi \citep{Abt1995}. Rejecting as a Be star -- this is a chemically peculiar star exhibiting rotational modulation. \\

{\it TIC 234813367 = AX Mon:} \citet{Puss2002} classify this as an interacting binary system with a K giant and a B(e) star that is accreting matter, noting the P Cygni profile and unusual emission variations over the orbit. This system appears in numerous studies of populations of Be stars, but an interacting binary with an accretion disk around the hot component is incompatible with the decretion disks of classical Be stars. The TESS light curve is dominated by stochastic variation, but there are clear periodic signals, including one that stands out near 3.5 \cd. Many BeSS spectra show a clear P Cygni profile in H$\alpha$. \\

{\it TIC 234835218 = EM* GGA 395:} A Herbig Ae/Be star \citep{Li2002}. Very clearly in a strong nebula, and the SED indicates a large amount of cool dust. Characterized as having strong stochastic variability and a single frequency near 2.6 \cd in TESS. \\

{\it TIC 238791674 = CD-49 3441:} A Herbig Ae/Be star according to \citep{The1994}. High amplitude stochastic variability is the most prominent feature in TESS. A 6.77 day signal is found in the KELT data \citep{Bartz2017}. Three low resolution BeSS spectra show strong H$\alpha$ emission about 7 times the continuum level (which would be abnormally high for a B8 classical Be star).  \\

{\it TIC 246189955 = HD 328990:} The SED is incompatible with that of a classical Be star, suggesting two stellar components (likely a cool giant, and a late B or early A main sequence star). There is only stochastic variability present, and the variability is dominated by low frequencies. The amplitude is very high ($\sim$10\% max - min) relative to other confirmed classical Be stars of its spectral type, A0e.  \\

{\it TIC 253212775 = V495 Cen:} An eclipsing interacting binary with a 33.5 day orbital period, a cool evolved star, and a hot mid-B dwarf with an accretion disk \citep{Rosales2018}. The TESS light curve is dominated by the orbital modulation, but there are also clear (possibly aperiodic) oscillations with amplitudes of about 1\% and frequencies between 0.5 -- 3 \cd.  \\

{\it TIC 253380837 = HD 113573:} The TESS data unambiguously show this to be an EB with an orbital period of about 1 day, and primary and secondary eclipses that are slightly different in depth, but both around 3\%. Although the field is somewhat crowded, the use of different sized apertures suggests the target star is the source of the EB signal. Besides the orbital period and its many harmonics, no other signals are present in the data.  \\

{\it TIC 289877581 = d Lup:} The TESS variability looks much more like rotation than pulsation. Given the 'p' in the spectral type (B3IVpe), this may have surface spots (chemical or magnetic) leading to modulation at the 0.48 \cd signal -- the strongest signal in the LC (assumed to be rotation). This star is included in catalogs of chemically peculiar stars \citep{Romanyuk2008}, supporting the plausibility of rotationally modulated brightness. The single BeSS spectrum is narrow lined in H$\alpha$, with no sign of emission. \citet{Arcos2018} analyze two epochs of spectra from 2013 and 2015, see no variability or emission in H$\alpha$, and fit blue photospheric absorption lines to arrive at a projected rotational velocity of $v$sin$i$ = 30 \kms, which would be unusually slow for a rapidly rotating classical Be star.   \\

{\it TIC 305090822 = HD 157273:}. Just one long sinusoidal signal exists, with P=15.72d (also very clear in KELT data). Clearly emission in H$\alpha$ in the 3 low-resolution BeSS spectra. The SED appears to be formed from two components. This is probably an interacting binary with an accretion disk and possibly ellipsoidal variation.   \\

{\it TIC 320228013 = HD 308829:} The TESS data show this to be a short-period EB (P$_{orb}\sim$ 6.77 d) with significant asymmetric out-of-eclipse variability possibly caused by some combination of a reflection effect and rotation of an inhomogeneous stellar surface. This system is a cluster member (Cl* IC 2944 THA 51). The relatively dense region of the sky makes blending in TESS problematic, so it is not yet certain that the EB signal can be attributed to HD 308829. The SED has peculiarities. This is a known X-ray source (NRS2013), included in the ROSAT all-sky bright source catalog \citep{Voges1999} and also observed with XMM-Newton, as discussed in \citet{Naze2013}.  \\

{\it TIC 333670665 = V863 Cen:} A known magnetic He-strong star \citep{Shultz2019}. The TESS variability looks more like rotation or binarity than pulsation with a period of about 1.3 d, and a weak harmonic. 3 BeSS spectra show narrow lines with no sign of emission from 2012, 2017, and 2018. \\

{\it TIC 380117288 = AI Cru:} A very short period EB (P$_{orb}\sim$ 1.4 d) with $\sim$50\% primary depth. Previously known to be an eclipsing binary \citep{Kreiner2004}.  \\

{\it TIC 381641106 = CSI-62-12087:} Listed as a WR star, MR 41 \citep{Roberts1962}. The TESS light curve is purely stochastic at a high amplitude and low frequency.   \\

{\it TIC 394728064 = DR Cha:} A clear EB with primary and secondary eclipses separated by about 20 days. No BeSS spectra exist. None of the six entries in the catalog of \citet{Skiff2009} indicate emission. It is unclear what the reason for the historical Be star classification of this system, but perhaps its binary nature has lead to confusion.   \\

{\it TIC 399669624 = 2 Ori:} A1Vne. This is not a classical Be star, but is rather a $\lambda$ Boo star \citep{Abt1995,Murphy2017}, which have peculiar abundance patterns and often pulsate in $\gamma$ Dor and/or $\delta$ Scuti modes. Many spectra from BeSS show a weak, symmetric, double-peaked H$\alpha$ emission profile that is roughly stable over 13 years, which, given the $\lambda$ Boo classification is probably a circumstellar accretion or debris disk or similar. This system has a high rotation rate, with $v$sin$i$ = 211 \kms from \citet{Glebocki2005}, or $v$sin$i$ = 261 \kms from \citet{Eiff2012}. \\

{\it TIC 408757239 = V716 Cen:} Short-period ellipsoidal variable and EB, with P$_{orb}\sim$ 1.5 d. A known EB \citep{Kreiner2004}.   \\

{\it TIC 467065657 = HD 97253:} Prominent stochastic variability with high amplitude. Classified as having a spectral type O5III(f) from GOSS survey \citep{Sota2014}, which is incompatible with being a classical Be star.   \\

{\it TIC 443616529 = phi Leo:} This system may have a debris disk, as there is some evidence for exocomets in the system \citep{Eiroa2016}. With a spectral type of A7IVne, this is too late to belong to the class of classical Be stars. The strongest signal by far is a single frequency at 6.48 \cd, plus some lower level variability that may be stochastic, being strongest at the lowest frequencies. Unclear why this is in lists of Be stars, although a debris disk may cause peculiarities (shell features) that can be confused with a gaseous decretion disk. No BeSS spectra from 2007 - 2020 show any sign of emission, nor do the three spectra on BeSOS. This is a high proper motion star.  \\

{\it TIC 455463415 = HD 135160:} A very strange EB with a strong reflection effect, a primary eclipse depth of about 2-2.5\%, and an orbital period of about 6 days. There are odd ``bumps'' in the LC not synchronized with the orbit, repeating roughly every 8 days. This is a known SB2, confirmed in \citep{Wang2018}. None of the references or spectral types in the 8 entries in the \citet{Skiff2009} catalog include mention of emission, so the reason for this being included in the  \citet{Jaschek1982} catalog of Be stars is unclear.  \\

\subsection{Possible non-classical Be stars rejected from the sample} \label{sec:maybe-not-Be-rejected}

{\it TIC 11411724 = StHA 52:} Embedded in a strong reflection nebula, NGC 2023. Unusual frequency spectrum. Four frequencies of similar strength spread out over 0.5 \cd, centered at 8.75 \cd, and other high frequencies between 10 -- 11.5 \cd. Not in BeSS, and no emission features in APOGEE. Given the strength and size of the reflection nebula in the visible, it is unclear exactly what is contributing to the TESS photometry. This is not included as a Be star. \\

{\it TIC 140132301 = HD 72126:} Extremely strong, stable, single frequency at 3.04 \cd, with a semi-amplitude of about 10\%. Very unusual LC and frequency spectrum for a Be star. No other Be stars in the sample have such a simple frequency spectrum with only one extremely strong frequency.  
There is some diffuse nebula in the vicinity, so perhaps this is mis-classified as a Be star. No BeSS spectra. SED does not appear to show any fluctuation in the IR/radio. In the Be star catalog of \citet{Jaschek1982}, but is not included as a Be star here.  \\

{\it TIC 151131426 = HV Lup:} There is one dominant signal with a double-wave pattern at P = 5.66 days with a $\sim$10\% semi-amplitude, plus stochastic variability. It is possible that this is an interacting binary with a 5.66 day orbital period. The behavior of the light curve and frequency spectrum most closely resembles other interacting binaries rather than a classical Be star. No BeSS spectra are available. \\

{\it TIC 213153401 =  HD 154538:} The TESS LC looks similar to that of Sigma Ori E. Two BeSS spectra are too low-resolution to provide a detailed view, but they are both clearly in absorption. This system could be an EB, but without any ellipsoidal variation and a very short period of $f = 2$ \cd and equal primary/secondary depths (or no secondary at all). Very close visual double, so hard to say from which star the signals are coming from. It is possible that the fainter of the pair is the one with emission: see \citet{Skiff2009}.   \\

{\it TIC 22825907 = HD 148877:} Dominated by a 10 d period with an amplitude of 4\%. Looks like rotation and/or binarity + stochastic variability. KELT data show a strong sinusoidal signal at 9.95 d, which is possibly double-waved at 19.91 d. SED appears to have 2 components. This is more likely a mass-transfer binary than a classical Be star. The stochastic variability peaks at a frequency of about 3 \cd. \\

{\it TIC 23091719 = NW Pup:} B2IVne. Lower frequency variability, including stochastic features, dominates the TESS LC. No obvious groups, which is unusual despite the clear variability. No sign of emission, and narrow-lined, including a narrow He 6678 line in one high-resolution BeSS spectrum from 2020-04-15. Unclear if this is truly a classical Be star given the narrow lines and lack of emission in 2019-2020. No sign of emission in 2014 in two epochs of BeSOS spectra, and their best-fit model gives $v$sin$i$ = 50 km s$^{-1}$. Far-radio excess in the SED. Included in the catalog of \citet{Egret1981}, where it is listed as a He-abnormal chemically peculiar star, and in the \citet{Renson2009} catalog of Ap and Am stars, being listed as B3 He var. BeSS spectral type and emission-line designation seems to come from only \citet{Hiltner1969}. Very narrow-lined in the blue in \citet{Chauville2001}. \\

{\it TIC 26175330 = 17 Sex:} A1Ve. No signals in TESS. Extremely deep and narrow H$\alpha$, dropping to about 0.175 times the continuum level. No sign of any emission in many BeSS spectra, including a professional spectrum from ELODIE. Listed as a Herbig Ae/Be star on Simbad, and classified as an A-shell star in \citet{Montesinos2009}. Likely falsely classified as a Be star because of shell features in this A-type star. \\

{\it TIC 269087549 = 19 Mon:} Possibly a hybrid SPB/$\beta$ Cephei pulsator, with a pair of $\beta$ Cephei-like pulsations centered around 5 \cd, with a combination frequency near 10 \cd. Known $\beta$ Cephei star in Simbad. H and He line profiles on BeSS look like significant deformed, either from line-profile variations from pulsation, and/or binarity. No sign of any emission from 2003 - 2020 in hundreds of spectra in BeSS. In the \citet{Jaschek1982} catalog of Be stars. In the \citet{Skiff2009} catalog, there are 11 entries, and only one of them hints at emission, with the note "em?" from \citet{Irvine1975}. Found to be an SB2 based on 9 spectra \citep{Chini2012}. Assuming this is not a classical Be star, and that the potential "em?" classification from 1975 was the result of being an SB2. \\

{\it TIC 284230347 = HD 55806:} Very high amplitude low-frequency variation, 3\% semi-amplitude, plus stochastic variation. A CoRoT Be star. In \citet{Fremat2006}, they note for HD 55806 that they could not find any set of fundamental parameters allowing a simultaneous fit of the observed He and Mg spectral lines. All of the He lines show unusual line shapes, probably related to the presence of a close companion. It is the only star in their sample of 64 to show this behavior. The low-frequency signal is also very apparent in the KELT data, being single-waved at 6.996 d or maybe double-waved at 13.98 d. This suggests a close binary, and perhaps an accretion disk scenario. The TESS data seem more similar to other interacting binaries compared to ``normal'' classical Be stars. \\

{\it TIC 319854805 = HD 47359:} Suspicious and interesting case with features that look like simple flickers with precurser phases and no apparent change in the frequency spectrum, and a pair of high frequency signals at 11.8 and 13.3 \cd. Just one dominant frequency and a very low amplitude harmonic of it, plus obvious low-frequency signals are present. Three BeSS spectra show weak H$\alpha$ emission, sometimes double-peaked, sometimes single-peaked, with E/C$\sim$1.2. Both the 12 d `flicker' signal and the f = 1.545 \cd are very clear in KELT. The P=12.13d signal is interesting with how regular it is in KELT and how much it resembles the TESS signal, and maybe is related to rotation or binarity. This probably is not a normal Be star flicker. Listed as B0.5Vp in  \citep{Yudin2001}. With $v$sin$i$ = 443 $\pm$ 40 km s$^{-1}$ \citep{Fremat2006}, this would have to be very close to edge-on, yet the H$\alpha$ profile is clearly not that of a shell star. The nature of this object is unclear, and it is not included in the statistics of this sample. \\

{\it TIC 322104948 = HD 306989 = V644 Cen:}. Suspicious LC. Clear 25 d period in KELT, and also apparent in TESS, but also with stochastic variability. Many high frequency signals in TESS are evident. No BeSS spectra. SED is structured in a way that suggests two components. YSO candidate from  \citep{Marton2016}. Possibly a very long period (200 years) eclipsing binary with an eclipse duration of 17+ years according to \citet{OConnell1951, Davies1987}. Likely not a classical Be star.  \\

{\it TIC 342257745 = HD 322422:} Unusual LC with a single low-frequency signal dominating ($f$ = 0.39 \cd), probably a weak group at 1.61 \cd, and some high frequency signals at 6.91, 8.3, and 13.4 \cd. The low-frequency signal is not apparent in the lower-precision KELT data. Embedded in a nebula. Listed as having emission in 7 references in \citet{Skiff2009}. $v$sin$i$ = 170 $\pm$ 11 km s$^{-1}$ from \citet{Zorec2016}. Possibly a Herbig Ae/Be star -- listed in the Extreme emission line objects (EELOs) table \citep{The1994}, which are usually most likely LBVs, B[e] stars, HAEBEs, PNs, or Symbiotics.   \\

{\it TIC 376077639 = V862 Ara:} B7IIIe. Clear case of sum and difference frequencies and harmonics of the main group. No emission in the 2 BeSS spectra with high enough resolution. The Be designation does not seem to have much evidence. The only reference for being an emission line star at all is from Hipparcos photometry, where it earned the VSX designation "BE:", where the colon indicates uncertainty \citep{Samus2017}. There does not seem to be reasonable evidence to claim this is a Be star.   \\

{\it TIC 455809360 = CD-61 4751:} Suspicious LC dominated by slow stochastic variation, with an SED that looks like 2 components. Behavior is more reminiscent of an interacting binary compared to that of Be stars, but it is unclear what the nature of this system is. No BeSS spectra are available.   \\

{\it TIC 457546452 = HD 126986:} Low frequency periodic double-waved LC with 4\% amplitude -- almost definitely binary or rotation related (with a 7.2 or 14.4 day period). Unlikely to be a classical Be star just based on TESS data. No BESS spectra. SED seems to have 2 components. Assume this is not a classical Be star.  \\

\subsection{Possible non-classical Be stars kept in the sample} \label{sec:maybe-not-Be-not-rejected}

{\it TIC 51288359 = HD 151083:} EB with a 0.901 d (or 1.8 d) period, plus higher frequency signals near 6.25 \cd and 11.5 \cd, and some stochastic variability. Eclipse depths are approximately 2\% -- 3\%. While such a short period EB would be incompatible with a classical Be star, a situation where such an EB is blended with a Be star cannot be ruled out, and there is little information on this target available in the literature and it lacks BeSS spectra. \\

{\it TIC 140031673 = HD 71510:} B2Ve. There appear to be groups but their pattern and delineation is unclear. None of the spectra in BeSOS (2014, 2015) or BeSS (2014/15/17/19) show H$\alpha$ in emission. $v$sin$i$ = 150 km s$^{-1}$ from BeSOS. In the \citet{Jaschek1982} catalog of Be stars. Apparently a known visual binary \citep[\textit{e.g.}][]{Wackerling1970}. None of the 6 references in \citet{Skiff2009} list emission. The visual binary pair is listed as B3V + G3V \citep{Pallavicini1992}, but that does not exclude the B-type star being a Be star. There is an IR nebula centered on the target \citep{Bodensteiner2018}. This star is of unclear nature. Perhaps the IR excess from the surrounding nebula is the reason for its Be designation, but without further evidence this cannot be excluded as a classical Be star. This is included in the statistics for the sample. Variability in TESS is largely stochastic, but there do seem to be coherent signals in the traditional g mode regime. \\

{\it TIC 147244857 = HD 70234:} Unusual LC for a Be star. Has only two isolated frequencies with constant power at 1.55 and 2.27 /cd causing a beating pattern with a very short envelope. First reported as an emission line star in \citet{Henize1976}, where it is listed as B9III(e). One low resolution BeSS spectrum clearly shows H$\alpha$ in emission at E/C$=$2.5. \\

{\it TIC 155573117 = CD-27 5181:} This looks more like an EB or a sig Ori E type case, with two single strong signals at exact harmonics (0.6 and 1.2 \cd), which resemble eclipses of about 1\% depth, plus stochastic variability. No BeSS spectra are available. Included in the Be star catalog of \citet{Jaschek1982}, and is apparently of a relatively early spectral type, being classified as OB+e in \citet{Stephenson1971}. The same photometric frequency is reported in \citep{Bartz2017}. Without a spectrum, we cannot determine whether or not this is a classical Be star, and is thus not rejected from the sample.  \\

{\it TIC 191312952 = HD 129772:} One dominant signal at 3.15 \cd. Unclear if it is an unresolved narrow group, or a single frequency modulated in amplitude (clearly seen in the LC, but could be beating). There are multiple harmonics of this signal/group. In the \citet{Renson2009} catalog of Ap, HgMn, and AM stars, where it is listed as having a spectral type of B8 Ca, suggesting an abundance of calcium, but its classification is also of ``doubtful nature''. No BeSS spectra are available. Insufficient evidence to rule this out as a classical Be star.  \\

{\it TIC 21249978 = V868 Ara:} Unusual LC. Strong sinusoidal signal with period of 22.51 days (possibly double-waved at twice this period). Therefore unlikely to be a classical Be star. One high-resolution BeSS spectrum shows H$\alpha$ E/C$\sim$3, with a deformed triple-peaked profile, and H$\beta$ is mostly filled in. The SED has considerable structure and may indicate two components. This is most likely either an ellipsoidal variable (but not eclipsing) with mass transfer and an accretion disk, or (perhaps less likely) a classical Be star with a decretion disk in a 45 d (or 22.5 d) binary. There seem to be two frequency groups centered around 4.5 and 9.0 \cd, plus the low-frequency orbital modulation and stochastic variation. Has a somewhat high proper motion, as 12.772 and -6.280 \citep{Gaia2018}. \\

{\it TIC 215983126 = HD 144970:} The star is very reddened, but is apparently B0V or B0Ve \citep{Cannon1993, Feast1961}. No BeSS spectra available. TESS data show only stochastic variability and possible short duration flares. Could be some other type of object, but this is inconclusive and so is included as a Be star in this work.  \\

{\it TIC 216158265 = HD 155436:} There is only \Bcep type pulsation, with nothing short of 6 \cd, but many signals are present between 6.5 -- 11 \cd. In the \citet{Jaschek1982} catalog of Be stars, and listed as having a spectral type of B0.5IIIn(e)	in \citet{Garrison1977}. One low resolution BeSS spectrum shows H$\alpha$ in emission with E/C$\sim$1.8. While the TESS frequency spectrum is unusual in not having any power at low frequencies, there is nothing to suggest this is not a Be star.  \\

{\it TIC 216875138 = HD 156172:} A known \Bcep pulsator \citep{Pigulski2008}, confirmed by the TESS data which show a pair of signals near 7 \cd, with a clear harmonic of the stronger of the two. There may be weak frequency groups in the TESS data, near 1 and 2 \cd. Perhaps this is a binary, similar to \Bcep itself, where the primary is a high frequency p mode pulsator and the secondary is a later type Be star with a decretion disk. \\

{\it TIC 256994805 = V715 Mon = HD 49567 = HR 2517:} Only a weak disk is present in one professional BeSS spectrum. This system is apparently is a high mass X-ray binary \citep{Khalak1998} and has also has been observed to show ``flares'' \citep{Sterken1996b}, which are strange since they are brighter in the bluer bands (Be outburts are generally of higher amplitude in redder bands). \\

{\it TIC 282808223 = HD 50820 = HR 2577:} A binary B3IVe+K2II system with a period of 58 years where the Balmer emission lines of the B star are variable in a fashion unrelated to the orbit \citep{Hendry1982}. The TESS LC shows mostly slow, gradual, low-amplitude variation plus mild stochastic variation. Included as a Be star in this work, although it is possible variability from the K giant contributes to the TESS data. \\

{\it TIC 307225534 = V767 Cen:} Variability at the 15\% level with a timescale of around 30 days, plus some much lower amplitude frequency groups at around 0.8, 1.7, 3.5, and 5.2 c/d. X-ray source. Absorption lines seem narrow in BeSS, and H$\alpha$ emission is single-peaked and about E/C$\sim$2.5 usually, but up to 4. B2IIIep from \citet{Slettebak1982}, $v$sin$i$ = 70 km s$^{-1}$ from \citet{Fremat2005}. Unclear how to proceed with this. The brightness drops by 15\% in about 10 days, which is a bit unusual. If this is a classical Be star, and there is no convincing evidence that it is not, then it is at a very low inclination angle due to having a low $v$sin$i$, single peaked emission, and narrow absorption lines. Its X-ray flux makes this an interesting system for further study.  \\

{\it TIC 308951795 = HD 306145:} Unusual LC, with many short, small ``bumps'' that are not obviously related to pulsation or flickers. There are also definitely high and maybe very high frequencies, and stochastic variation out to high frequencies.  No emission in the single low-resolution BeSS spectrum. Very close visual double (clearly seen in 2MASS). The strongest signal is at 0.25 \cd, or P=3.9 days (also apparent in KELT). The variability in TESS is very unusual, but cannot be ruled out as a classical Be star based on existing data. \\

{\it TIC 315679257 = HD 146596:} Just low frequency signals ($\sim$1\% in amplitude) and stochastic variability. No BeSS spectra are available. Has H$\alpha$ emission from HARPS according to \citet{Rainer2016}, and is therefore classified as a Be star, but without estimates of stellar parameters. The SED shows an excess in the mid-IR with a clear down-turn. The H$\alpha$ line (from \url{http://sisma.brera.inaf.it/index.jsp}) looks typical for a Be star, with E/C$\sim$2.3, and V/R$\sim$1. Possibly some weak and very narrow emission in other non-Balmer lines (near 4922 \AA, 5020 \AA, and some others), which appear to have P Cygni profiles, like near 4922 \AA, 5020 \AA. The nature of this object is unclear, but cannot be ruled out as a classical Be star.  \\

{\it TIC 316792722 = HD 99771:} The two lowest frequency groups appear to be in the typical configuration, at 0.61 and 1.23 \cd. There is a third group near 2.19 \cd, and then many isolated frequencies with exact harmonics. It is possible that the signal near 4.96 \cd is split, and there may be a slightly unusual SED excess at long wavelengths. This system is embedded in a large cloud. Included in the \citet{Jaschek1982} catalog of Be stars, but there do not seem to be any more recent works or spectra that are relevant. No signals in KELT, as expected, due to their low TESS amplitudes (maximum of 0.4 ppt). \\

{\it TIC 322233181 = HD 306962:} Dominated by stochastic variability, and a single signal at 5.97 \cd. Also in SVM selection of WISE YSO Candidates \citep{Marton2016}. Not enough evidence to discount this being a classical Be star, but it is suspect. \\

{\it TIC 381747495 = HD 105753:} Odd LC. Dominated by low frequency stochastic variation, and a somewhat strong group near 8.5 \cd. Clear long term trends in KELT data. There is no information to rule this out as a classical Be star, but the light curve and frequency spectrum are suspicious. No similar cases are seen in the sample. No BeSS spectra.  \\

{\it TIC 405520863 = 39 Cru:} Looks more like rotation or binarity. Only a single signal at 1.30 c/d, and a few weak harmonics. Two BeSS spectra have H$\alpha$ emission with E/C$\sim$3.5, and a symmetric double peak, and two spectra on BeSOS from 2014 and 2015 have the same profile. Unusual in its simplicity and constant H$\alpha$ emission. There is no obvious evidence against this being a classical Be star, but its TESS light curve is remarkably simple. \\

{\it TIC 451280762 = HD 99146:} Remarkably rich frequency spectrum in TESS, with high levels of stochastic variation, and many clearly periodic signals out to $\sim$16 \cd that are not organized in a way that resembles the typical groups. No BeSS spectra are available. Clear long term variation and possibly outbursts in KELT. The frequency spectrum is highly unusual, but there is no evidence to this not being a classical Be star. \\

{\it TIC 466715331 = HD 308217:} The TESS LC is dominated by stochastic variation and low frequencies, but there is a frequency 'group' near 6.2 \cd which looks like p modes, and also an isolated signal at 8.44 \cd. There are events that vaguely resemble flickers (with a quasi-period of about 10 days), but could be related to rotation. The single BeSS spectrum is in absorption. Cannot rule this out as a classical Be star.  \\

{\it TIC 427400331 = HD 290662:}. Has a very high-freq group at 75 \cd with 0.6 ppt amplitude, plus typical lower-frequency groups centered at 1.05 and 2.12 \cd and with amplitudes of around 0.25 - 1.5 ppt. This is listed in the \citet{Renson2009} catalog of Ap and Am stars, where this high frequency group could then plausibly be roAp or $\delta$ Scuti pulsation. \citet{Renson2009}) gives a spectral type of B9 Fe. 1 low-resolution BeSS spectrum does not obviously show any emission. \citet{Skiff2009} also notes an A0Vp designation from 1971 saying that Fe II lines prominent. It is unclear what the nature of this star is, and there is insufficient evidence to reject this as a Be star.   \\

{\it TIC 440399815 = HD 113605:} Low level stochastic variability is the main feature in TESS. In KELT there is a fairly strong periodic signal at $P$ = 32.087 days, which could be related to binarity. Without any spectra, the nature of this object is unclear, but there is no convincing evidence against this being a classical Be star.

\subsection{Be star systems of interest} \label{sec:interesting}

{\it TIC 14088298 = HD 33453:} B8Vne. Two very narrow groups with strong beating patterns. Groups are at 3.28 and 6.53 \cd, which is one of the widest separations in the whole sample.  Normal looking symmetric H$\alpha$ spectrum in BeSS. The extreme group separation may make this interesting to study (perhaps it has a relatively high rotational frequency). \\

{\it TIC 65803653 = 27 CMa:} Looks like a composite spectrum with a single signal at 0.76 \cd and its first harmonic, plus two groups at 1.33 and 2.69 \cd. There are also isolated signals around 10.9 and 13.5 \cd, plus a pair of signals near 5.16 and 5.90 \cd. Shell star with strongly varying disk, including strong asymmetries from 7 BeSS spectra between 2006 - 2019. Good candidate for binarity. Known \Bcep pulsator \citep{Stankov2005}. \citet{Balona1994} find the same \Bcep frequency -- they report 10.893 \cd, and a low frequency signal of 0.796 \cd. \\

{\it TIC 118842700 = QV Tel:} B3IIIpe. One strong low frequency signal (around 0.1 \cd), and strong stochastic variability. No obvious patterns in the LC. BeSOS H$\alpha$ is symmetric and double-peaked with E/C$\sim$1.5. There are two narrow C II lines to the right of H$\alpha$. \citet{Arcos2018} estimate $v$sin$i$ = 50 km s$^{-1}$, which even seems too high comparing the fit to the data. H$\alpha$ is definitely variable in the 17 BeSS spectra from 2011 - 2020. This is HR 6819, the potential triple (BH + B) + Be system \citep{Rivinius2020}, which can explain the narrow lines since the giant B star (with narrow lines) contributes about 50\% of the total visible flux. \\

{\it TIC 127493611 = omi Pup:} BIVnne. Strong, clear, coherent sinusoidal signal at $f$ = 0.495 \cd (amplitude $\sim$6 ppt), and a few high frequency signals near 8.5, 10.2, and 14.5 \cd (amplitudes between 0.1 and 0.3 ppt). H$\alpha$ is almost flat-topped in BeSS, but slanted down to the blue. H$\alpha$ does not change much over 2 years in BeSOS spectra. Likely a Be + sdO binary with a 28.9 d period \citep{Koubsky2012}. Helium emission can change very rapidly, with clear variability seen from night to night in some BeSS spectra. No sdO spectrum detected in \citep{Wang2018}. Despite having no frequency groups, this system seems to be able to support a rather strong disk at all observed times. \\

{\it TIC 139385056 = FY CMa:} B0.5 IVe. Famous Be+sdO binary. Actually a highly unusual LC compared to the rest of the sample. There is low frequency stochastic variability. The most prominent feature is a pair of signals centered at 8.50 and 8.99 \cd, of equal strength of about 2.2 ppt. There are many frequencies between 3 -- 9 \cd, which may form some doublets or triplets.  \\

{\it TIC 281047621 = ome Ori:} Variable disk, double-peaked and usually roughly symmetric in BeSS spectra. Sometimes He wing emission. Unusually simple LC, with a very strong peak at 1.07 \cd, and a harmonic at 2.14 \cd. Small frequency groups at around 0.5 and 1.54 \cd. Looks like it is ionizing some nearby gas. Hard to say if there are groups or just single frequencies. Assuming they are groups, since the amplitude clearly, but slowly and mildly, varies for the main 1.07 \cd signal. Perhaps hosts a weak magnetic field \citep{Neiner2012c}. Interestingly, the rotation frequency derived in \citet{Neiner2003} is 0.765 \cd ($P$ = 1.307 d). This is obviously different from the main TESS signal. However, there is some weak power in TESS at 2 $\times$ 0.765 c/d though (and a small group near 0.5 \cd). \\

{\it TIC 401635731 = V1012 Cen:} Extremely unusual and interesting flicker-looking events that are decidedly non-sinusoidal (and have an exact harmonic). Single low-resolution BeSS spectrum shows H$\alpha$ in emission at E/C$\sim$3. Strong 5.49 d signal in KELT, which is also the dominant signal in TESS. This signal also seems to have a beat envelope, in both KELT + TESS. Also a signal at 2.33 \cd, which is strongest when the beat envelope of the low-freq signals are low, but is present throughout, and does seem relatively strong near peak brightness.  \\

{\it TIC 439164152 = CPD-63 2495:}. O9.5Ve. Has one eclipse near the end of the dataset, reaching a depth of about 0.5\%. No optical BeSS spectra, but 3 IUE spectra exist. High Mass X-ray Binary \citep[HMXRB,][]{Lutovinov2013}. \citet{Chernyakova2014} gives an interesting study of the object from its 2010-2011 periastron passage. Known to have an orbit of $P_{orb}$ = 1236.9 d, e = 0.87 \citep{Johnston1992,Johnston1994}. Seems like a real classical Be star, and not a supergiant. Probably the eclipse is unrelated to the HMXRB and may be a blend or some other source. With an orbital period of about 3.4 years, an eclipse is extremely unlikely. \\

{\it TIC 468095832 = 2E 1118.7-6138 = WRAY 15-793:} HMXRB and a gamma-ray source, with a spectral type of O9.5III/Ve \citep{Janot1981}. Dramatic variability is seen in KELT, with events having amplitudes up to 2 magnitudes, and duration of $\sim$100 days.  \\

{\it TIC 21249978 = V868 Ara:} See Sec.~\ref{sec:maybe-not-Be-not-rejected}.

\clearpage 

\section{Table of sample} \label{sec:appendix_sample}

\startlongtable
\begin{deluxetable*}{ccccccccccc} 
\tabletypesize{\footnotesize}
\tablenum{2}
\tablecaption{Table of the sample. \label{tbl:table_sample}\\ The TIC numbers, common ID, and spectral type (as listed on the BeSS database) are given, along with the following information. \textit{qual.:} Indicates the availability of TESS data of reasonable quality. Values of 0 mean the system was either not observed by TESS in Cycle 1, or there were significant problems with the data. \textit{Be:} A value of ``Y'' means this a classical Be star, ``S'' means the classical Be designation may be suspect, but there is insufficient evidence to reject it, ``U'' means this is rejected as a classical Be star but the nature of the system is not necessarily known, and ``N'' marks systems known to be something other than a classical Be star. For any star with insufficient data or that is rejected from the sample, the remaining fields are filled with ``/''. \textit{signals:} Values of ``S'' indicate the presence of stochastic variability, ``I'' when isolated signals exist, ``L'' for when low-frequency signals dominate, ``H'' when high frequency signals exist, ``V'' when very high frequency signals exist, ``F'' for systems with flickers (and ``F+'' when an enhancement in frequency groups accompanies one or more flickers), and ``G'' if there are frequency groups. \textit{N$_{groups}$:} gives the number of groups identified in the light curve. \textit{typical groups:} values of ``Y'' indicate the system has groups in the typical configuration, ``?'' means this is unclear, and ``N'' means there are not groups in the typical configuration. The final three columns give the central frequency of $g1$ and $g2$ in systems with groups having the typical configuration, and note which is the stronger of these two groups (or ``sim.'' if they are of similar strength).}
\tablehead{ 
\colhead{TIC} & \colhead{Common} & \colhead{ST} & \colhead{data} & \colhead{Be} & \colhead{signals}  & \colhead{N$_{groups}$} & \colhead{typical} & \colhead{$f_{g1}$} & \colhead{$f_{g2}$}  & \colhead{stronger}\\
\colhead{ID}   & \colhead{ID}   & \colhead{}   & \colhead{quality}      & \colhead{}   &        \colhead{}  & \colhead{}             & \colhead{groups}  & \colhead{(\cd)}      & \colhead{(\cd)} & \colhead{group}
} 
\startdata 
1748132 & HD 75740 & A0IIIe & 0 & / & / & / & / & / & / & / \\ 
3178733 & HD 147580 & B7Ve & 0 & / & / & / & / & / & / & / \\ 
4827953 & V647 Mon & B1Vne & 1 & Y & S/G & 1 & N & -- & -- & -- \\ 
5528993 & HD 89884 & B5IIIe & 1 & Y & G & 4 & Y & 1.22 & 2.45 & $g2$ \\ 
6110321 & SS 120 & B8e: & 1 & Y & I/H/G & 2 & Y & 2.76 & 5.42 & $g2$ \\ 
10176636 & V757 Mon & B3IV & 1 & Y & G & 4 & Y & 0.84 & 1.60 & $g2$ \\ 
10536200 & CD-53 6689 & B2.5IVne & 1 & Y & L/F+/G & 3 & Y & 1.17 & 2.18 & sim. \\ 
11411724 & StHA 52 & B1.5V & 1 & U & / & / & / & / & / & / \\ 
11559798 & OY Hya & B5Ve & 1 & Y & I/L/G & 2 & Y & 2.35 & 5.08 & sim. \\ 
11972111 & HD 84567 & B0.5IIIne & 1 & Y & S/I/L & 0 & N & -- & -- & -- \\ 
14088298 & HD 33453 & B8Vne & 1 & Y & G & 2 & Y & 3.28 & 6.53 & $g1$ \\ 
14498757 & OT Gem & B2Ve & 1 & Y & S/L/F+/G & 3 & Y & 2.12 & 4.01 & $g2$ \\ 
14709809 & RY Gem & A2Ve & 1 & N & / & / & / & / & / & / \\ 
16688664 & CP-45 8706 & B9 & 0 & / & / & / & / & / & / & / \\ 
16902823 & HD 159489 & B1Ve & 1 & Y & S/L/F/G & 3 & Y & 1.05 & 2.06 & sim. \\ 
19727094 & HD 148692 & B7Ve & 0 & / & / & / & / & / & / & / \\ 
21249978 & V868 Ara & B6Vne & 1 & S & S/I/L/G & 4 & N & -- & -- & -- \\ 
22123033 & HD 85860 & B4Ve & 1 & Y & S/I/H/G & 5 & Y & 1.88 & 4.30 & $g1$ \\ 
22688271 & HD 162568 & B2IIIne & 1 & Y & G & 4 & Y & 2.12 & 4.09 & $g1$ \\ 
22825907 & HD 148877 & B6Ve & 1 & U & / & / & / & / & / & / \\ 
23037766 & NV Pup &  B2Ve & 1 & Y & G & 4 & Y & 1.04 & 2.06 & $g2$ \\ 
23091719 & NW Pup &  B2IVne & 1 & U & / & / & / & / & / & / \\ 
23321922 & HD 163007 & B4Vne & 1 & Y & G & 2 & Y & 0.55 & 1.10 & $g1$ \\ 
25256180 & V728 Mon & B1.5IVe & 1 & Y & S/L/G & 3 & Y & 1.61 & 3.23 & $g1$ \\ 
25883580 & HD 330950 & B1Ve & 1 & Y & S/L/F+/G & 3 & Y & 1.05 & 2.18 & sim. \\ 
26175330 & 17 Sex &  A1Ve & 1 & U & / & / & / & / & / & / \\ 
26311013 & HD 149128 & B7Ve & 0 & / & / & / & / & / & / & / \\ 
28885389 & HD 149298 & B2IIe & 1 & Y & L/F+/G & 3 & Y & 1.16 & 2.44 & sim. \\ 
29051733 & CD-46 4657 & A1IIe & 1 & Y & S/G & 5 & Y & 1.33 & 2.80 & $g1$ \\ 
29122883 & GW Vel & B2Vne & 1 & Y & S/L/F+/G & 3 & Y & 2.39 & 3.61 & $g1$ \\ 
29628522 & IY CMa &  B3Ve & 1 & Y & G & 4 & Y & 1.73 & 3.54 & sim. \\ 
29690468 & CD-45 4676 & B0.5IIIe & 1 & Y & S/I/L/H/G & 3 & Y & 0.74 & 1.44 & sim. \\ 
29786229 & HD 57551 & B8IIIe & 1 & Y & G & 2 & Y & 0.71 & 1.83 & $g2$ \\ 
30051402 & OU Vel & B2Vne & 1 & Y & H/G & 5 & Y & 1.61 & 2.64 & $g2$ \\ 
30444949 & V958 Cen & B5Ve & 1 & Y & G & 3 & Y & 1.25 & 2.25 & $g2$ \\ 
30562668 & HD 76838 & B2IVe & 1 & N & / & / & / & / & / & / \\ 
31062736 & IU Vel & B2.5Vne & 1 & Y & G & 3 & Y & 1.59 & 3.15 & $g1$ \\ 
31071126 & CD-46 4821 & Be & 1 & Y & I/G & 3 & Y & 2.31 & 4.80 & sim. \\ 
31449841 & HD 149610 & B5Ve & 1 & Y & G & 3 & Y & 2.26 & 4.59 & $g1$ \\ 
31753296 & HD 160648 & B6IIIe & 0 & / & / & / & / & / & / & / \\ 
32667467 & HD 51452 & B0IVe & 1 & Y & S & 0 & N & -- & -- & -- \\ 
33287617 & HD 53032 & A2e & 1 & Y & G & 3 & Y & 0.77 & 1.61 & $g1$ \\ 
33362448 & V749 Mon & B4IVe & 1 & Y & G & 3 & Y & 2.45 & 4.90 & $g1$ \\ 
33993533 & HD 43544 &  B2.5Ve & 1 & Y & G & 4 & Y & 2.12 & 4.10 & $g2$ \\ 
35311214 & HD 149729 & B2Vne & 1 & Y & S/L/F+/G & 3 & Y & 2.33 & 4.50 & $g2$ \\ 
35735021 & HD 149814 & B6IVe & 1 & Y & I & 0 & N & -- & -- & -- \\ 
38118168 & HD 78482 & B8Ve & 1 & Y & G & 2 & N & -- & -- & -- \\ 
42753726 & CW Cir & B0.5Vne & 1 & Y & S/L/F/G & 2 & N & -- & -- & -- \\ 
42889751 & HD 45626 & B7pshe & 1 & Y & G & 5 & Y & 0.85 & 1.70 & $g1$ \\ 
43708176 & HD 135354 & B8Vne & 1 & Y & G & 2 & ? & 2.18 & 4.31 & sim. \\ 
46028220 & f Car &  B3Vne & 1 & Y & G & 3 & Y & 1.95 & 3.53 & sim. \\ 
46211527 & HD 136250 & B7Vne & 1 & Y & I/H/G & 2 & Y & 2.50 & 4.99 & $g1$ \\ 
46226875 & HD 75661 & B2Vne & 1 & Y & S/L/F+/G & 2 & Y & 2.03 & 3.46 & $g2$ \\ 
46705199 & gam Cir &  B5IVe & 1 & Y & G & 4 & Y & 2.30 & 4.37 & sim. \\ 
47296054 & eps PsA &  B8Ve & 1 & Y & G & 2 & Y & 0.84 & 1.64 & sim. \\ 
51033628 & HD 72043 & B8e & 1 & Y & I/H/G & 2 & Y & 2.12 & 4.18 & $g1$ \\ 
51288359 & HD 151083 & B2Vne & 1 & S & S/I/H & 0 & N & -- & -- & -- \\ 
52638345 & 10 CMa &  B2IIIe & 1 & Y & L/F+/G & 4 & Y & 0.61 & 1.31 & sim. \\ 
52665242 & HD 47054 &  B7IIIe & 1 & Y & G & 2 & Y & 1.14 & 2.30 & sim. \\ 
52686322 & HP CMa &  B2IIIe & 1 & Y & L/G & 3 & Y & 1.30 & 2.37 & sim. \\ 
52748770 & HD 47160 & B8IVe & 1 & Y & G & 1 & N & -- & -- & -- \\ 
52929072 & HD 71042 & B2.5ne & 1 & Y & G & 4 & Y & 1.27 & 2.63 & $g1$ \\ 
53063082 & HZ CMa &  B6Vnpe & 1 & N & / & / & / & / & / & / \\ 
53301259 & CD-44 4392 & B2IVe & 1 & Y & L/F+/G & 3 & Y & 1.84 & 3.42 & $g1$ \\ 
53327951 & V733 Mon & B2Vpe & 1 & Y & L/G & 2 & N & -- & -- & -- \\ 
53992511 & UU PsA &  B4IVne & 1 & Y & G & 5 & Y & 1.50 & 3.09 & $g2$ \\ 
55295028 & HD 33599 & B2Vpe & 1 & Y & I/G & 3 & Y & 0.86 & 1.80 & $g1$ \\ 
56179720 & 56 Eri &  B2Ve & 1 & Y & G & 3 & Y & 0.73 & 1.46 & $g2$ \\ 
56447057 & HD 151743 & B8IVe & 1 & Y & I/G & 3 & Y & 1.77 & 3.55 & $g1$ \\ 
61566852 & HD 179253 & B7Ve & 0 & / & / & / & / & / & / & / \\ 
61678783 & HD 56670 & B0.5Ve & 1 & Y & L/G & 4 & Y & 0.88 & 1.77 & $g1$ \\ 
65672662 & V763 Mon & B5e & 1 & Y & S/I/L/H/G & 4 & Y & 1.93 & 3.87 & sim. \\ 
65803653 & 27 CMa &  B3IIIe & 1 & Y & I/H/G & 3 & Y & 1.33 & 2.71 & $g1$ \\ 
65903024 & ome CMa &  B2IVe & 1 & Y & L/G & 5 & Y & 0.62 & 1.26 & $g1$ \\ 
68003064 & HD 62367 & B9e & 1 & Y & G & 2 & Y & 0.78 & 1.65 & $g1$ \\ 
70966650 & HD 140336 & B1:IIInne & 1 & Y & G & 3 & Y & 1.67 & 3.84 & $g1$ \\ 
71132174 & 228 Eri &  B2Vne & 1 & Y & F+/G & 5 & Y & 1.38 & 2.80 & $g1$ \\ 
71727949 & V696 Mon &  B2Vne & 1 & Y & S/G & 3 & N & -- & -- & -- \\ 
73039225 & V695 Mon & B2.5Ve & 1 & Y & G & 4 & Y & 1.13 & 2.25 & $g2$ \\ 
73759997 & V864 Ara & B7Vnnpe & 1 & Y & S/L/F+/G & 2 & N & -- & -- & -- \\ 
75047606 & HD 79621 &  B9Ve & 1 & Y & I/G & 2 & Y & 1.65 & 3.36 & sim. \\ 
75210369 & HD 79811 & B5Ve & 1 & Y & G & 2 & Y & 1.83 & 3.68 & $g1$ \\ 
75581184 & HD 80156 & B8.5IVe & 1 & Y & I/H/V/G & 5 & Y & 2.69 & 5.42 & $g1$ \\ 
75708438 & HD 140883 & B6Vnne & 1 & Y & S/G & 2 & N & -- & -- & -- \\ 
76336597 & QQ Vel & B5nne & 1 & Y & L/G & 3 & Y & 0.82 & 2.11 & sim. \\ 
76556351 & HD 140843 & B5IIIe & 1 & Y & G & 3 & Y & 0.70 & 1.29 & $g1$ \\ 
80719034 & HD 67985 & B8Vne & 1 & Y & I/H/V/G & 2 & ? & 2.98 & 5.96 & $g2$ \\ 
80899869 & HD 67978 & B2Vnne & 1 & Y & S/L/F+/G & 4 & Y & 1.53 & 2.72 & $g2$ \\ 
81584371 & FV CMa &  B2Vnne & 1 & Y & L/G & 3 & Y & 0.77 & 1.52 & $g1$ \\ 
81615536 & HD 69168 & B2Ve & 1 & Y & L/G & 3 & N & -- & -- & -- \\ 
81859368 & HD 69404 & B2Vnne & 1 & Y & I/H/G & 3 & Y & 2.19 & 4.32 & $g1$ \\ 
82370215 & HD 70064 & B5Ve & 1 & Y & I/G & 2 & N & -- & -- & -- \\ 
84513533 & HD 141926 & B2nne & 1 & Y & G & 3 & Y & 1.12 & 2.34 & $g1$ \\ 
85001695 & V1063 Sco & B4IIIe & 1 & Y & G & 3 & Y & 1.99 & 3.99 & $g1$ \\ 
89593655 & HD 142237 & B2Vne & 1 & Y & G & 5 & Y & 1.61 & 3.21 & $g1$ \\ 
90296023 & FY Vel & B2IIpshe & 1 & N & / & / & / & / & / & / \\ 
90836305 & HD 142349 & B5IVe & 1 & Y & L/F+/G & 3 & Y & 1.53 & 3.15 & sim. \\ 
91689592 & BD+12 938 & none & 1 & Y & G & 3 & Y & 2.23 & 4.61 & $g2$ \\ 
111384365 & NO CMa &  B5IIIne & 0 & / & / & / & / & / & / & / \\ 
112679272 & HD 163435 & B9IIIe & 1 & Y & S/G & 2 & N & -- & -- & -- \\ 
118842700 & QV Tel &  B3IIIpe & 1 & Y & S/I/L & 0 & N & -- & -- & -- \\ 
121345684 & V848 Ara & B2IVe & 1 & Y & S/L/F/G & 3 & Y & 1.10 & 2.20 & sim. \\ 
121689850 & HD 153199 & B2.5Ve & 1 & Y & L/G & 3 & Y & 0.87 & 1.65 & $g2$ \\ 
121905577 & HD 153295 & B2e & 1 & Y & S/I/H/G & 3 & ? & 0.98 & 1.76 & sim. \\ 
123036723 & HD 60794 & B4IIIe & 1 & Y & G & 6 & Y & 2.35 & 4.75 & $g1$ \\ 
123545883 & V743 Mon & B7IIIe & 1 & N & / & / & / & / & / & / \\ 
123828144 & HD 62894 & B8e & 1 & Y & I/H/V/G & 0 & N & -- & -- & -- \\ 
124328133 & HD 51404 & B1.5Ve & 1 & Y & G & 3 & Y & 2.58 & 5.24 & $g1$ \\ 
124407255 & V747 Mon & B3Ve & 1 & Y & G & 4 & Y & 1.49 & 2.85 & $g1$ \\ 
124921846 & HD 44506 &  B1.5Ve & 1 & Y & L/G & 4 & Y & 0.78 & 1.63 & $g2$ \\ 
125221242 & HD 154170 & B9IIIe & 0 & / & / & / & / & / & / & / \\ 
125406140 & BD-06 1895 & Be & 1 & Y & S/I & 0 & N & -- & -- & -- \\ 
125759847 & HD 154314 & B9IIIe & 1 & Y & S/I/H/G & 3 & Y & 2.43 & 4.85 & $g1$ \\ 
126125347 & HD 134484 & B9IIIe & 1 & Y & S/I/G & 2 & ? & 0.92 & 1.74 & $g2$ \\ 
126460390 & HD 134671 & B7Ve & 1 & Y & S & 0 & N & -- & -- & -- \\ 
126899506 & HD 298298 & B0e & 1 & Y & G & 4 & Y & 1.22 & 2.39 & sim. \\ 
127493611 & omi Pup &  B1IVnne & 1 & Y & S/I/L/H & 0 & N & -- & -- & -- \\ 
131104893 & HD 63988 & B8Ve & 1 & Y & G & 3 & Y & 0.61 & 1.44 & $g2$ \\ 
134653865 & V817 Cen &  B3IVe & 1 & Y & L/F+/G & 3 & Y & 0.56 & 1.08 & $g2$ \\ 
138187342 & HD 136556 & B2.5Vne & 1 & Y & L/F/G & 4 & Y & 1.90 & 3.77 & sim. \\ 
139385056 & FY CMa &  B1IIe & 1 & Y & I/H/G & 5 & N & -- & -- & -- \\ 
139472176 & HD 14850 & B8Ve & 1 & Y & G & 2 & Y & 0.77 & 1.50 & $g1$ \\ 
139870215 & HD 71823 & B3Vne & 1 & Y & S/F+/G & 3 & Y & 1.33 & 2.55 & $g2$ \\ 
140001327 & HD 72014 & B1.5Vnne & 1 & Y & S/I/L/H/V/F/G & 4 & Y & 1.57 & 3.30 & sim. \\ 
140006471 & HD 72067 &  B2Vne & 1 & Y & I/L/F+/G & 4 & Y & 2.09 & 4.09 & $g2$ \\ 
140031673 & HD 71510 &  B2Ve & 1 & S & S/L & 0 & N & -- & -- & -- \\ 
140132301 & HD 72126 & B2nne & 1 & U & / & / & / & / & / & / \\ 
140300162 & NR Vel & B2Ve & 1 & Y & S/L/G & 3 & Y & 0.81 & 1.56 & $g1$ \\ 
141037731 & HD 136968 & B5Vne & 1 & Y & I/H/G & 3 & Y & 2.71 & 5.28 & $g1$ \\ 
141898455 & CD-45 4394 & B2Vne & 1 & Y & S/L/F+/G & 3 & Y & 1.77 & 3.44 & $g2$ \\ 
141973945 & bet Hya &  B9IIIspe & 1 & N & / & / & / & / & / & / \\ 
143543729 & HD 64716 & B6Ve & 1 & Y & S/I/L/G & 1 & N & -- & -- & -- \\ 
144028101 & mu Lup &  B8Ve & 1 & Y & I/L/H/V & 0 & N & -- & -- & -- \\ 
145373620 & HD 75551 & B2Vne & 1 & Y & S/L/F+/G & 3 & Y & 1.42 & 2.84 & sim. \\ 
145506624 & CD-47 4412 & A5e & 1 & Y & G & 3 & Y & 0.58 & 1.20 & $g1$ \\ 
146036695 & r Pup &  B1.5IIIe & 1 & Y & L/G & 3 & Y & 1.04 & 1.96 & sim. \\ 
146454560 & HD 137836 & B8Vne & 1 & Y & S/I/G & 2 & Y & 0.75 & 1.76 & $g1$ \\ 
147244857 & HD 70234 & B9IIIe & 1 & S & I & 0 & N & -- & -- & -- \\ 
151131426 & HV Lup & B1Vne & 1 & S & S/L & 0 & N & -- & -- & -- \\ 
151300497 & V1075 Sco &  O8Ve & 1 & N & / & / & / & / & / & / \\ 
153541389 & V1390 Ori & B2Ve & 1 & Y & S/G & 4 & Y & 1.62 & 3.19 & $g1$ \\ 
155573117 & CD-27 5181 & Be & 1 & S & I & 0 & N & -- & -- & -- \\ 
156627345 & HD 250980 & B0e & 1 & Y & G & 2 & Y & 0.71 & 1.41 & $g1$ \\ 
158734620 & HD 38856 & B5Ve & 1 & Y & G & 3 & Y & 2.88 & 5.62 & sim. \\ 
159059628 & HD 39557 & B5IIIe & 1 & Y & G & 3 & Y & 1.50 & 3.02 & $g1$ \\ 
160401972 & V1018 Cen & B2pe & 1 & Y & G & 4 & Y & 1.40 & 2.75 & sim. \\ 
166703712 & V774 Cen & B3Vne & 1 & Y & I/L/F+/G & 3 & Y & 2.30 & 4.50 & sim. \\ 
167107487 & HD 43264 & B9IIIe & 1 & Y & I/G & 2 & Y & 0.76 & 1.52 & $g1$ \\ 
167110617 & HD 43285 & B5IVe & 1 & Y & I/H/V/G & 4 & N & -- & -- & -- \\ 
168115492 & TYC 4812-2496-1 & none & 1 & Y & I & 0 & N & -- & -- & -- \\ 
168505959 & HD 112107 & B9.5Vne & 1 & Y & S/I/L/G & 2 & N & -- & -- & -- \\ 
168627067 & HD 54464 & B2.5IIIe & 1 & Y & G & 2 & Y & 1.46 & 2.85 & $g1$ \\ 
171895762 & V828 Ara & B2IVne & 1 & Y & L/F+/G & 3 & Y & 1.56 & 3.00 & sim. \\ 
171896106 & HD 153262 & B2Vnne & 1 & Y & G & 2 & N & -- & -- & -- \\ 
173381049 & HD 59197 & B6Ve & 1 & Y & I & 0 & N & -- & -- & -- \\ 
173956099 & HD 60669 & B8IIIe & 1 & Y & G & 2 & Y & 2.13 & 4.29 & sim. \\ 
174008063 & z Pup &  B3Vne & 1 & Y & G & 7 & Y & 0.87 & 1.80 & $g2$ \\ 
174121793 & HD 155438 & B6IVne & 1 & Y & I/G & 2 & Y & 1.22 & 2.44 & sim. \\ 
175523591 & V392 Pup &  B7Ve & 1 & Y & G & 4 & Y & 2.52 & 5.00 & $g1$ \\ 
177204351 & HD 53667 & B0IIIe & 1 & Y & S & 0 & N & -- & -- & -- \\ 
177853845 & HD 54858 & A0IIe & 1 & Y & G & 5 & Y & 1.83 & 3.64 & $g1$ \\ 
178719204 & HD 70340 & A2Vnnpe & 1 & Y & L/G & 1 & N & -- & -- & -- \\ 
180092491 & DO Cru & B2Ve & 1 & Y & S/L/F+/G & 3 & Y & 1.61 & 3.20 & $g2$ \\ 
181506336 & HD 74867 & B7IVe & 1 & Y & G & 2 & Y & 1.78 & 3.58 & $g1$ \\ 
181600430 & HD 75081 & B9Ve & 1 & Y & I/G & 3 & Y & 1.30 & 2.63 & $g1$ \\ 
186979808 & CD-29 6963 & Be & 1 & Y & G & 2 & Y & 0.62 & 1.63 & $g2$ \\ 
187458882 & HD 57682 & O9Ve & 1 & N & / & / & / & / & / & / \\ 
189271286 & HD 81753 & B6Ve & 1 & Y & G & 3 & Y & 1.53 & 3.00 & $g1$ \\ 
190393155 & HD 75925 & B4Vnne & 1 & Y & S/I/H/V/G & 3 & N & -- & -- & -- \\ 
190558492 & HD 139790 & B2IIIne & 1 & Y & S/G & 4 & Y & 1.02 & 1.96 & sim. \\ 
191312952 & HD 129772 & B8IIIpe & 1 & S & I & 0 & N & -- & -- & -- \\ 
200519617 & V1374 Ori & B8e & 1 & Y & S/I/G & 3 & Y & 0.96 & 1.91 & sim. \\ 
201373582 & HD 42054 &  B4IVe & 0 & / & / & / & / & / & / & / \\ 
203452834 & HL Lib & B9IVe & 1 & Y & I & 0 & N & -- & -- & -- \\ 
206792542 & MWC 527 & B3 & 1 & Y & G & 5 & Y & 1.52 & 3.01 & $g2$ \\ 
206840215 & BD+08 1343 & A2 & 1 & Y & I & 0 & N & -- & -- & -- \\ 
207020262 & BD+08 1366 & (B8?) & 1 & Y & I/G & 5 & ? & 0.75 & 1.59 & sim. \\ 
207043035 & HD 258782 & B9IVe & 0 & / & / & / & / & / & / & / \\ 
207043038 & HD 45995 & B2Vnne & 1 & Y & S/G & 4 & Y & 1.11 & 2.08 & $g1$ \\ 
207162338 & HD 145040 & B8IVe & 0 & / & / & / & / & / & / & / \\ 
207176480 & HD 19818 & B9.5Vne & 1 & N & / & / & / & / & / & / \\ 
207580161 & HD 119835 & B8.5IIIne & 1 & U & / & / & / & / & / & / \\ 
207633164 & HD 119958 & B8Ve & 1 & Y & S/I/G & 2 & N & -- & -- & -- \\ 
207881543 & HD 120330 & B2.5Vnne & 1 & Y & G & 4 & Y & 1.64 & 3.55 & $g1$ \\ 
208495676 & HD 146531 & B3Ve & 1 & Y & S/L/G & 3 & Y & 2.22 & 4.46 & sim. \\ 
208580496 & HD 146463 & B3Vnne & 1 & Y & G & 4 & Y & 1.91 & 3.87 & $g1$ \\ 
209829040 & HD 122450 & B2.5ne & 1 & Y & S/L & 0 & N & -- & -- & -- \\ 
210283000 & HD 123042 & B8.5IVe & 0 & / & / & / & / & / & / & / \\ 
210506327 & HD 123296 & B8.5Ve & 1 & Y & S/L & 0 & N & -- & -- & -- \\ 
211750828 & HD 153222 & B1IIe & 1 & Y & S/L/F+/G & 3 & Y & 2.12 & 3.88 & $g1$ \\ 
212551379 & HD 153879 & B1.5Vne & 1 & Y & G & 3 & Y & 1.67 & 3.26 & $g1$ \\ 
212755033 & HD 154154 & B2Vnne & 1 & Y & G & 3 & Y & 1.91 & 3.95 & $g1$ \\ 
212806475 & HD 154111 & B5IVe & 1 & Y & G & 3 & Y & 1.48 & 2.07 & sim. \\ 
213153401 & HD 154538 & B3Ve & 1 & U & / & / & / & / & / & / \\ 
215511795 & HD 144555 & B1Vne & 1 & Y & F+/G & 4 & Y & 1.53 & 2.76 & $g1$ \\ 
215983126 & HD 144970 & B0e & 1 & U & / & / & / & / & / & / \\ 
216059854 & HD 145107 & B5IIIe & 0 & / & / & / & / & / & / & / \\ 
216158265 & HD 155436 & B2.5e & 1 & S & I/H/G & 1 & N & -- & -- & -- \\ 
216649528 & V1076 Sco & B2.5Vnne & 1 & Y & G & 3 & Y & 2.22 & 4.61 & sim. \\ 
216775415 & HD 156008 & B8IVe & 1 & Y & S/G & 2 & N & -- & -- & -- \\ 
216875138 & HD 156172 & O9IIe & 1 & S & I/H & 0 & N & -- & -- & -- \\ 
217134952 & HD 156398 & B9.5V+... & 1 & Y & S/G & 1 & N & -- & -- & -- \\ 
217597679 & HD 157099 & B3Vne & 1 & Y & G & 4 & Y & 1.66 & 3.47 & $g1$ \\ 
217667184 & iot Ara &  B2IIIne & 1 & Y & G & 4 & Y & 1.86 & 3.69 & sim. \\ 
218139464 & V750 Ara & B2Vne & 1 & Y & G & 3 & Y & 0.83 & 1.63 & $g1$ \\ 
218854809 & V830 Ara & B2IIpe & 1 & Y & S/L/G & 3 & Y & 1.38 & 2.44 & sim. \\ 
220114410 & TYC 158-270-1 & B8III & 1 & Y & G & 2 & Y & 1.47 & 2.91 & $g1$ \\ 
220322383 & 15 Mon &  B1Ve & 1 & N & / & / & / & / & / & / \\ 
220728104 & HD 147302 & B2IIIne & 1 & Y & S/L/G & 2 & Y & 2.01 & 4.27 & $g1$ \\ 
221040358 & V376 Nor & B9IVe & 1 & Y & S/L & 0 & N & -- & -- & -- \\ 
221092292 & HD 148027 & B9IVe & 1 & Y & S/G & 2 & N & -- & -- & -- \\ 
221155177 & HD 148028 & B5IVe & 1 & Y & I & 0 & N & -- & -- & -- \\ 
223994987 & HD 147756 & B2Vne & 1 & Y & G & 4 & Y & 1.43 & 3.05 & sim. \\ 
224244458 & bet Scl &  B9.5IVmnpe & 1 & N & / & / & / & / & / & / \\ 
224802716 & HD 148001 & B5IIIe & 1 & Y & G & 2 & N & -- & -- & -- \\ 
225735711 & OZ Nor & B2IIe & 1 & Y & S/L/F/G & 2 & N & -- & -- & -- \\ 
226712279 & HD 148567 & B2IIne & 0 & / & / & / & / & / & / & / \\ 
229210549 & V1059 Sco & B2IIe & 0 & / & / & / & / & / & / & / \\ 
229355590 & HD 259631 & B5e & 1 & Y & G & 4 & Y & 1.64 & 3.12 & $g1$ \\ 
229356387 & HD 259597 & B0.5Vnne & 1 & Y & G & 3 & Y & 1.08 & 2.26 & $g1$ \\ 
231038852 & HD 149568 & B6Vne & 0 & / & / & / & / & / & / & / \\ 
231942365 & HD 43789 & B6.5Ve & 1 & Y & S/I & 0 & N & -- & -- & -- \\ 
232367669 & HD 42406 & B4IVe & 0 & / & / & / & / & / & / & / \\ 
233563508 & HD 150288 & B2Ve & 1 & Y & G & 4 & Y & 1.22 & 2.49 & sim. \\ 
233969306 & EM* RJHA 40 & B3Ib & 1 & Y & G & 3 & Y & 1.05 & 2.28 & sim. \\ 
234230792 & V739 Mon & B0.5IVe & 1 & Y & S/I/H/G & 4 & Y & 1.37 & 2.80 & sim. \\ 
234422456 & HD 150422 & B1.5ne & 1 & Y & S/G & 3 & Y & 1.39 & 2.75 & $g1$ \\ 
234752189 & HD 150625 & B8e & 1 & Y & G & 3 & N & -- & -- & -- \\ 
234752466 & HD 150533 & B0e & 1 & Y & S/I/H & 0 & N & -- & -- & -- \\ 
234806470 & V725 Mon & B0.5IVe & 1 & Y & G & 3 & Y & 1.09 & 2.18 & $g1$ \\ 
234813367 & AX Mon & B2IIIpshev & 1 & N & / & / & / & / & / & / \\ 
234835218 & Cl* NGC 2244 PS 26 & B7Ve & 1 & N & / & / & / & / & / & / \\ 
234853418 & AS 128 & B5 & 1 & Y & I/H/G & 3 & ? & 2.43 & 4.60 & sim. \\ 
234887704 & Cl* NGC 2244 PS 543 & B8Ve & 1 & Y & S/I/H & 0 & N & -- & -- & -- \\ 
234929785 & HD 259440 & B0pe & 1 & Y & S/G & 3 & Y & 0.86 & 1.65 & sim. \\ 
234933368 & EM* GGA 399 & B3Ve & 1 & Y & L/F/G & 3 & Y & 0.76 & 1.54 & sim. \\ 
234933597 & HD 46484 & B0.5IVe & 1 & Y & S/I/H/G & 1 & N & -- & -- & -- \\ 
235376820 & HD 50581 & A0IVe & 1 & Y & I & 0 & N & -- & -- & -- \\ 
235488755 & HD 266894 & Be & 1 & Y & S/G & 3 & ? & 0.81 & 1.45 & sim. \\ 
235540713 & HD 51506 & B2.5IVe & 1 & Y & S/G & 3 & Y & 2.26 & 4.00 & $g2$ \\ 
237059039 & HD 151113 & B2Ve & 1 & Y & G & 4 & Y & 1.30 & 2.70 & $g1$ \\ 
237532447 & BD+00 1654 & B8III & 1 & Y & I & 0 & N & -- & -- & -- \\ 
237651093 & HD 50696 & B1.5IIIe & 1 & Y & I/H/V/G & 3 & Y & 2.16 & 4.35 & $g1$ \\ 
237668110 & V744 Mon & B1.5Ve & 1 & Y & I/G & 3 & Y & 2.34 & 4.62 & $g2$ \\ 
237840594 & BD+04 1529 & B9V & 1 & Y & I/G & 3 & Y & 1.77 & 3.54 & sim. \\ 
238791674 & CD-49 3441 & B8e & 1 & N & / & / & / & / & / & / \\ 
246189955 & HD 328990 & A0e & 1 & N & / & / & / & / & / & / \\ 
246796068 & HD 152004 & B3IVne & 1 & Y & G & 3 & Y & 1.51 & 2.75 & sim. \\ 
247589847 & BD+13 976 & A0 & 1 & Y & I/H & 0 & N & -- & -- & -- \\ 
247699160 & HD 152541 & B7Ve & 1 & Y & I/G & 1 & N & -- & -- & -- \\ 
250213274 & HD 130534 & B7Ve & 1 & Y & G & 2 & N & -- & -- & -- \\ 
253110498 & HD 112512 & B7Ve & 1 & Y & S/I/G & 3 & Y & 0.96 & 1.89 & sim. \\ 
253177102 & HD 112825 & B1.5IVe & 1 & Y & S/L/F+/G & 2 & Y & 1.30 & 2.60 & $g2$ \\ 
253212775 & V495 Cen & Be & 1 & N & / & / & / & / & / & / \\ 
253304151 & HD 113260 & B8.5IIIe & 1 & Y & S/I/G & 2 & Y & 1.37 & 2.73 & $g1$ \\ 
253380837 & HD 113573 & B4Ve & 1 & N & / & / & / & / & / & / \\ 
254340855 & HD 139431 & B2Vne & 1 & Y & L/F+/G & 3 & Y & 1.99 & 3.45 & $g2$ \\ 
256994805 & V715 Mon & B3IIIe & 1 & S & I/G & 2 & Y & 0.50 & 1.00 & $g2$ \\ 
257065501 & HD 49585 & B0.5IVe & 1 & Y & L/G & 3 & Y & 1.40 & 2.80 & sim. \\ 
258306961 & HD 143700 & B1.5Vne & 1 & Y & S/L/G & 3 & Y & 1.58 & 3.12 & $g2$ \\ 
258704817 & CO Cir &  B2.5Ve & 1 & Y & L/F+/G & 3 & Y & 1.25 & 2.40 & sim. \\ 
259449942 & V378 Pup &  B2.5Ve & 1 & Y & G & 4 & Y & 0.93 & 1.88 & $g2$ \\ 
260640910 & mu Pic &  B9Ve & 1 & Y & I/G & 4 & Y & 2.29 & 4.40 & $g1$ \\ 
261862960 & mu02 Cru &  B5Vne & 0 & / & / & / & / & / & / & / \\ 
261992444 & HD 112147 & Be & 1 & Y & S/L/F/G & 3 & Y & 1.49 & 3.69 & sim. \\ 
261993378 & lam Cru &  B3Vne & 1 & Y & G & 5 & Y & 3.25 & 6.11 & sim. \\ 
263805130 & HD 256577 & B2IVpe & 1 & Y & S/L/G & 1 & N & -- & -- & -- \\ 
263875550 & HD 126527 & B9Ve & 0 & / & / & / & / & / & / & / \\ 
264459943 & 25 Ori &  B1Vpe & 1 & Y & L/F+/G & 3 & Y & 1.37 & 2.73 & sim. \\ 
264631173 & V1372 Ori & B2Vne & 1 & Y & I/G & 2 & Y & 2.80 & 5.56 & $g1$ \\ 
265555524 & HD 43913 & Be & 1 & Y & I/G & 1 & Y & 0.99 & 1.97 & $g1$ \\ 
266322419 & HD 64109 & B8e & 0 & / & / & / & / & / & / & / \\ 
266656195 & BT CMi & B2Vne & 1 & Y & L/F+/G & 3 & Y & 1.90 & 3.60 & sim. \\ 
267105647 & HD 145301 & B9IVne & 0 & / & / & / & / & / & / & / \\ 
268351249 & HD 102423 & B9.5Ve & 1 & Y & I/L & 0 & N & -- & -- & -- \\ 
269087549 & 19 Mon &  B1Ve & 1 & U & / & / & / & / & / & / \\ 
269870654 & HD 86272 & B5Vne & 1 & Y & G & 4 & Y & 2.38 & 4.85 & sim. \\ 
270934336 & OR Vel & B3Vne & 1 & Y & G & 1 & N & -- & -- & -- \\ 
270991656 & HD 108051 & B8IIIe & 0 & / & / & / & / & / & / & / \\ 
271190727 & HD 75658 & B2.5ne & 1 & Y & G & 4 & Y & 0.84 & 1.92 & sim. \\ 
271379300 & HD 70461 & B6Ve & 1 & Y & G & 3 & Y & 1.30 & 2.70 & $g1$ \\ 
271614171 & HD 138131 & B6Vne & 1 & Y & G & 3 & Y & 1.59 & 3.30 & sim. \\ 
274590119 & HD 143140 & B8IIIe & 0 & / & / & / & / & / & / & / \\ 
274907223 & HD 86689 & A3ne & 1 & Y & G & 4 & Y & 1.88 & 3.75 & sim. \\ 
275927504 & HD 143545 & B1.5Vne & 1 & Y & G & 4 & Y & 1.54 & 3.22 & $g1$ \\ 
278313134 & HD 139314 & B2e & 1 & Y & S/L/F+/G & 2 & N & -- & -- & -- \\ 
279307075 & HD 139491 & B9IVe & 0 & / & / & / & / & / & / & / \\ 
279430029 & HD 53048 & B6Vne & 1 & Y & I & 0 & N & -- & -- & -- \\ 
280117704 & CoRoT 102667801 & B7V & 0 & / & / & / & / & / & / & / \\ 
280643259 & EM* RJHA 51 & B5Ib & 1 & Y & S/I/L/G & 3 & Y & 0.92 & 1.74 & $g1$ \\ 
280835427 & HD 83597 & B2Ve & 1 & Y & S/L/F+/G & 3 & Y & 1.45 & 2.69 & sim. \\ 
281047621 & ome Ori &  B2IIIe & 1 & Y & G & 3 & ? & 1.07 & 2.11 & sim. \\ 
281643356 & V907 Cen & B2.5Ve & 1 & Y & L/F+/G & 3 & Y & 1.06 & 2.34 & sim. \\ 
281717549 & HD 49787 & B1Ve & 1 & Y & L/G & 3 & Y & 1.69 & 3.13 & sim. \\ 
281741629 & BG Phe & B5e & 1 & Y & L/F+/G & 3 & Y & 1.04 & 1.96 & $g2$ \\ 
282207882 & BD-05 1837 & B7Ib/II & 1 & Y & I/H/G & 4 & Y & 1.35 & 2.88 & $g2$ \\ 
282306169 & HD 50209 & B8IVe & 1 & Y & I/G & 1 & N & -- & -- & -- \\ 
282808223 & HD 50820 & B3IVe & 1 & S & S/L/G & 1 & N & -- & -- & -- \\ 
282899554 & HD 50891 & B0.5Ve & 1 & Y & G & 2 & Y & 0.94 & 1.78 & $g2$ \\ 
283206453 & V746 Mon & B1.5IVe & 1 & Y & I/G & 4 & Y & 0.63 & 1.39 & $g2$ \\ 
284230347 & HD 55806 & B7IIIe & 1 & U & / & / & / & / & / & / \\ 
285187855 & HD 138477 & B7IIIe & 1 & Y & S/L/G & 1 & N & -- & -- & -- \\ 
285699059 & HD 74401 & B1IIIne & 1 & Y & I/H & 0 & N & -- & -- & -- \\ 
285916305 & HD 65930 & B2IIIe & 1 & Y & S/I/L/H & 0 & N & -- & -- & -- \\ 
286159211 & HD 69651 & B9Vne & 1 & Y & I & 0 & N & -- & -- & -- \\ 
289877581 & d Lup &  B3IVpe & 1 & N & / & / & / & / & / & / \\ 
290523979 & HD 306657 & B8e & 1 & Y & G & 3 & Y & 2.19 & 4.32 & sim. \\ 
290525745 & HD 306793 & B3Ve & 1 & Y & G & 4 & Y & 2.19 & 4.58 & $g2$ \\ 
290682353 & V911 Cen & Be & 1 & Y & S/L/F+/G & 1 & N & -- & -- & -- \\ 
290682406 & V855 Cen & Be & 1 & Y & G & 3 & Y & 1.13 & 2.13 & $g1$ \\ 
291385725 & HD 254647 & Bpe & 1 & Y & S/I/H & 0 & N & -- & -- & -- \\ 
291429783 & HD 127112 & B8Ve & 0 & / & / & / & / & / & / & / \\ 
292362364 & BD+01 1699 & B5/7Ib: & 1 & Y & G & 3 & Y & 0.57 & 1.21 & $g2$ \\ 
292550464 & HD 127515 & B7.5Vne & 1 & Y & I & 0 & N & -- & -- & -- \\ 
292674898 & V1008 Cen & B2.5Vne & 1 & Y & S/L/F+/G & 2 & Y & 1.50 & 3.00 & sim. \\ 
293290586 & HD 129842 & B8.5IVe & 1 & Y & S/I/L & 0 & N & -- & -- & -- \\ 
294125876 & HD 42477 & A0Vnne & 0 & / & / & / & / & / & / & / \\ 
294308857 & HD 253084 & B5e & 0 & / & / & / & / & / & / & / \\ 
294355648 & 69 Ori &  B5Vne & 0 & / & / & / & / & / & / & / \\ 
294553294 & CQ Cir & B1Ve & 1 & Y & G & 3 & Y & 1.20 & 2.50 & $g1$ \\ 
294852531 & HD 298339 & B2Vne & 1 & Y & I & 0 & N & -- & -- & -- \\ 
295099096 & QR Vel & B2Vne & 1 & Y & G & 4 & Y & 1.22 & 2.31 & sim. \\ 
295581098 & HD 298377 & B3Vne & 1 & Y & L/F+/G & 3 & Y & 1.18 & 2.50 & $g2$ \\ 
296969980 & tet Cir &  B4Vnpe & 1 & Y & G & 4 & Y & 2.11 & 4.45 & $g1$ \\ 
299022962 & HD 83043 & B2.5Vne & 1 & Y & L/F+/G & 4 & Y & 1.43 & 2.74 & $g2$ \\ 
299035061 & HD 120417 & B9IVne & 0 & / & / & / & / & / & / & / \\ 
299315060 & MWC 521 & B0V & 0 & / & / & / & / & / & / & / \\ 
300524274 & HD 146381 & B6IIIe & 1 & Y & I/G & 2 & N & -- & -- & -- \\ 
301435200 & HD 307350 & B2Ve & 1 & Y & I/H/V/G & 1 & N & -- & -- & -- \\ 
302348015 & BD+13 895 & B8 & 1 & Y & G & 4 & Y & 2.24 & 4.58 & sim. \\ 
302962039 & E Car &  B2IVe & 1 & Y & S/I/L/F/G & 1 & N & -- & -- & -- \\ 
305090822 & HD 157273 & B8.5Vne & 1 & N & / & / & / & / & / & / \\ 
305629042 & HD 103514 & B6e & 1 & Y & G & 3 & Y & 0.82 & 1.58 & $g2$ \\ 
305896243 & HD 103574 & B2Ve & 1 & Y & G & 3 & Y & 0.90 & 1.85 & sim. \\ 
306419817 & HD 103872 & B5e & 1 & Y & G & 4 & Y & 0.58 & 1.25 & sim. \\ 
307100246 & HD 104011 & Be & 1 & Y & L/F+/G & 3 & ? & 0.86 & 1.86 & sim. \\ 
307225534 & V767 Cen &  B2Ve & 1 & Y & S/L/F+/G & 4 & Y & 0.78 & 1.66 & sim. \\ 
308748912 & HD 68423 & B8Ve & 1 & Y & -- & 0 & N & -- & -- & -- \\ 
308760236 & HD 106730 & B0.5e & 1 & Y & L/F/G & 3 & Y & 0.85 & 1.88 & sim. \\ 
308951795 & HD 306145 & B2Vne & 1 & S & S/I/L/H/V/G & 1 & N & -- & -- & -- \\ 
309184049 & HD 106960 & B8IIIe & 1 & Y & G & 2 & Y & 1.59 & 3.22 & $g1$ \\ 
310645192 & V471 Car & B5ne & 1 & Y & L/F+/G & 4 & Y & 1.25 & 2.67 & $g1$ \\ 
311334494 & HD 107609 & B8.5IVe & 1 & Y & S/I/L/G & 1 & N & -- & -- & -- \\ 
312909005 & HD 116875 & B8Ve & 0 & / & / & / & / & / & / & / \\ 
312931751 & HD 116827 & B3.5Vne & 1 & Y & G & 3 & Y & 2.17 & 4.24 & sim. \\ 
313074666 & V967 Cen & B0IIe & 1 & Y & I/L/G & 1 & N & -- & -- & -- \\ 
313955481 & V969 Cen & B3Ve & 1 & Y & G & 3 & Y & 1.04 & 2.04 & $g2$ \\ 
314416691 & HD 146357 & B4IIIe & 1 & Y & L/G & 4 & Y & 1.55 & 3.00 & $g2$ \\ 
314868712 & HD 146444 & B2Vne & 1 & Y & L/G & 3 & Y & 0.95 & 1.86 & $g1$ \\ 
315036866 & HD 117357 & O9.5Ve & 1 & Y & I/H/G & 2 & ? & 0.78 & 2.00 & sim. \\ 
315207705 & HD 146501 & B6Vne & 1 & Y & I/G & 3 & Y & 2.91 & 5.80 & sim. \\ 
315679257 & HD 146596 & B5IVe & 1 & S & S/I/L/G & 0 & N & -- & -- & -- \\ 
315830698 & HD 99354 & B1IIIne & 1 & Y & I & 0 & N & -- & -- & -- \\ 
316279926 & WRAY 15-1119 & Be & 0 & / & / & / & / & / & / & / \\ 
316792722 & HD 99771 & B7Vne & 1 & S & I/G & 3 & Y & 0.61 & 1.23 & $g1$ \\ 
317666530 & HD 118094 & B8Vne & 0 & / & / & / & / & / & / & / \\ 
317795621 & HD 100199 & B0.5IIIne & 1 & Y & S/I/G & 4 & Y & 0.93 & 1.81 & sim. \\ 
319152335 & HD 57386 & B1.5Vnnpe & 1 & Y & S/I/H/G & 4 & ? & 1.84 & 3.10 & sim. \\ 
319730932 & BD+00 1516 & B9 & 1 & Y & I & 0 & N & -- & -- & -- \\ 
319854805 & HD 47359 & B0IVe & 1 & U & / & / & / & / & / & / \\ 
320228013 & HD 308829 & B5e & 1 & N & / & / & / & / & / & / \\ 
322104948 & V644 Cen & B3Ve & 1 & U & / & / & / & / & / & / \\ 
322233181 & HD 306962 & Oe & 1 & S & S/I/L/H & 0 & N & -- & -- & -- \\ 
323325708 & HD 102189 & B5e & 1 & Y & G & 4 & Y & 1.20 & 2.33 & $g2$ \\ 
323348984 & HD 37330 & B6Ve & 1 & Y & G & 4 & Y & 3.02 & 6.02 & $g1$ \\ 
323612875 & HD 102369 & B8.5IIIe & 1 & Y & I/L & 0 & N & -- & -- & -- \\ 
323613746 & HD 102352 & B2Vne & 1 & Y & F+/G & 3 & Y & 1.80 & 3.20 & $g1$ \\ 
324252945 & HD 121195 & B8IVne & 0 & / & / & / & / & / & / & / \\ 
324268119 & V801 Cen & B1Vne & 1 & Y & S/I/H/F+/G & 4 & Y & 1.50 & 3.07 & sim. \\ 
324940394 & HD 148907 & B6Ve & 1 & Y & I/H/G & 2 & N & -- & -- & -- \\ 
325170579 & j Cen &  B3Vne & 1 & Y & L/F+/G & 4 & Y & 1.76 & 3.39 & sim. \\ 
325237845 & HD 102742 & B3Ve & 1 & Y & L/F+/G & 4 & Y & 1.25 & 2.11 & $g2$ \\ 
325273334 & V987 Cen & B5Vne & 1 & Y & I/H/G & 3 & Y & 1.85 & 3.73 & $g1$ \\ 
326022066 & HD 164050 & B8Ve & 1 & Y & G & 5 & Y & 2.08 & 4.17 & $g1$ \\ 
327913813 & HD 122282 & B6pshe & 1 & Y & S/G & 3 & Y & 1.31 & 2.61 & $g2$ \\ 
329059929 & HD 122691 & Be & 1 & Y & I/G & 1 & N & -- & -- & -- \\ 
329060936 & HD 122669 & Be & 1 & Y & G & 2 & Y & 0.94 & 2.10 & $g1$ \\ 
330435560 & HD 123131 & B6.5Ve & 1 & Y & I/G & 1 & N & -- & -- & -- \\ 
330719215 & HD 123207 & B6Vne & 0 & / & / & / & / & / & / & / \\ 
333670665 & V863 Cen &  B6IIIe & 1 & N & / & / & / & / & / & / \\ 
334776134 & HD 91120 &  B8.5IVe & 1 & Y & S/G & 3 & Y & 1.33 & 2.52 & $g1$ \\ 
335092511 & HD 124489 & B8IVe & 0 & / & / & / & / & / & / & / \\ 
335444921 & CD-44 9840 & B8V & 0 & / & / & / & / & / & / & / \\ 
336343562 & LS Mus &  B2IVne & 1 & Y & L/F+/G & 5 & Y & 1.83 & 3.73 & sim. \\ 
338532011 & CD-60 4299 & B2Ve & 1 & Y & G & 4 & Y & 0.61 & 1.26 & $g1$ \\ 
338738989 & HD 152060 & B2Ve & 1 & Y & I/H & 0 & N & -- & -- & -- \\ 
340633517 & HD 63453 & B9Vne & 1 & Y & I/G & 2 & Y & 2.01 & 4.01 & $g1$ \\ 
340878553 & HD 141689 & B2.5ne & 1 & Y & S/I/G & 2 & N & -- & -- & -- \\ 
341040849 & HD 64831 & B8Vne & 1 & Y & G & 3 & Y & 1.39 & 2.80 & $g1$ \\ 
342201547 & HD 69026 & B1.5Ve & 1 & Y & S/L/F+/G & 3 & Y & 2.12 & 3.67 & sim. \\ 
342257745 & HD 322422 & B1Ve & 1 & U & / & / & / & / & / & / \\ 
343376491 & MQ TrA & Oe & 1 & Y & S/L/F+/G & 2 & Y & 1.50 & 3.00 & $g2$ \\ 
344341901 & HD 150930 & B7IIIe & 1 & Y & S/G & 1 & N & -- & -- & -- \\ 
347225198 & HD 152358 & Bpshe & 1 & Y & S/G & 2 & Y & 0.86 & 1.85 & sim. \\ 
347529796 & HD 152505 & B2.5Vne & 0 & / & / & / & / & / & / & / \\ 
347665026 & V846 Ara & B3Vnpe & 1 & Y & G & 3 & Y & 2.16 & 4.23 & sim. \\ 
350472809 & HD 147230 & B7Ve & 1 & Y & I/G & 2 & Y & 1.31 & 2.55 & sim. \\ 
355553423 & HD 79206 & B3.5Vne & 1 & Y & I/G & 7 & Y & 1.39 & 2.82 & sim. \\ 
355653322 & eps Tuc &  B8Ve & 1 & Y & I & 0 & N & -- & -- & -- \\ 
356521959 & HD 77147 & B8Ve & 1 & Y & S/I/G & 2 & Y & 1.27 & 2.54 & $g1$ \\ 
358467471 & HD 65663 & B8Ve & 1 & Y & I & 0 & N & -- & -- & -- \\ 
358818468 & HD 80459 & B6Vne & 1 & Y & G & 5 & Y & 1.86 & 3.71 & $g1$ \\ 
363236260 & HD 84523 & B4Ve & 1 & Y & G & 5 & Y & 0.54 & 1.10 & $g1$ \\ 
363748801 & HD 149671 &  B7IVe & 1 & Y & G & 4 & Y & 1.56 & 3.13 & $g1$ \\ 
364398342 & V374 Car &  B2IVnpe & 1 & Y & L/F+/G & 3 & Y & 1.24 & 2.39 & $g2$ \\ 
369312341 & HD 111054 & B7IVe & 1 & Y & I/G & 2 & Y & 1.47 & 2.96 & $g1$ \\ 
372913367 & V373 Car & Be & 1 & Y & I/H & 0 & N & -- & -- & -- \\ 
374690639 & HD 87366 & B9IIIe & 1 & Y & I/L & 0 & N & -- & -- & -- \\ 
374753420 & HD 87543 & B7IVne & 1 & Y & I/G & 4 & Y & 0.56 & 1.17 & sim. \\ 
375232307 & lam Pav &  B2Ve & 1 & Y & I/L/G & 1 & N & -- & -- & -- \\ 
376077639 & V862 Ara &  B7IIIe & 1 & U & / & / & / & / & / & / \\ 
376480349 & HD 104582 & B8.5Ve & 1 & Y & G & 2 & Y & 1.10 & 2.10 & $g1$ \\ 
378812108 & HD 104552 & B4e & 1 & Y & G & 4 & Y & 1.14 & 2.17 & sim. \\ 
378996846 & DG Cru & B2Vne & 1 & Y & G & 3 & Y & 0.84 & 1.70 & $g1$ \\ 
380117288 & AI Cru & B2IVe & 0 & / & / & / & / & / & / & / \\ 
380224370 & Cl* NGC 4103 SBW 5 & B8e & 0 & / & / & / & / & / & / & / \\ 
381257866 & HD 105675 & Be & 1 & Y & S/L/G & 3 & N & -- & -- & -- \\ 
381641106 & CSI-62-12087 & Be & 1 & N & / & / & / & / & / & / \\ 
381747495 & HD 105753 & B2.5e & 1 & S & S/L/H/G & 2 & N & -- & -- & -- \\ 
382396569 & CD-62 623 & Be & 1 & Y & G & 4 & Y & 0.64 & 1.33 & $g1$ \\ 
382611398 & HD 106262 & B8.5Ve & 0 & / & / & / & / & / & / & / \\ 
382750740 & CD-59 4161 & O9.5Ve & 1 & Y & I/F+/G & 4 & Y & 1.22 & 2.45 & $g2$ \\ 
382795601 & HD 144965 & B3Vne & 1 & Y & G & 3 & Y & 2.62 & 5.40 & $g2$ \\ 
383162375 & HD 155352 & B2Ve & 1 & Y & L/F+/G & 3 & Y & 2.17 & 3.71 & $g2$ \\ 
383285551 & HD 77032 & B5Vne & 1 & Y & G & 4 & Y & 1.66 & 3.25 & $g2$ \\ 
383287960 & HD 76985 & B5Vne & 1 & Y & G & 1 & N & -- & -- & -- \\ 
383542260 & V807 Cen & B4e & 1 & Y & S/I/L/H/F+/G & 3 & Y & 1.95 & 3.80 & sim. \\ 
383763886 & CD-44 9598 & B6Ve & 1 & Y & L/G & 3 & Y & 0.78 & 1.60 & $g2$ \\ 
384471407 & HD 78328 & B9.5IIIe & 1 & Y & I/V/G & 1 & N & -- & -- & -- \\ 
385626507 & HD 79778 & B2Vne & 1 & Y & L/F+/G & 3 & Y & 0.82 & 1.65 & $g2$ \\ 
385873047 & FW CMa &  B2Vne & 1 & Y & L/G & 4 & Y & 1.02 & 1.88 & sim. \\ 
386024312 & HD 80284 & B5Vnne & 1 & Y & I/H/G & 6 & Y & 2.34 & 4.66 & $g1$ \\ 
389217256 & HD 126693 & B4Vne & 1 & Y & L/G & 3 & Y & 2.31 & 4.47 & sim. \\ 
389576843 & HD 169999 & B8Vne & 1 & Y & I/G & 5 & Y & 2.26 & 4.47 & $g1$ \\ 
390071999 & DK Cru & B2IVne & 1 & Y & S/F+/G & 3 & ? & 2.04 & 3.69 & sim. \\ 
390441711 & V518 Car &  B4Ve & 0 & / & / & / & / & / & / & / \\ 
394060863 & V1010 Cen & B2Vne & 1 & Y & L/G & 3 & Y & 1.06 & 2.13 & sim. \\ 
394728064 & DR Cha &  B4IVe & 1 & N & / & / & / & / & / & / \\ 
395429445 & HD 127756 & B1.5Vne & 1 & Y & G & 4 & Y & 0.89 & 1.89 & $g1$ \\ 
399669624 & 2 Ori &  A1Vne & 1 & N & / & / & / & / & / & / \\ 
399990331 & HD 128477 & B8.5IVe & 1 & Y & I/L & 0 & N & -- & -- & -- \\ 
400136687 & HD 76568 & B1Vnne & 1 & Y & I/G & 8 & N & -- & -- & -- \\ 
401635731 & V1012 Cen & B3Vne & 1 & Y & I/L/F/G & 1 & Y & 2.36 & 4.80 & $g1$ \\ 
405387285 & HD 106793 & B8.5IVe & 1 & Y & I/G & 5 & Y & 2.01 & 4.02 & $g1$ \\ 
405520863 & 39 Cru &  B6IVe & 1 & S & I & 0 & N & -- & -- & -- \\ 
405577964 & HD 111077 & Be & 1 & Y & G & 5 & Y & 1.56 & 3.21 & $g1$ \\ 
406341244 & V952 Cen & B2Vnne & 1 & Y & G & 3 & Y & 1.23 & 2.50 & $g1$ \\ 
408382023 & I Hya &  B5Ve & 1 & Y & I/H/G & 3 & Y & 1.55 & 3.06 & $g1$ \\ 
408757239 & V716 Cen & B5Ve & 1 & N & / & / & / & / & / & / \\ 
409358619 & V795 Cen &  B4Vne & 1 & Y & H/G & 4 & ? & 2.58 & 5.75 & sim. \\ 
409791884 & HD 71072 & B4IIIe & 1 & Y & G & 4 & Y & 0.82 & 1.72 & sim. \\ 
410599289 & HD 145846 & B1.5ne & 1 & Y & S/L/F+/G & 4 & Y & 1.39 & 2.94 & $g1$ \\ 
411643280 & HD 127782 & B8Ve & 1 & Y & G & 2 & Y & 1.78 & 3.53 & sim. \\ 
411933674 & TYC 1283-1360-1 & none & 0 & / & / & / & / & / & / & / \\ 
412140640 & HD 147274 & B8IIIe & 0 & / & / & / & / & / & / & / \\ 
413296783 & HD 125106 & A0Ve & 0 & / & / & / & / & / & / & / \\ 
413350702 & HD 125015 & B7Vne & 0 & / & / & / & / & / & / & / \\ 
419585280 & HD 110699 & B6Vne & 1 & Y & I/G & 2 & Y & 1.13 & 2.36 & $g1$ \\ 
420003853 & HD 117411 & B8Ve & 1 & Y & I & 0 & N & -- & -- & -- \\ 
421217840 & eps Aps &  B4Ve & 0 & / & / & / & / & / & / & / \\ 
423528378 & zet Crv &  B8Ve & 0 & / & / & / & / & / & / & / \\ 
424884512 & HD 140946 & B2.5Vnne & 1 & Y & F+/G & 3 & Y & 1.72 & 3.68 & sim. \\ 
425215861 & lam Eri &  B2IVne & 1 & Y & S/H/G & 4 & Y & 1.39 & 2.73 & $g1$ \\ 
425657630 & kap01 Aps &  B2Vnpe & 0 & / & / & / & / & / & / & / \\ 
426683285 & CD-56 6163 & Oe & 1 & Y & I/L/H/G & 4 & Y & 1.35 & 2.85 & $g1$ \\ 
427395049 & 43 Ori &  O9.5Vpe & 0 & / & / & / & / & / & / & / \\ 
427396133 & HD 37115 & B6Ve & 1 & Y & I/H/G & 1 & N & -- & -- & -- \\ 
427400331 & HD 290662 & A0Vpe & 1 & S & L/V/G & 2 & Y & 1.05 & 2.12 & sim. \\ 
427451587 & HD 37149 & B8Ve & 1 & Y & G & 2 & Y & 3.10 & 6.23 & sim. \\ 
429143022 & HD 101142 & B5IIIe & 1 & Y & I/G & 2 & Y & 1.20 & 2.40 & $g2$ \\ 
430248409 & V364 Nor & B1Vne & 1 & Y & S/L/F/G & 2 & Y & 2.15 & 3.79 & sim. \\ 
433936219 & BZ Cru &  B0.5IVpe & 1 & Y & S/I/L/H/G & 3 & Y & 1.52 & 2.93 & $g2$ \\ 
434244626 & HD 110625 & B8.5IIIe & 1 & Y & G & 3 & Y & 1.01 & 1.99 & sim. \\ 
434254190 & BQ Cru & Be & 0 & / & / & / & / & / & / & / \\ 
434998753 & HD 111363 & B3IIIe & 1 & Y & G & 3 & Y & 1.10 & 2.11 & $g1$ \\ 
435101969 & HD 111408 & B7IVe & 0 & / & / & / & / & / & / & / \\ 
436378066 & HD 312101 & B8e & 1 & Y & I/G & 2 & Y & 2.45 & 4.94 & sim. \\ 
436802660 & HD 81354 & B4Ve & 1 & Y & S/I/L/F+/G & 2 & Y & 1.80 & 3.33 & $g2$ \\ 
437371531 & V480 Car & B2.5Ve & 1 & Y & S/L/F/G & 3 & Y & 0.77 & 1.60 & $g2$ \\ 
437790952 & 120 Tau &  B2IVe & 0 & / & / & / & / & / & / & / \\ 
438103655 & HD 44637 & B2Vpe & 1 & Y & S/I/L & 0 & N & -- & -- & -- \\ 
438306275 & PZ Gem & O9pe & 1 & Y & S/I/L & 0 & N & -- & -- & -- \\ 
438877747 & V946 Cen & B6IIIne & 1 & Y & G & 2 & Y & 0.87 & 1.65 & sim. \\ 
439164152 & CPD-63 2495 & B2e & 1 & Y & S/L/G & 1 & N & -- & -- & -- \\ 
439331111 & HD 83060 & B2Vnne & 1 & Y & G & 4 & Y & 1.38 & 2.63 & sim. \\ 
439397894 & 2 Cet &  B9IVne & 1 & Y & S/G & 3 & Y & 1.43 & 2.87 & $g1$ \\ 
439430953 & HD 83032 & B7IIIe & 1 & Y & I/G & 2 & Y & 0.82 & 1.85 & $g1$ \\ 
440399815 & HD 113605 & B2.5ne & 1 & S & S/L & 0 & N & -- & -- & -- \\ 
441752709 & HD 84361 & B2.5Ve & 1 & Y & L/F+/G & 4 & Y & 1.46 & 2.66 & $g2$ \\ 
442240473 & HD 84511 & Bpshe & 1 & N & / & / & / & / & / & / \\ 
442759685 & HD 84777 & B8Vne & 1 & Y & I/G & 2 & Y & 0.94 & 1.91 & $g1$ \\ 
443126408 & FR CMa &  B1Vpe & 1 & Y & G & 4 & Y & 0.99 & 1.93 & $g2$ \\ 
443175014 & HD 85083 & B5IIIe & 1 & Y & L/F+/G & 3 & Y & 2.30 & 4.20 & $g2$ \\ 
443317662 & HD 45260 & B8e & 1 & Y & G & 3 & Y & 2.75 & 5.24 & sim. \\ 
443616529 & phi Leo &  A7IVne & 1 & N & / & / & / & / & / & / \\ 
444676711 & HD 85495 & B4IIIe & 1 & Y & G & 4 & Y & 0.71 & 1.43 & $g1$ \\ 
445988959 & HD 53416 & B8e & 0 & / & / & / & / & / & / & / \\ 
447970564 & CD-45 4826 & Be & 1 & Y & G & 5 & Y & 0.89 & 1.80 & $g1$ \\ 
449365065 & V1369 Ori & B5Vpe & 0 & / & / & / & / & / & / & / \\ 
449459169 & V1171 Cen & B3IVe & 1 & Y & G & 4 & Y & 1.04 & 2.27 & $g1$ \\ 
450276053 & V338 Car & B9e & 1 & N & / & / & / & / & / & / \\ 
450609787 & HD 108295 & B2.5Vne & 1 & Y & G & 4 & Y & 2.05 & 4.10 & sim. \\ 
450726617 & HD 108418 & B5Vne & 0 & / & / & / & / & / & / & / \\ 
450826315 & CPD-61 3212 & Be & 1 & Y & G & 4 & Y & 0.91 & 1.84 & sim. \\ 
451280762 & HD 99146 & B3Ve & 1 & S & S/I/H & 0 & N & -- & -- & -- \\ 
451445219 & HD 99467 & B9IIIe & 1 & Y & I/L/G & 5 & Y & 0.69 & 1.44 & $g1$ \\ 
451875177 & HD 304395 & B8e & 1 & Y & G & 3 & Y & 1.71 & 3.37 & sim. \\ 
452163402 & A Cen &  B9Ve & 1 & Y & -- & 0 & N & -- & -- & -- \\ 
454081779 & HD 102383 & B6Vne & 1 & Y & I/G & 2 & Y & 1.64 & 3.30 & $g1$ \\ 
454559349 & CU Cir & B3Vne & 0 & / & / & / & / & / & / & / \\ 
454778372 & CV Cir & B1.5IVne & 1 & Y & L/F+/G & 3 & Y & 1.24 & 2.45 & sim. \\ 
455389730 & HD 131168 & B2Ve & 1 & Y & I/L/H/G & 3 & Y & 3.00 & 5.80 & $g1$ \\ 
455463415 & HD 135160 &  B0.5Ve & 1 & N & / & / & / & / & / & / \\ 
455809360 & CD-61 4751 & Oe & 1 & U & / & / & / & / & / & / \\ 
456446715 & HD 136860 & B8IVe & 1 & Y & I/H & 0 & N & -- & -- & -- \\ 
457546452 & HD 126986 & B9IVne & 1 & U & / & / & / & / & / & / \\ 
459065293 & HD 93563 &  B8.5IIIe & 0 & / & / & / & / & / & / & / \\ 
460703129 & DQ Cru & B2IIIe & 1 & Y & I & 0 & N & -- & -- & -- \\ 
460976315 & HD 137380 & B9IIIe & 1 & Y & S/I & 0 & N & -- & -- & -- \\ 
463103957 & QY Car &  B2IVnpe & 1 & Y & L/F+/G & 3 & Y & 1.31 & 2.47 & sim. \\ 
464028809 & J Vel &  B3IIIe & 1 & Y & I/L/G & 3 & ? & 0.45 & 0.80 & sim. \\ 
464643264 & CCDM J19106-6003AB & B9II/III & 0 & / & / & / & / & / & / & / \\ 
466353921 & HD 95972 & B2Vnne & 1 & Y & S/I/G & 20 & N & -- & -- & -- \\ 
466527712 & HD 306016 & B2Ve & 1 & Y & I/L/F+/G & 3 & Y & 1.56 & 3.21 & $g2$ \\ 
466628058 & HD 96357 & B4Ve & 1 & Y & G & 3 & Y & 1.56 & 3.09 & $g1$ \\ 
466715331 & HD 308217 & B3Ve & 1 & S & S/I/L/H & 0 & N & -- & -- & -- \\ 
466880435 & HD 306085 & B2Ve & 1 & Y & G & 4 & Y & 0.75 & 1.50 & $g1$ \\ 
467027607 & HD 306111 & Oe & 1 & Y & S/I/L/H/V/G & 3 & Y & 1.83 & 3.69 & $g1$ \\ 
467059272 & V353 Car & B2Ve & 1 & Y & L/F+/G & 3 & Y & 1.89 & 3.52 & $g2$ \\ 
467065657 & HD 97253 & O5IIIe & 1 & N & / & / & / & / & / & / \\ 
467199933 & HD 97382 & B6e & 1 & Y & G & 5 & Y & 0.51 & 0.95 & sim. \\ 
467408384 & HD 306209 & B1Ve & 1 & Y & S/L/F+/G & 3 & Y & 1.60 & 2.80 & $g2$ \\ 
467409163 & HD 306205 & B0IIe & 1 & Y & G & 5 & Y & 1.37 & 2.68 & $g1$ \\ 
467456980 & HD 306267 & Oe & 1 & Y & G & 4 & Y & 1.63 & 3.37 & sim. \\ 
467821546 & HD 306299 & B2Ve & 1 & Y & I/G & 3 & Y & 1.98 & 4.04 & $g2$ \\ 
468074636 & HD 98624 & B1Vne & 1 & Y & S/I/H & 0 & N & -- & -- & -- \\ 
468095832 & 2E 1118.7-6138 & O9.5IVe & 1 & Y & S/I/H & 0 & N & -- & -- & -- \\ 
468195912 & HD 98876 & B5Vne & 1 & Y & I/L/G & 3 & Y & 1.71 & 3.44 & $g1$ \\ 
\enddata
\end{deluxetable*}

\bibliography{main}{}
\bibliographystyle{aasjournal}

\end{document}